\documentclass[10pt,twoside,a4paper,fleqn]{report}

\usepackage[english,mt]{ethidsc} 
\usepackage{tabulary}
\usepackage{booktabs}
\usepackage{amsmath}
\usepackage{relsize}
\usepackage{graphics}

\usepackage{tikz}
\usepackage{collcell}
\usepackage{etoolbox}

\usepackage{adjustbox}

\usepackage{pgfplotstable}

\usepackage[font=small, skip=4pt]{caption}

\newcommand{\vect}[1]{\boldsymbol{#1}}
\DeclareMathOperator*{\argmin}{arg\,min} 
\DeclareMathOperator*{\argmax}{arg\,max} 

\providecommand{\norm}[1]{\lVert#1\rVert}

\title{Machine Learning and\\
Deep Learning methods for predictive modelling from Raman spectra in bioprocessing}

\studentA{Semion Rozov}
\ethidA{12-938-973}
\emailA{rozovs@ethz.ch}

\supervision{Dr. Michael Sokolov \\
Dr. Fabian Feidl \\
Prof. Dr. Gonzalo Guillen Gosalbez}

\date{February 2020}

\begin{document}

\maketitle 							


\pagenumbering{roman} 				



\pagestyle{fancy}               	
\pagenumbering{arabic}				

\cleardoublepage
\chapter*{Abstract}
\addcontentsline{toc}{chapter}{Abstract}

In chemical processing and bioprocessing, conventional online sensors are limited to measure only basic process variables like pressure and temperature, pH, dissolved O and CO$_2$ and viable cell density (VCD). The concentration of other chemical species is more difficult to measure, as it usually requires an at-line or off-line approach. Such approaches are invasive and slow compared to on-line sensing. It is known that different molecules can be distinguished by their interaction with monochromatic light, producing different profiles for the resulting Raman spectrum, depending on the concentration. Given the availability of reference measurements for the target variable, regression methods can be used to model the relationship between the profile of the Raman spectra and the concentration of the analyte.
\\This work focused on pretreatment methods of Raman spectra for the facilitation of the regression task using Machine Learning and Deep Learning methods, as well as the development of new regression models based on these methods.
\\In the majority of cases, this allowed to outperform conventional Raman models in terms of prediction error and prediction robustness.
\cleardoublepage

\thispagestyle{empty}
\tableofcontents
\listoffigures
\listoftables

\cleardoublepage

\chapter{Introduction}\label{sec:myintroduction}

In chemical processing and bioprocessing, conventional online sensors are limited to measure only basic process variables like pressure and temperature, pH, dissolved O and CO$_2$ and viable cell density (VCD) \cite{Vojinovi2006, Zhao2015,Glassey2011}. The concentration of other chemical species is more difficult to measure, as it usually requires an at-line or off-line approach. Such approaches are invasive and slow compared to on-line sensing.\\
Modern Raman spectroscopy, on the other hand, offers a basis for real-time sensing of process variables. \cite{Biechele2015, vanDerMaaten2008}.
Furthermore, such methods can be implemented into closed-loop process-control strategies such as Model Predictive Control \cite{Matthews2016, Berry2015,Craven2014}.
Successful monitoring of important process variables such as VCD, glucose and lactate concentrations as well as product titer and others has been reported. \cite{FeidlF19}\\
The fundamental idea of this method is that different molecules can be distinguished by their interaction with monochromatic light (inelastic scattering), producing different profiles for the resulting Raman spectrum, depending on the concentration. Given the availability of reference measurements for the target variable, regression methods can be used to model the relationship between the profile of the Raman spectra and the concentration of the analyte. Together with the real-time acquisition of the spectrum, this allows for an online measurement of multiple concentrations of interest simultaneously. \cite{https://doi.org/10.3929/ethz-b-000381363,FeidlF19,Gautam2015}.\\
The regression task can be easily solved for univariate problems where the spectrum is produced by single molecules for pure target solutions. However, for more complex macro molecules, such as proteins and mixtures of process variables, the solution is non-trivial because of the overlapping contributions of each component to the measured Raman spectrum. Furthermore, this problem gains additional complexity due to the possible interactions of the species and impurities affecting the spectrum.\\
Successful development of Raman models requires the use of elaborated spectral preprocessing techniques, wavelength selection as well as outlier removal. \cite{Gautam2015,Lasch2012,CHEN2005,deGroot2001}. The careful selection of the training and test sets also plays a crucial role for the robust calibration of the models and their generalization capability. \cite{GALVAO2005}\\
Regarding model choice, previous studies have reported successful application of Machine Learning methods, such as Gaussian Processes (GP), Support Vector Regression (SVR) and Random Forests (RF) on spectral data, which have shown advantages compared to the Partial Least Squares method normally used for solving such problems exhibiting high multicollinearity. \cite{https://doi.org/10.3929/ethz-b-000381363,FeidlF19,Cui2017,Tange2015}.\\
Deep Learning methods such as Convolutional Neural Networks were also previously considered for solving this type of problem. \cite{https://doi.org/10.3929/ethz-b-000381363,FeidlF19,Acquarelli2017}. However, their efficient training proved to be challenging for tasks with limited training sample availability. \cite{Tange2017}. This is a possible manifestation of the curse of dimensionality, which usually designates the increasing difficulty of fitting nonlinear models when the dimensionality of the training data is increased. Possible ways of countering this effect, which were tested in this work for the case of regression on Raman spectral data, are synthetic data augmentation and input feature selection based on a feature importance analysis.\\
Another solution for dimensionality reduction, is proposed by applying regression on low-dimensional latent representations obtained with Variational Autoencoders, which were previously reported to show promising results for regression on Magnetic Resonance Image data.\cite{DBLP:journals/corr/abs-1904-05948}\\
Further improvement methods leveraged fused predictions from multiple simultaneously trained models.

\section{Goals of the project}\label{sec:goals}

This work focused on pretreatement methods of Raman spectra for the facilitation of the regression task using Machine Learning (ML) and Deep Learning (DL) methods, as well as the development of new regression models based on these methods. The goal was to outperform conventional Raman models in terms of prediction error and prediction robustness. The described methods were tested on two distinct datasets for the prediction of four different target variables, essential to the monitoring of bioprocesses.

\bigbreak

First, the evaluation metrics which are used to compare the different models will be presented. Then the two datasets will be briefly described, followed by the introduction of the used preprocessing methods established by \cite{https://doi.org/10.3929/ethz-b-000381363,FeidlF19} as well as newly proposed methods for data augmentation and variable selection. Next, the different ML and DL models will be introduced, describing the different models' working principles and hyperparemeter choice.
Finally, the results will be presented for each target variable in a different section, jointly comparing the models performance on both datasets and for different improvement settings.

\cleardoublepage

\chapter{Experimental Section and Setup }\label{sec:experimentalsection}
This chapter describes the datasets used in the experiments, the pretreatment methods and scoring methods used in this work.

\section{Evaluation criteria and model selection}\label{sec:criteria}
\subsection{Metrics and Loss Functions}\label{sec:metrics}
An error metric is a scoring method designed to measure a predictive model's performance and penalize errors. Thus choosing a metric has direct impact on model comparison and selection.\\
The scoring method used for the training of a model is called loss function, or also sometimes, cost function. It is the objective function that needs to be minimized by a numerical optimization algorithm during the training iterations.\\
Normally, the metric and loss function are either the same or closely related. The reason for picking a different loss from the metric is usually computational efficiency considerations. Furthermore, additional terms may be added to the optimization objective, for example for outlier filtering or regularization. Both will be discussed in further sections.
\\
Two similar models can have contradicting scores for two different metrics, depending on the loss function which was used for the optimization (see Tab. \ref{tab:lossfunctions} for a list commonly used in regression tasks).
This will be demonstrated shortly in a simple toy example.
In search and optimization, the \emph{No Free Lunch (NFL) Theorem} \cite{wolpert_macready_1997} is often interpreted as follows :

\begin{quote}
   \emph{"A general-purpose, universal optimization strategy is impossible. The only way one strategy can outperform another is if it is specialized to the structure of the specific problem under consideration"} \cite{ho_pepyne_2002} 
\end{quote}{}


\begin{table}[h]
\centering
\caption{Loss functions for regression problems}
\label{tab:lossfunctions}
\begin{tabulary}{\textwidth}{L R @{} >{${}}c<{{}$} @{} L}

\toprule
Mean squared error & MSE &= &$\displaystyle\frac{1}{n}\sum_{t=1}^{n}e_t^2$   \\
\midrule
Root mean squared error & RMSE &= &$\displaystyle\sqrt{\frac{1}{n}\sum_{t=1}^{n}e_t^2}$ \\
\midrule
Mean absolute error & MAE &= &$\displaystyle\frac{1}{n}\sum_{t=1}^{n}|e_t|$ \\
\midrule
Mean absolute percentage error & MAPE &= &$\displaystyle\frac{100\%}{n}\sum_{t=1}^{n}\left |\frac{e_t}{y_t}\right|$\\
\bottomrule

\end{tabulary}

\end{table}

For an exemplary case with the following data :
\begin{verbatim}
    0.0001 0.0002 0.0003 0.0004 50000 0.0004 0.0004 0.0003 0.0002 0.0001
\end{verbatim}
when fitting a simple regression model with intercept $\alpha$, the Mean Squared Error (MSE) solution would be $\alpha = 5000.00023$ whereas the Mean Absolute Error (MAE) solution would be $\alpha = 0.0003$. \\
MSE leads to faster convergence as it penalizes the errors quadratically (larger gradients compared to MAE), however, from the above example it can be seen that it is very sensitive to outliers. In that respect MAE is more robust. Fortunately, there is a loss function which combines the best of both:
\bigbreak
\begin{minipage}[h]{\textwidth}
\centering

\begin{equation}
Huber Loss = \left\{\begin{array}{cl}
\frac{1}{2} \left[e_t\right]^2 & \text{for }|e_t| \le \delta, \\
\delta \left(|e_t|-\delta/2\right) & \text{otherwise.}
\end{array}\right.
\end{equation}
\end{minipage}
\bigbreak

As can be seen from the above formula, this requires correct estimation of $\delta$, which is quite difficult in practice, as it defines whether to consider a point an outlier or not. Originally, default parameters were used, which should normally depend on the target variable (because of different ranges).

In the course of this project, all of the above loss functions were tested and it was decided to settle for MSE instead of the Huber Loss as the main loss function for optimization due to it's universal nature, but also because it was decided to treat the issue of outliers separately (see subsection \ref{subsec:outlierremoval})
For metrics, the Root Mean Squared Error (RMSE) and the Normalized Root Mean Squared Error (N RMSE) were used, with:

\bigbreak
\begin{minipage}[h]{\textwidth}

\centering
\begin{equation}
    \text{N RMSE} = \displaystyle\frac{\text{RMSE}}{\sigma_{\text{N RMSE}}}
\end{equation}

\end{minipage}

\bigbreak
The reason for this choice is the enhanced interpretability of such an error, as it is decoupled from the variance of the sample set. Thus performance can be compared across multiple targets.

\newpage
\subsection{Model selection and evaluation}
\begin{figure}[h]
  \begin{minipage}[h]{0.98\textwidth}
    \includegraphics[width = \textwidth]{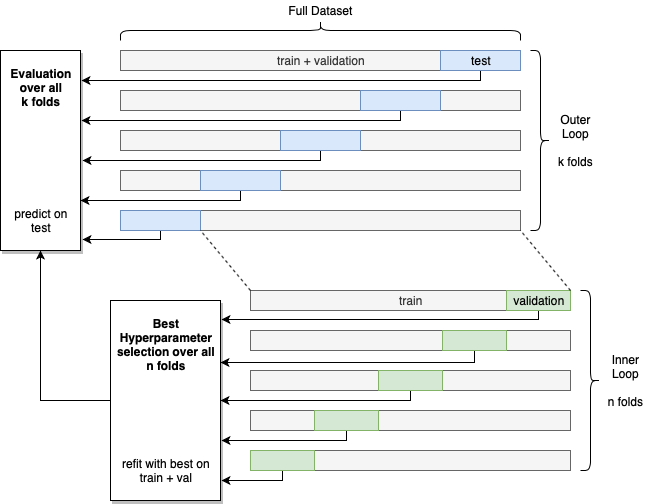}
  \end{minipage}
\caption{5-Fold Nested Cross-validation used to train and test models. The inner loop used for hyper parameter selection, the outer loop is used for model evaluation}
\label{fig:nestedcv}
\end{figure}

Due to the scarcity of data for training and testing, technique called \emph{Nested Cross-validation} was used to check if a model performing well on the validation set also performs well on the test set.\cite{Varma2006} This process is depicted in Fig. \ref{fig:nestedcv}: First, the entire dataset is divided into $k=5$ folds.\footnote{$k=5$ for Glucose, Lactate and VCD targets. For Titer, due to counting of the relative fold lengths in entire batches instead of fractions of batches, close to equal-length folds could be better achieved for $k=6$}. The exact splits are described in chapter 6, sections \ref{sec:resultsTiter} and \ref{sec:resultsLactate}. One fold is held out for testing (test set = 20\% of entire set) while the rest was divided into $n=5$ folds. Here again, one fold was held out for validation (validation set = 20\% of the remaining set or 16\% of the entire set) and the rest is used for training (training set = 64\% of entire set). This is done iteratively n times for all validation folds. The aim of this inner cross-validation loop is to find the set of hyperparameters for which the model has best performance on the validation data. The best model is the refitted on the test + validation set and evaluated on the previously unseen test set. The whole procedure is iteratively repeated with the next test folds. This outer loop serves to provide a robust estimate of the model's performance on previously unseen data.\\

Some adjustments had to be made to the originally implemented algorithm\cite{https://doi.org/10.3929/ethz-b-000381363}: To make train and test splits more similar to each other, the dataset was shuffled batchwise with the following scheme:

\begin{enumerate}
    \item First, the length of a fold should be calculated from the relative split, call it \texttt{len\_fold}
    
    \item Then, starting from the first batch every \texttt{len\_fold}-th batch would be added to the test fold until it's full
    \item The same would be repeated starting from the 2nd batch until reaching the \texttt{(len\_fold-1)}-th batch, which would complete the last test fold.
\end{enumerate}

The reader is referred to the Python code in Appendix A. 1 for the exact implementation of the described shuffling algorithm.

Batchwise shuffling was only used to obtain the test folds for the outer cv-loop, to ensure realistic testing scenarios. In the inner cv-loop, on the other hand, the set was shuffled sample-wise.

As a result, every point in the data set is used for prediction. The obtained predictions over all folds were aggregated in two different ways:

\begin{enumerate}
    \item N RMSE over all measurements. Predictions were collected from all k iterations to calculate this error, therefore it measures the average performance over the whole dataset)
    \item N RMSE averaged over the k test folds. It is used for reference to the above measurement.
\end{enumerate}

Additionally, a confidence bound was calculated for these scores in form of the standard deviation (normalized) of the errors (absolute).
To obtain a better understanding of the predictive power of a model, the predictions were plotted together with the ground truth for every test fold.\\
The previously mentioned evaluation criteria were also calculated for the train sets to estimate potential overfitting of the models during model design.

\newpage
\section{Description of the datasets}\label{sec:descofthedatasets}
First of all, it is important to notice that the problem was formulated as a static regression task, the temporal evolution of the target variables was not taken into account, as the primary purpose of a Raman model is measurement and not forecasting.\\
The models were trained and evaluated on 1 large (DS - 2000 samples) and 1 small dataset (DS2 - 150 samples). The first consists of a wide variety of different batches with different peak concentrations, whereas the last two are assumed to consist of much more homogeneous batches (with similar run times and start and end values)
and the following target variables: concentrations (in \unitfrac{g}{L}) of Titer, Glucose, Lactate and Viable Cell Density (VCD, in \unitfrac{E5 cells}{mL}). A particular focus was given Titer models, mainly because of the convenient size and distribution of this dataset.\\

It was noticed that only Glucose seems to have a distribution which resembles a Gaussian distribution, whereas Titer and Lactate can be considered right tailed distributions. VCD can be described as a right tailed distribution with a local peak in the mid range.

\bigbreak


Model choice heavily depends on the size and distribution of the dataset. We will see in later sections that simple ML models fit well on small and homogeneous datasets, where some DL models struggle, which is due to the lack of training samples. Previously, it was assumed that DS13 would be large enough to produce a fit superior to ML models. However, after closer analysis, this proved to not always be the case: plotting the evolution of the target concentrations has shown that for the highest concentrations, only a few measurements were available. This suggests that some models might have difficulties to fit to data and correctly estimate target concentrations in this range.
Thus, the skewed nature of the distributions poses two challenges: On one hand, measurements forming the long tails might be erroneously considered outliers by a given estimator because of the low number of training examples in this area. On the other hand, augmenting the tails to compensate this lack of training examples might lead to the propagation of outliers.
We will see in upcoming sections, how the individual distributions could be transformed to achieve better results with DL models.


\newpage
\section{Preprocessing of the datasets}\label{sec:preprocessing}

\subsection{Outlier Removal}\label{subsec:outlierremoval}

At this point, it can be deemed necessary to cite a definition of the term \emph{outlier}: "An outlying observation, or \emph{outlier}, is one that appears to deviate markedly from other members of the sample in which it occurs" \cite{grubbs_1969}. In this work, this word refers to both measurements of the target variable and the corresponding spectral measurement, as detecting the exact origin of the outlier was not in the scope of this work. Depending on the nature and origin of the data, the dimension and extent of the above mentioned "deviation" from other members of the sample can be formulated in different ways, for example through the means of inter-quartile ranges or standard deviations. However, these conventional methods make the assumption that the underlying data is normally distributed, which, as we have seen in an earlier section, does not always hold. A different approach had to be found, which will be described in this section. 
\bigbreak

\begin{figure}[h]
    \centering
  \begin{minipage}[t]{0.98\textwidth}
        \includegraphics[width=\linewidth]{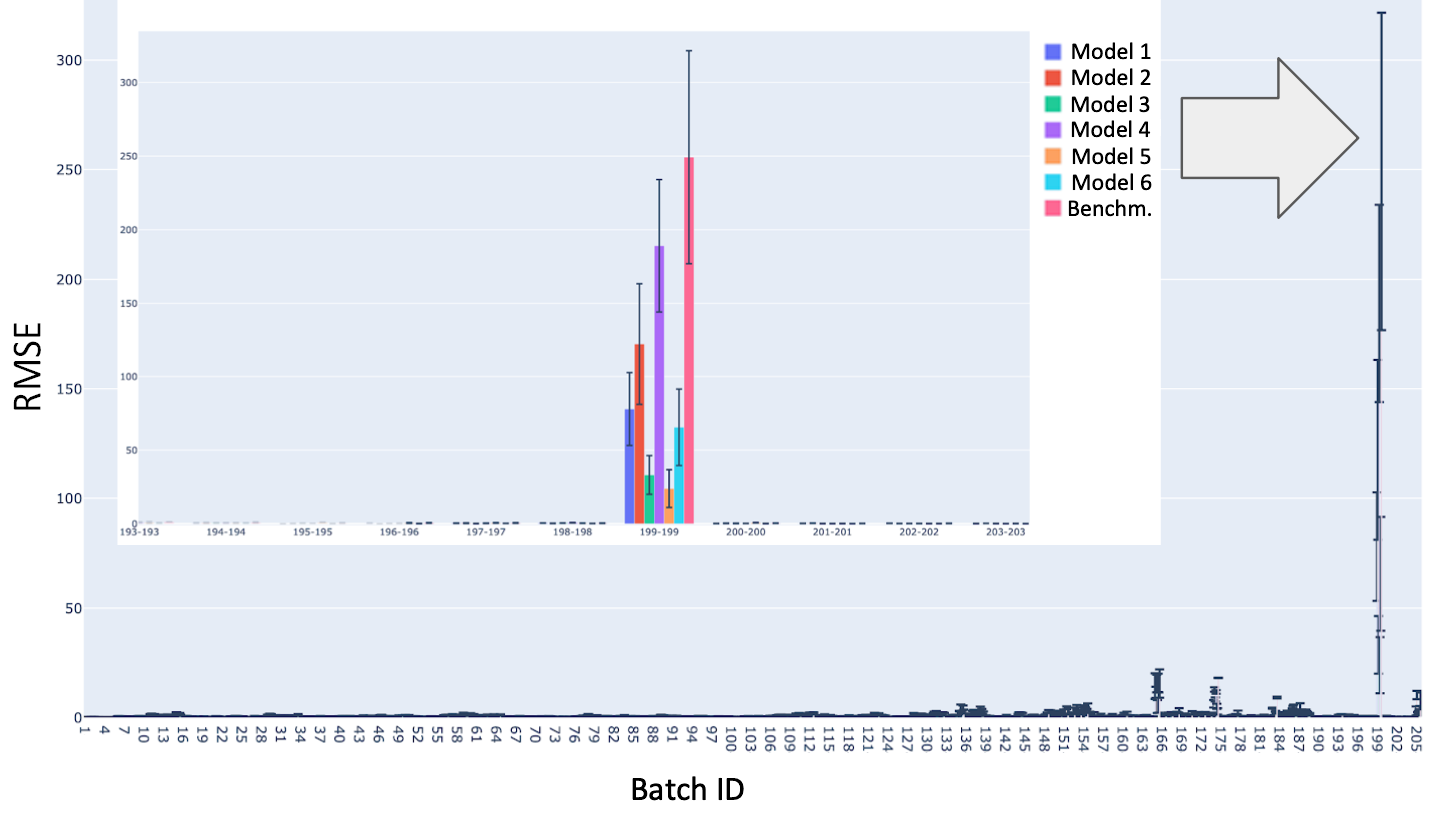}
        \caption{Batchwise evaluation of different models on DS13 for Glucose. Test scores shown are N RMSE. Note abnormaly high N RMSE values for all models on batch 199 in comparison to the performance on other batches.}
         \label{img:ds13Gluoutlierperbatch}
    \end{minipage}
\end{figure}

Throughout the course of this project, it was noticed that all evaluated types of models performed equally bad on certain batches. It was unclear if the spectral data of those batches was corrupted by an unwanted reaction or faulty process. Thus, it was decided to perform an additional cleanup of the data. The test performance on every batch was evaluated iteratively (tested one by one), while training on the remaining dataset. An example of the results is shown in Fig. \ref{img:ds13Gluoutlierperbatch}. The fact that also the benchmark model performed poorly on batch 199 suggests that the measurements originating from this batch can be treated as outliers and may be removed from the set. In the illustrated example, such a reasoning is obvious. However, for different target data the conclusions is not always that straightforward and require a more detailed analysis.

It was therefore decided to remove data points by setting different thresholds for the test performance of a reference model (an earlier Neural Network model). Examples are : Mean RMSE, 1-, 2- and 3-standard deviations of RMSEs. In the final evaluation the 1-std RMSE threshold was used, but tested only for Titer prediction.

\newpage
\subsection{Preprocessing pipeline}\label{subsec:pp_pipeline}
One of the first important steps in an ML process is the preprocessing of the data. In order to assure model robustness, this preprocessing has to be reproducible. There are two approaches to this problem. Either preprocess the data only once before the model fitting and selection process or implement it into the model itself as a preprocessing pipeline: this is a trade-off is between computation time and ease-of-use. It was decided to chose the later as some of the  preprocessing steps require prior fitting (for example estimating the scaling parameters), which means that this routine processes should be only done on the train set and not the test set, which is automatically managed by the \texttt{sklearn.pipeline.Pipeline} class of the scikit-learn library. A pipeline essentially consists of a list of transforms and a final estimator at the end. As a consequence, a pipeline is itself an estimator inheriting all the methods of the final estimator.
\bigbreak
The used preprocessing steps were taken from the previously established pipeline \cite{https://doi.org/10.3929/ethz-b-000381363} with some minor adjustment and is ordered as follows :

\begin{description}
  \item[Smoothing: Savitzky-Golay Filter]
    As any signal, the captured raw spectrum carries an amount of noise which should be removed before further analysis. A popular method for smoothing this kind of signal is the Savitzky-Golay filter. \cite{Savitzky1964} It essentially tries to locally fit a low degree polynomial of defined segment width (window length) and polynomial order by solving the linear least squares problem on a subset of the data points (on each segment). A closed-form solution exists for the case of equally spaced data points.\\
    Additional smoothing is performed by a derivative step. While the window length and polynomial order essentially define the smoothing resolution, the derivative order aims to remove baseline shift and background noise.\\
    From past experiments and parameter tuning it is known that the following set of parameters yields best results : $windowlength=17$, $polyorder=2$, $deriv=1$
    \\Unlike in the previous implementation, it was decided to put this filter at the beginning of the pipeline and before the Drop Columns Filter (see following paragraph) as otherwise it would create discontinuities which would be misinterpreted by the Savitzky-Golay Filter as informative features.

  \item[Bandwidth: Drop Columns Filter]
    The captured spectrum has a large bandwidth, parts of which do not necessarily capture information relevant for the Raman model. The lower range ($< 450nm$) as well as the upper range ($>3100nm$) of the spectrum are limited by the specifications of the measurement device, as beyond these values it essentially captures noise. Additional noise originates from the analyser window material in the $[1820 - 1880nm]$ and $[2530 - 2590nm]$ range. These wavelengths were removed from all analysed spectra. Additional variables were removed following a feature importance analysis using the Extreme Gradient Boosting (XGB) algorithm, which will be discussed in the Regression Models chapter. The feature importance analysis was not part of the pipeline itself (because of time-efficiency considerations). In order to stay consistent with the original preprocessing, an additional instance of the Drop Columns Filter was placed just after the first, which would handle this additional feature removal (by default, no arguments are passed to this second Drop Columns Filter, which then has no effect).

  \item[Dynamic Range:] Spectral data has a very large dynamic range which has to be scaled down for numerical stability reasons: 
    \begin{description}
        \item[Standard Normal Variate (SNV) Normalizer:] Scales the spectrum row-wise by calculating the mean and standard deviation over the wavelengths, subtracting the mean from each wavelenghth and dividing it by the standard deviation. Its purpose is to remove potential differences in measurements originating from different analysers.
        \item[Standard Scaler:] Scales the spectrum column-wise, similar to the method described above.
    \end{description}
\end{description}

\newpage
\subsection{Dimensionality reduction}

\begin{figure}[h]
  \begin{minipage}[h]{0.98\textwidth}
    \includegraphics[width = \textwidth]{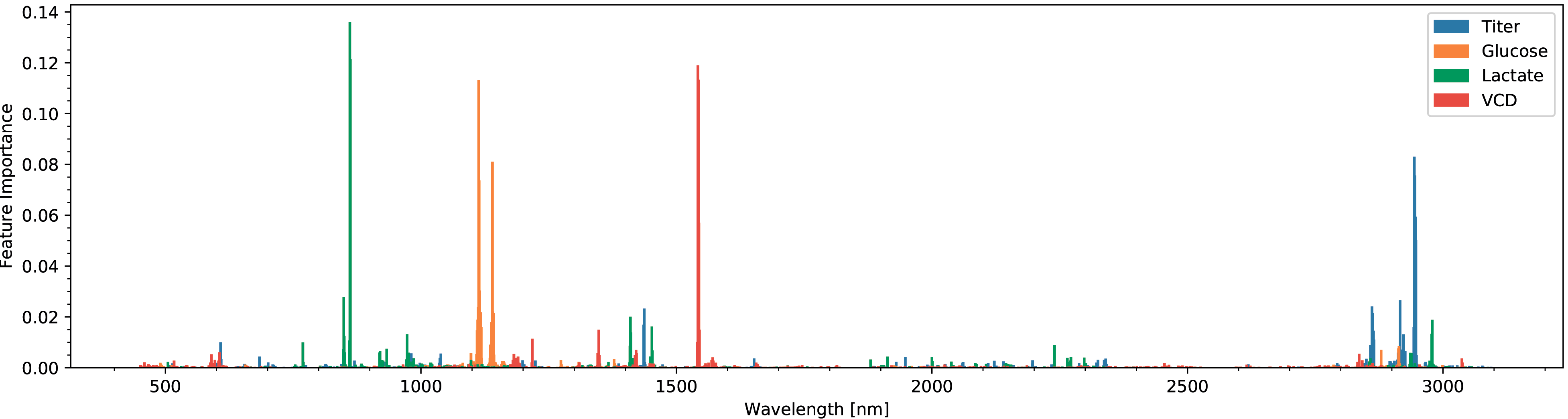}
  \end{minipage}
  \hfill
  \caption{Feature Importance of the target variables in DS13 determined with XGB.}
  \label{img:ds13varimportanceall}
\end{figure}

\begin{figure}[h]

  \begin{minipage}[h]{0.38\textwidth}
  a)
    \includegraphics[width = \textwidth]{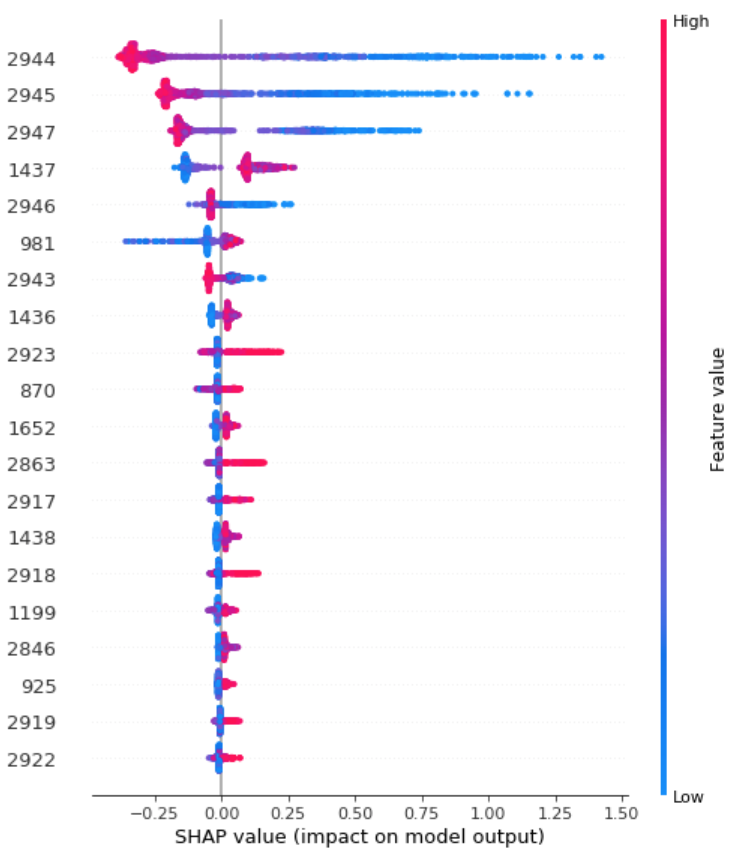}
  \end{minipage}
\begin{minipage}[h]{0.58\textwidth}
    \includegraphics[width = \textwidth]{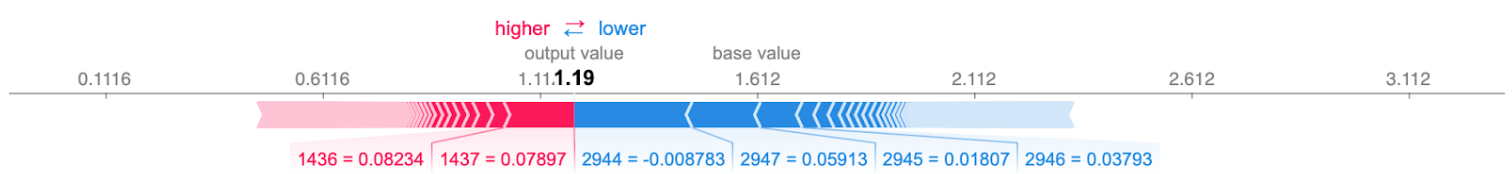}
    b)
    \bigbreak
      \hfill
    \includegraphics[width = \textwidth]{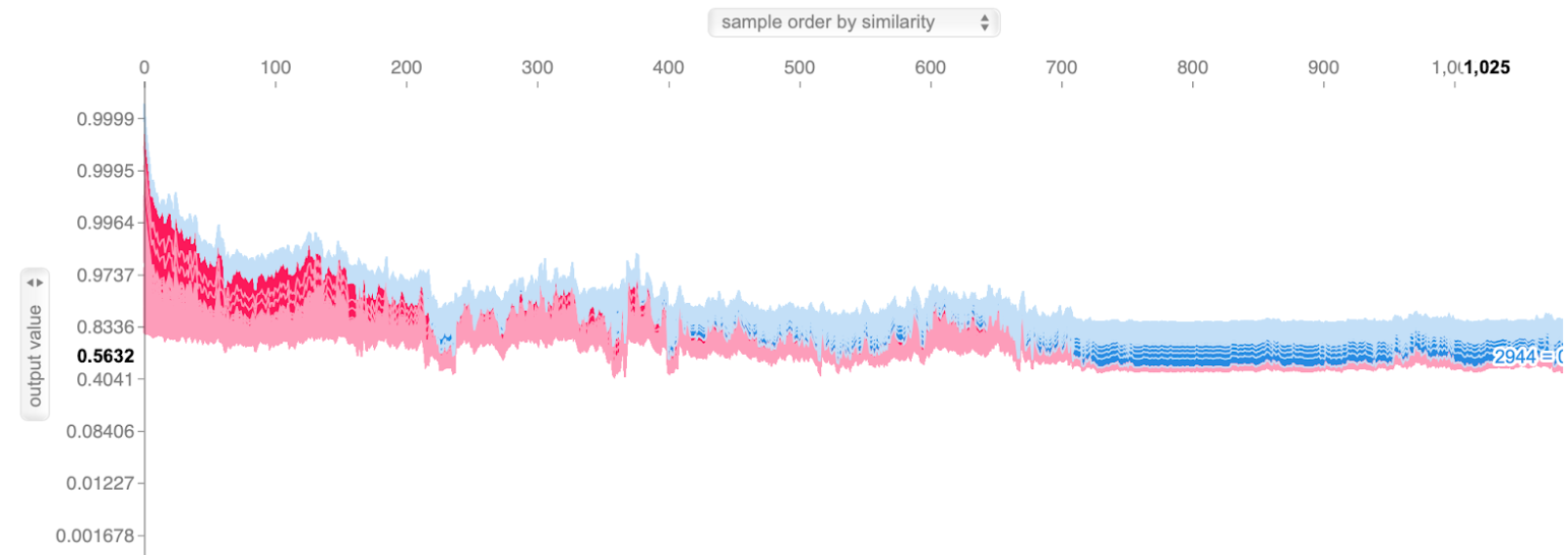}
    c)
      \hfill
 \end{minipage}
       \caption{a) Summary plot for top 20 most significant wavelengths for the predicted Titer concentration by the XGB Regressor.
       b) Force plot for a given measured spectrum
       c) Stacked Force plot for all measurements on DS13}
       \label{img:ds13varimportanceTiter}
\end{figure}

The spectrum preprocessed by the standard pipeline is formed by 2533 wavelengths. It was hypothesized that not all of these wavelengths carry information necessary for the regression towards the target concentrations of the molecules of interest. This was one reason for exploring different dimensionality reduction techniques. Another reason was the small size of the biggest dataset (roughly 2000 spectra for DS13), which added to the difficulty of the regression problem, as the number of variables (features), in our case the wavelengths, was higher than the size of the dataset. Given that the majority of studies suggest much larger estimates for the minimum number of samples per number of independent predictor variables, reducing the dimensionality of the regression problem appeared to be a necessary measure.\\
Previous studies have shown that the wavelengths are highly correlated, and therefore the dimensionality problem can be solved with Partial Least Squares Regression (PLS) \cite{wold_sjeriksson_2001} or Principal Components Regression
(PCR) \cite{jolliffe_1982} methods, which inherently contain a form of dimensionality reduction. These methods are well known and were successfully tested with the analysed datasets previously and are therefore not in the focus of this work, which essentially deals with ML and DL models.\\
While testing the Extreme Gradient Boosting Regressor (XGB) \cite{ChenG16}, which was not part of the previously used toolboxes for predictive modeling from Raman spectra in bioprocessing, it was noticed that it featured a method used for variable importance analysis. The method is described in more detail in chapter 3, section 1.4. The results of this analysis are plotted for the 4 target variables in Fig. \ref{img:ds13varimportanceall}. 4 distinct regions with a high density of peaks can be clearly seen, as well as a few local peaks. This shows that features from different peak-regions seem to be independent variables and are suitable for an efficient regression.
An interesting finding was that the same analysis as for DS1 yielded a different ranking of features for DS2, which means that such a variable importance analysis is process-dependent.\\
The obtained list of variables was then verified with a tool used for explaining the output of ML models which is called SHAP - \emph{SHapley Additive exPlanations} \cite{NIPS2017_7062}. Its authors propose a gametheoretic approach to explain any ML model's output. As of now, it is capable of explaining ensembles of decision trees such as XGB, but also deep learning models built with Tensorflow (TF), Keras or Pytorch
\cite{tensorflow2015-whitepaper, chollet2015keras,NEURIPS2019_9015}\footnote{Because of timing constraints, the later capability was not yet explored, however, it would be potentially interesting to extend the use of this toolbox onto the developed Keras/TF models in order to gain a better understanding of their inner workings and find bottlenecks in the current designs.}. \\
A detailed example of the XGB model's output explanation is shown in Fig. \ref{img:ds13varimportanceTiter} a) for Titer: This is a summary plot which summarizes the effect of the top 20 features (the rest is not shown because their effect on the SHAP value is negligible). The SHAP value is a measure of the impact of a single wavelength on the model's output. Each dot in the summary plot corresponds to the SHAP response of the model for the intensity at the corresponding wavelength (for a given spectrum measurement). The color of the dots implies the value of the intensity at this wavelength, higher intensities correspond to red dots and lower intensities correspond to blue dots. The thickness of the trace signifies the aggregation of different inputs for the same SHAP value.
The force plot in Fig. \ref{img:ds13varimportanceTiter} b) shows the impact of of the top most significant features for a selected spectral measurement. Aggregating and sorting such force plots by similarity over the whole dataset produces the stacked force shown in Fig. \ref{img:ds13varimportanceTiter} c). Projecting this onto the wavelength axis and and sorting by the sum of absolute SHAP values over all samples leads us back to the summary plot.

At this point it is important to remind the reader that this analysis only reveals the impact of the learned input features by the XGB Regressor on the predicted target concentration, which does not necessarily mean that the true relationship of these features with the measured target concentration is the one suggested by the SHAP explainer. It can be nevertheless hypothesized that there is \emph{a} relationship, and since at this point it is only of interest knowing which features contribute the most to this, it seems to be a useful tool for the pre-regression analysis and better understanding of the spectral data.

\newpage
\subsection{Data Augmentation}

In a DL context, augmenting a dataset means artificially generating additional training data from the original dataset. This technique became most popular for enhancing the training of Convolutional Neural Networks (CNN) by augmenting the dataset with different alterations of the same images. Usually, these alterations consists in adding noise to the training sample, rotation, scaling and similar transformations, while preserving the meaningful information carried by the training sample. Doing so enables the model to learn only the essential features and makes it more robust to noisy data. In general, this leads to the convergence of training and test error and helps to reduce overfitting. A detailed survey of current augmentation techniques and their taxonomy can be found in \cite{shorten_khoshgoftaar_2019}. In order to apply these techniques effectively, it is necessary to understand which portions of the data sample are meaningful for the underlying model and therefore should not be altered. For image data, this is often a simple task, as they are easy to interpret, in contrast to spectral data, especially for someone not specializing in the domain. The safest option to generate new samples is therefore to add Gaussian noise to the existing, without significantly altering the shape of the spectrum. The degree (augmentation factor) and shape of this noise are crucial design decisions for achieving superior results, and so is also the decision of where to introduce the noise in the overall pipeline.\\
The optimal augmentation factor can be determined by plotting it versus the error, which showed that it converges to a constant value. For experiments with Titer, this turned out to be a value between 1-3. The parameters for added noise can be estimated by comparing the spectra for two consecutive measurements of the target concentration. The noised spectra should be clearly distinguishable from each other and resemble the overall profile of the original signal. The noise model was taken from a previous implementation, which was based on \cite{shorten_khoshgoftaar_2019}. For the detailed algorithm, the reader can refer to the Python implementation in Appendix A. 2. (the default values gave good results and were used for all targets and datasets). Note that only the raw input features are augmented, the targets are not altered by augmentation. Performing the augmentation as the first step in the pipeline seemed to be a logical decision, similar to the reasoning behind placing the scaling steps at the end of the pipeline.\\
Furthermore, it is important to notice that augmentation is only beneficial on the training set, as performing it on the whole data set before the first CV split could potentially lead to information leakage into the test set (augmented spectra sampled from the same original spectrum could land in both the train and test set, which must be avoided). Hence, the augmentation takes place just after the CV split in the outer loop. An even stricter measure would be to place it after the CV split in the inner loop (to avoid leakage from the test set into to the validation set), this was not done because of implementation / time constrains, but should be explored in future development.

\subsection{Data Resampling}

In the previous section we have seen how data augmentation was applied to individual spectra, however it is not yet clear to which samples of the spectra it should be applied. The original implementation suggested taking a fraction of a random permutation of the original dataset for augmentation factors < 1, which corresponds to uniform sampling without replacement. For augmentation factors > 1 it supported only whole numbers as augmentation was applied to multiples of the full dataset (no sampling), which was not flexible enough to verify our assumption. For this reason, a custom sampling algorithm had to be implemented. \\
In \ref{sec:descofthedatasets} we have seen that the distributions, with exception of Glucose, are mostly tailed. It was assumed that models lacked training examples in these tailed regions, as their performance was weakest in these regions. The goal was to come up with a sampling scheme that would add importance to these areas. Prior to that, it was necessary to gain understanding about the probability density function of the target variables and the desired distribution. As we are dealing with finite datasets, they had to be modelled with histograms of adaptive bin size, according to the following developed algorithm:

\begin{enumerate}
    \item First, the target's histogram was created with the minimal bin size, which was gradually increased until every bin contained a predefined minimum number of samples.
    \item Then, the desired target distribution was constructed with the same bin boundaries and superimposed with the original distribution.
    \item Finally, each bin was sampled the number of times required to compensate for the difference in counts between the two distributions, the sample was augmented using the algorithm described in the previous subsection.
\end{enumerate}

\begin{figure}[ht]
  \begin{minipage}[h]{0.24\textwidth}
    \includegraphics[width =\textwidth]{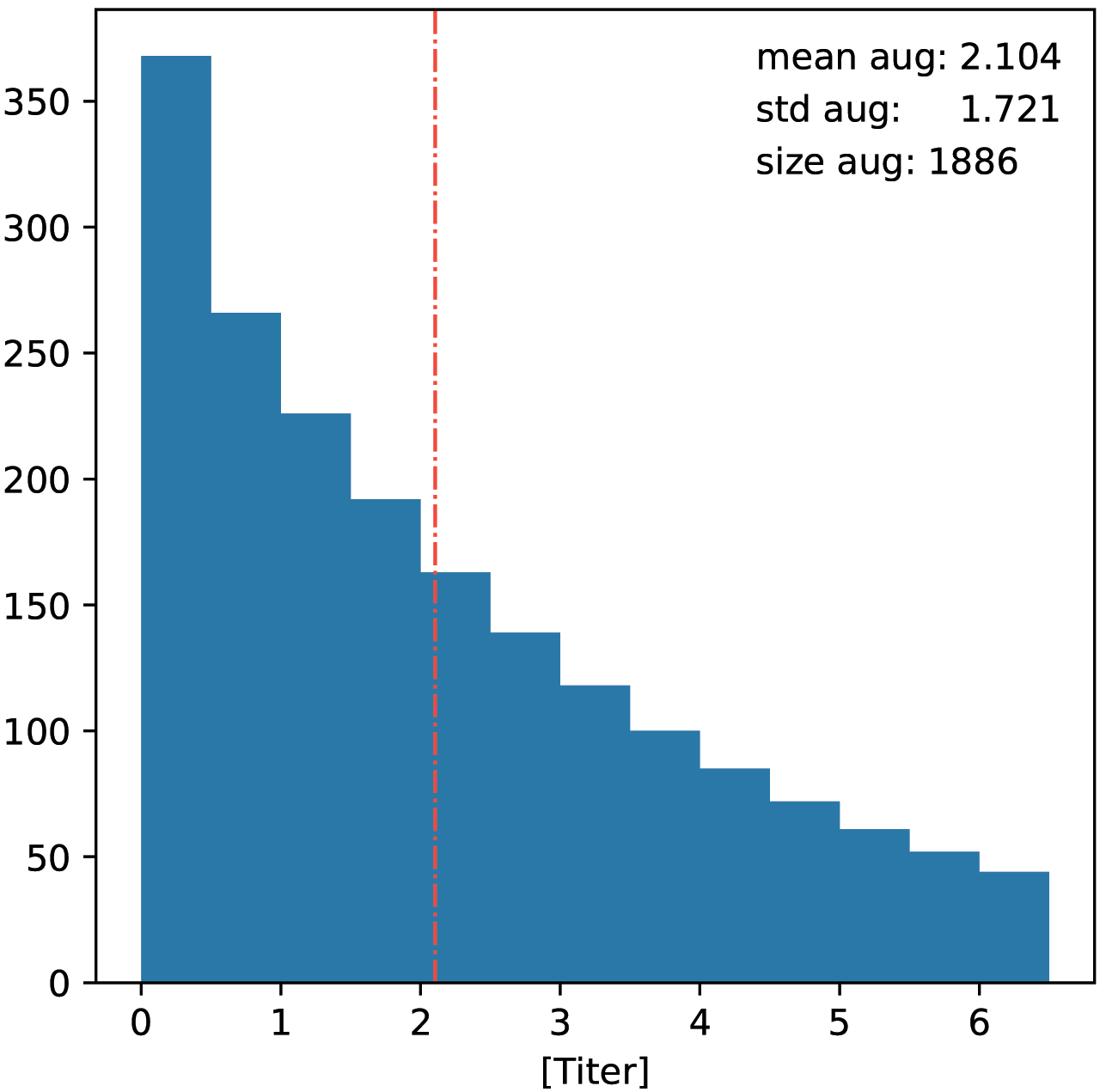}
  \end{minipage}
  \hfill
  \begin{minipage}[h]{0.24\textwidth}
    \includegraphics[width = \textwidth]{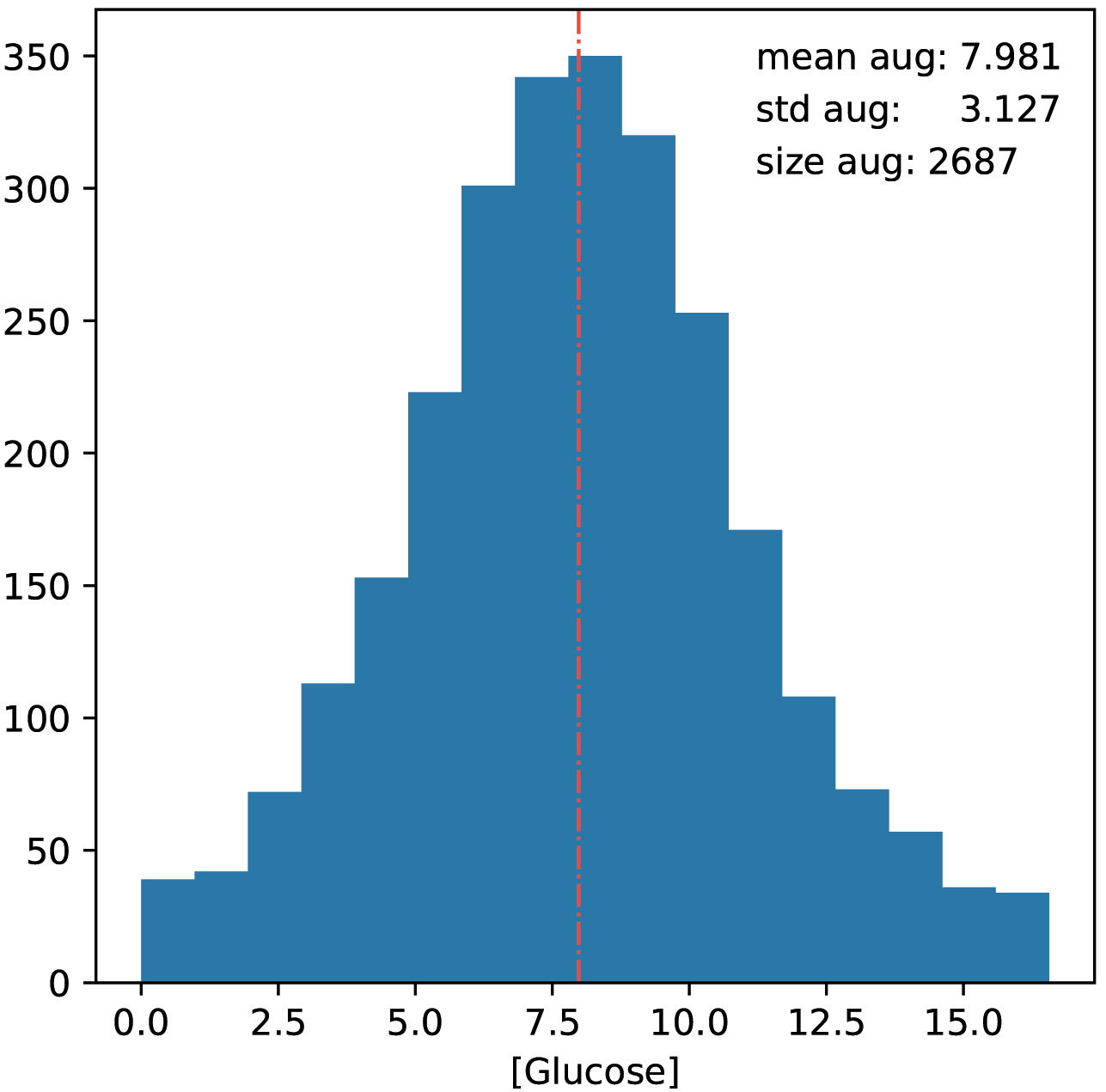}
  \end{minipage}
  \hfill
  \begin{minipage}[h]{0.24\textwidth}
    \includegraphics[width = \textwidth]{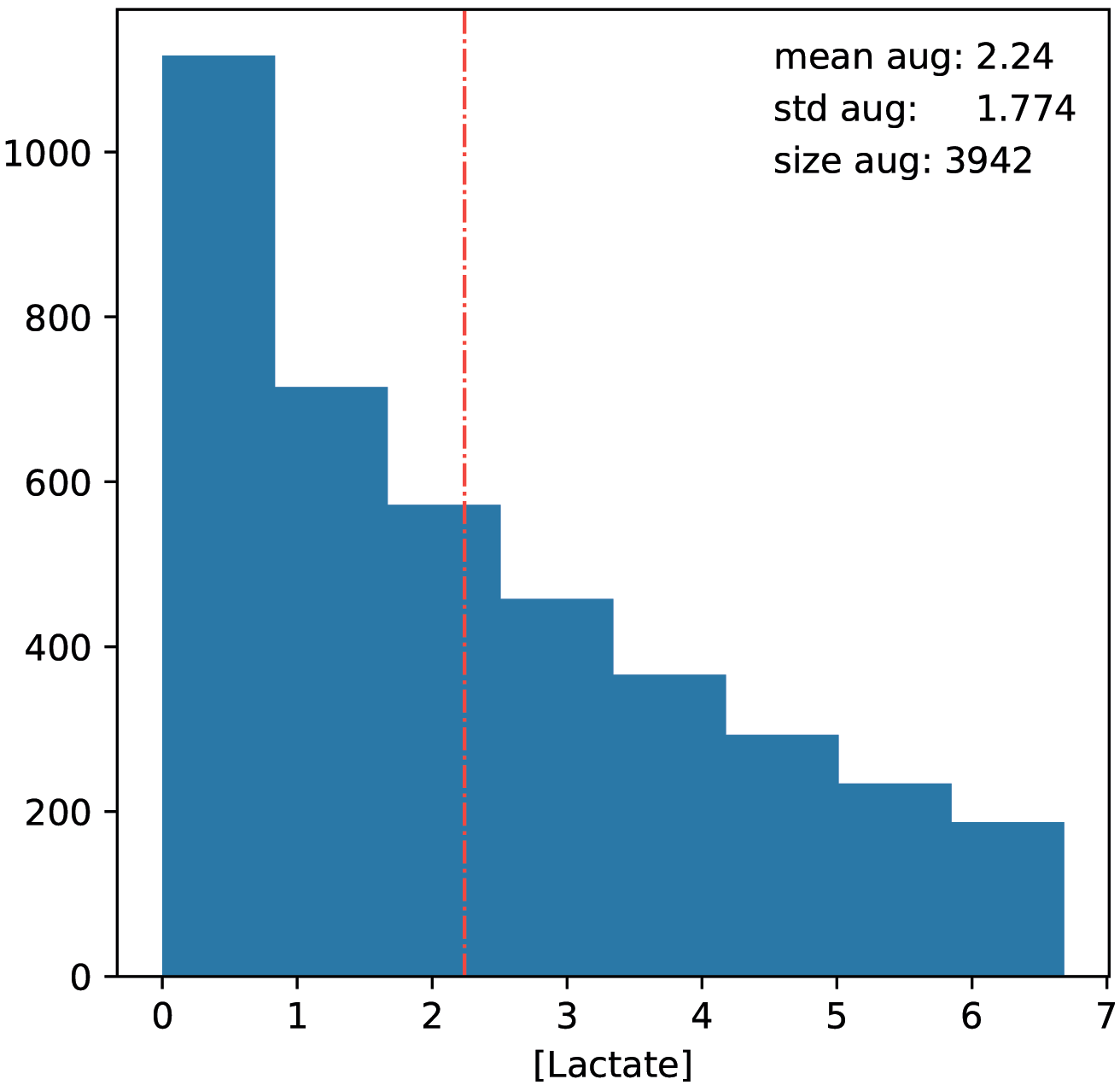}
  \end{minipage}
  \hfill
  \begin{minipage}[h]{0.24\textwidth}
    \includegraphics[width = \textwidth]{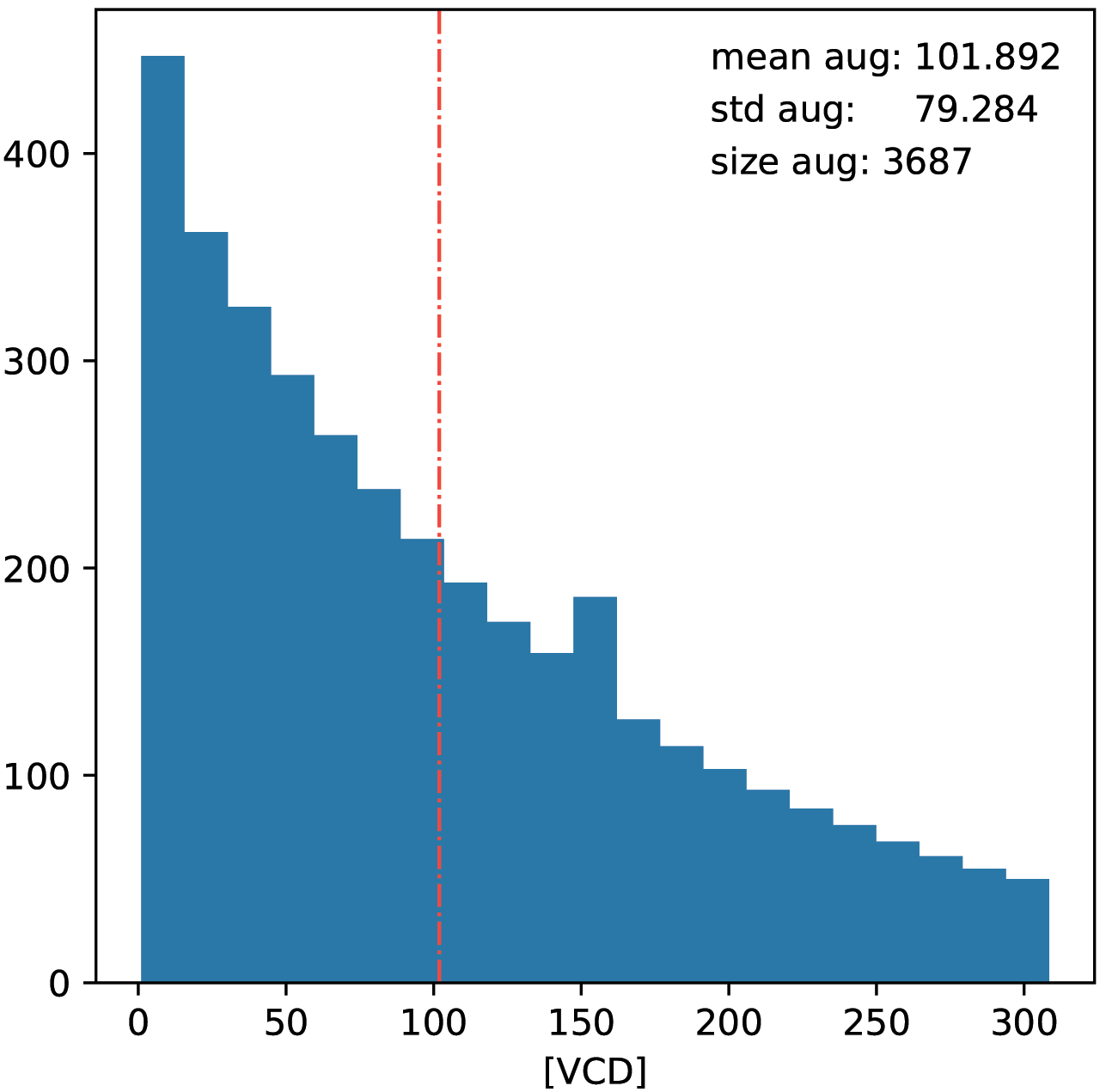}
  \end{minipage}
  \hfill
  \caption{Augmented distributions of Target Variables in DS1.}
  \label{img:ds13resampled}
\end{figure}

\begin{figure}[ht]
  \begin{minipage}[h]{0.24\textwidth}
    \includegraphics[width =\textwidth]{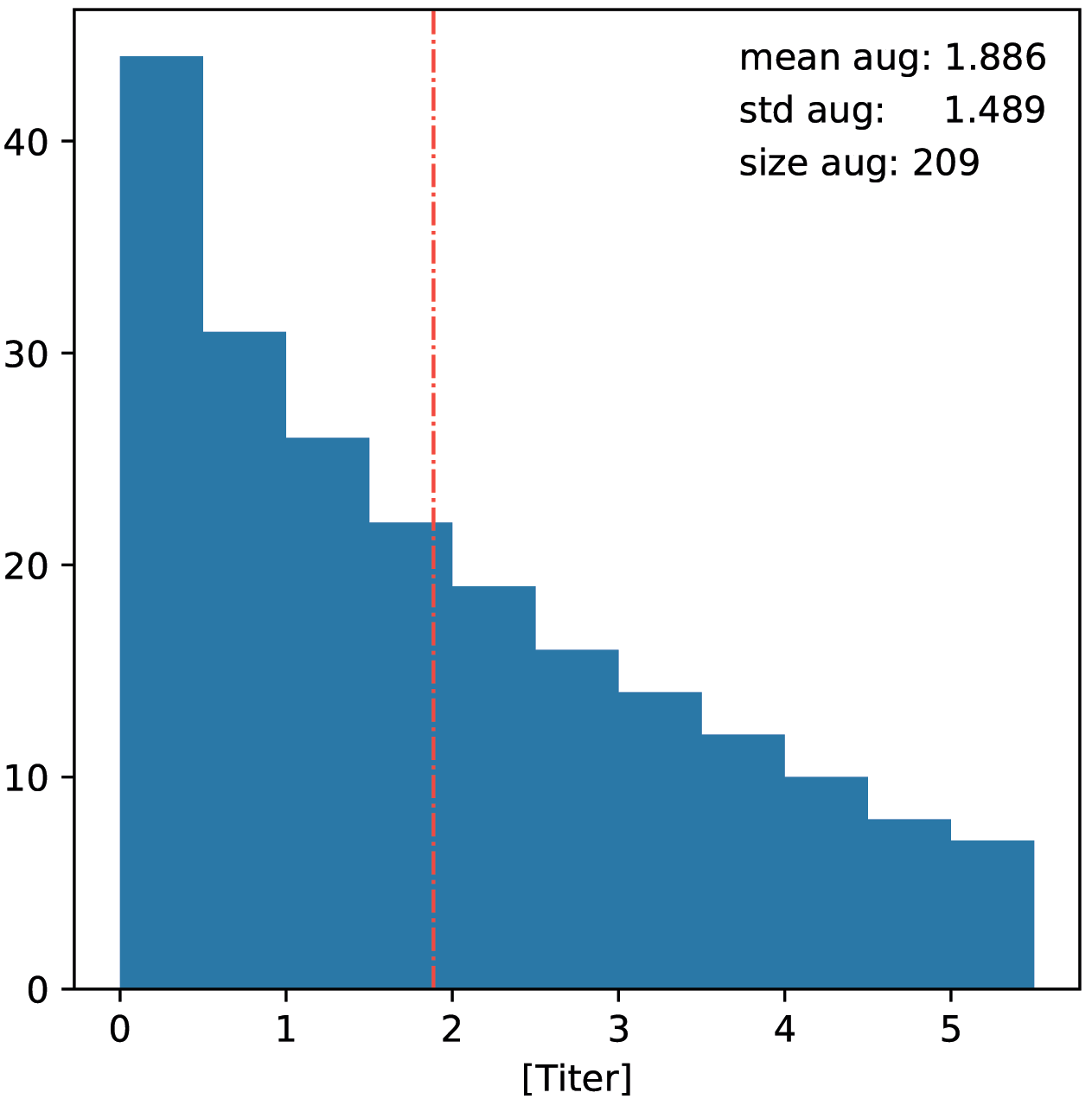}
  \end{minipage}
  \hfill
  \begin{minipage}[h]{0.24\textwidth}
    \includegraphics[width = \textwidth]{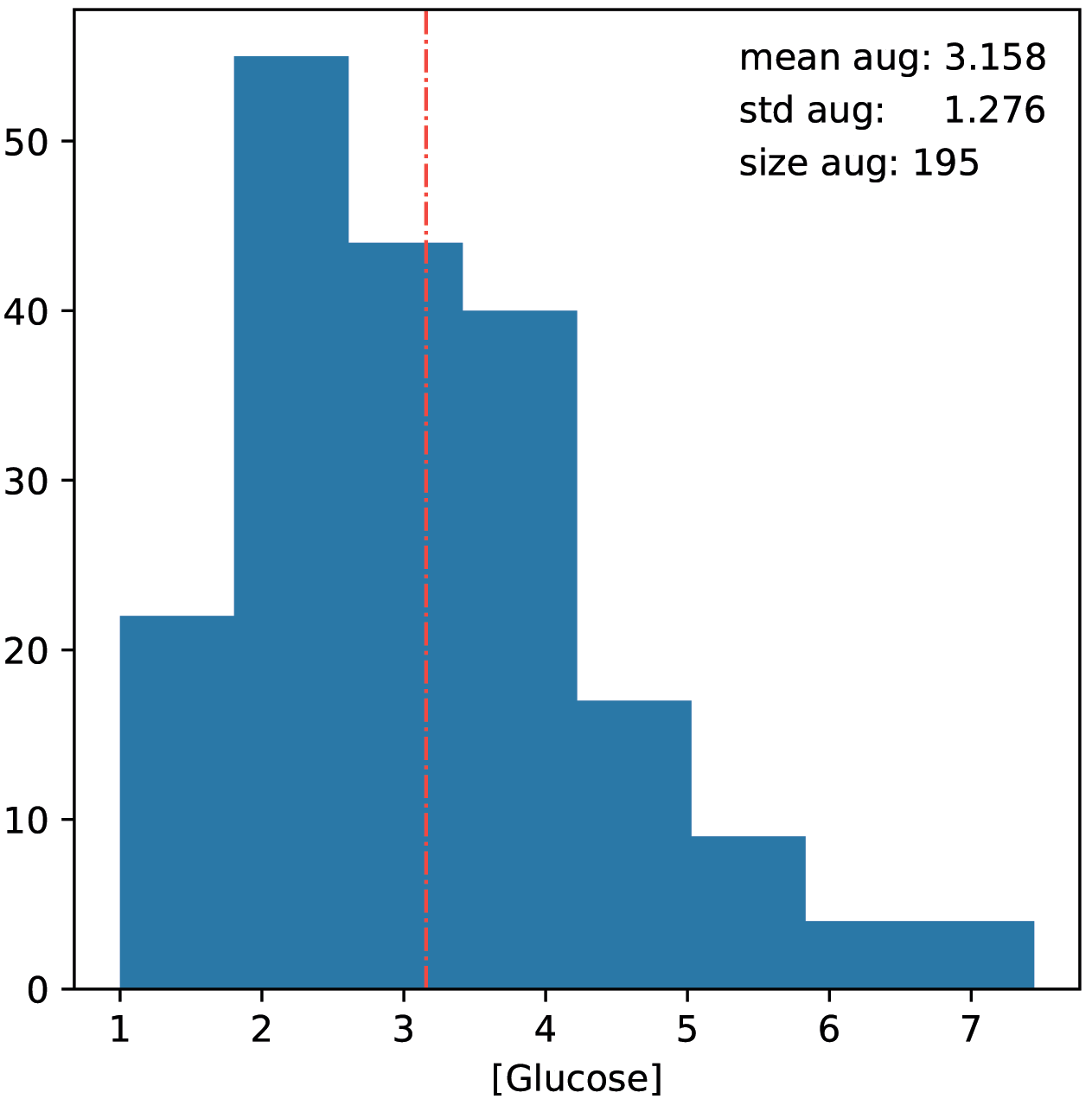}
  \end{minipage}
  \hfill
  \begin{minipage}[h]{0.24\textwidth}
    \includegraphics[width = \textwidth]{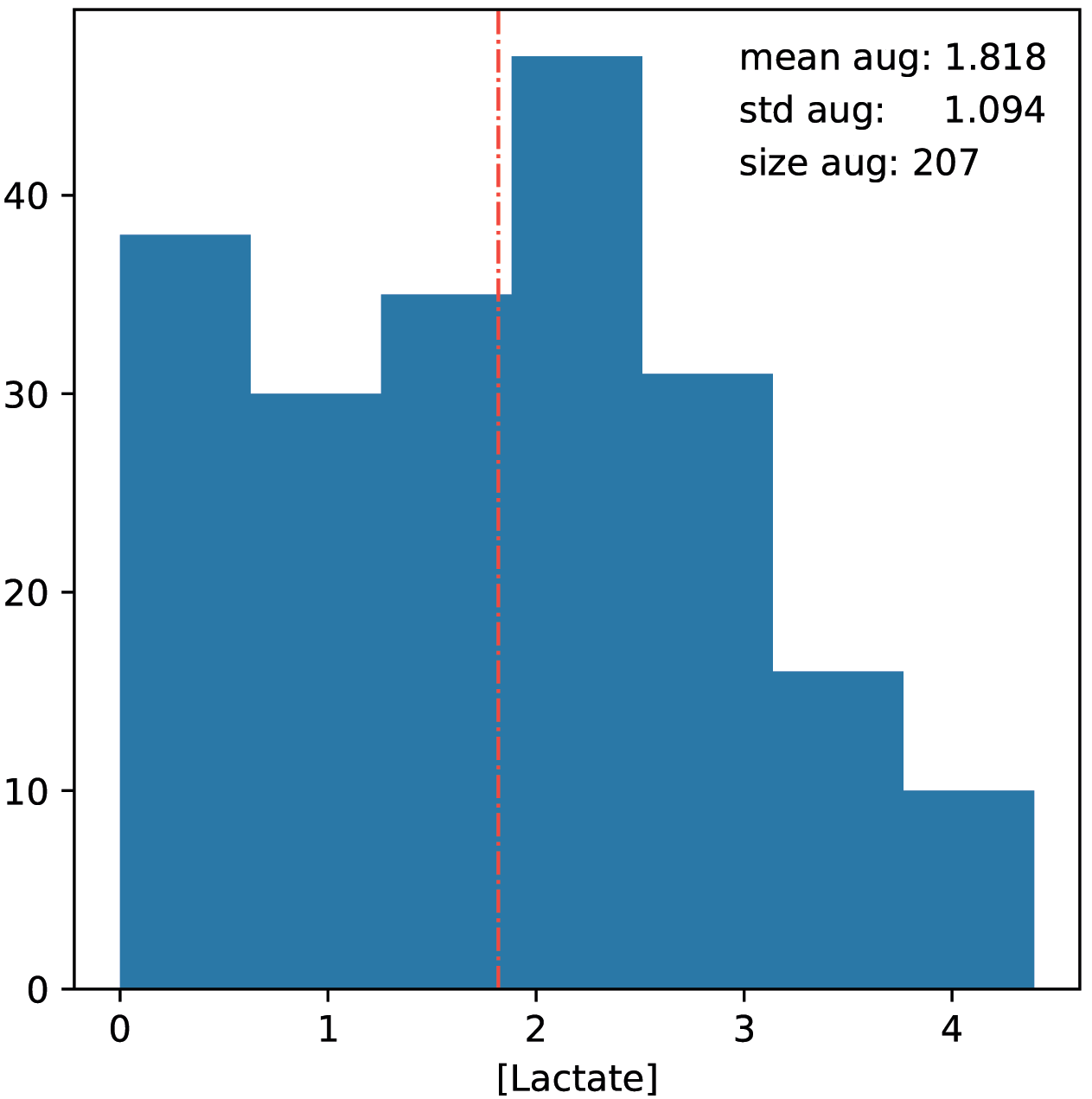}
  \end{minipage}
  \hfill
  \begin{minipage}[h]{0.24\textwidth}
    \includegraphics[width = \textwidth]{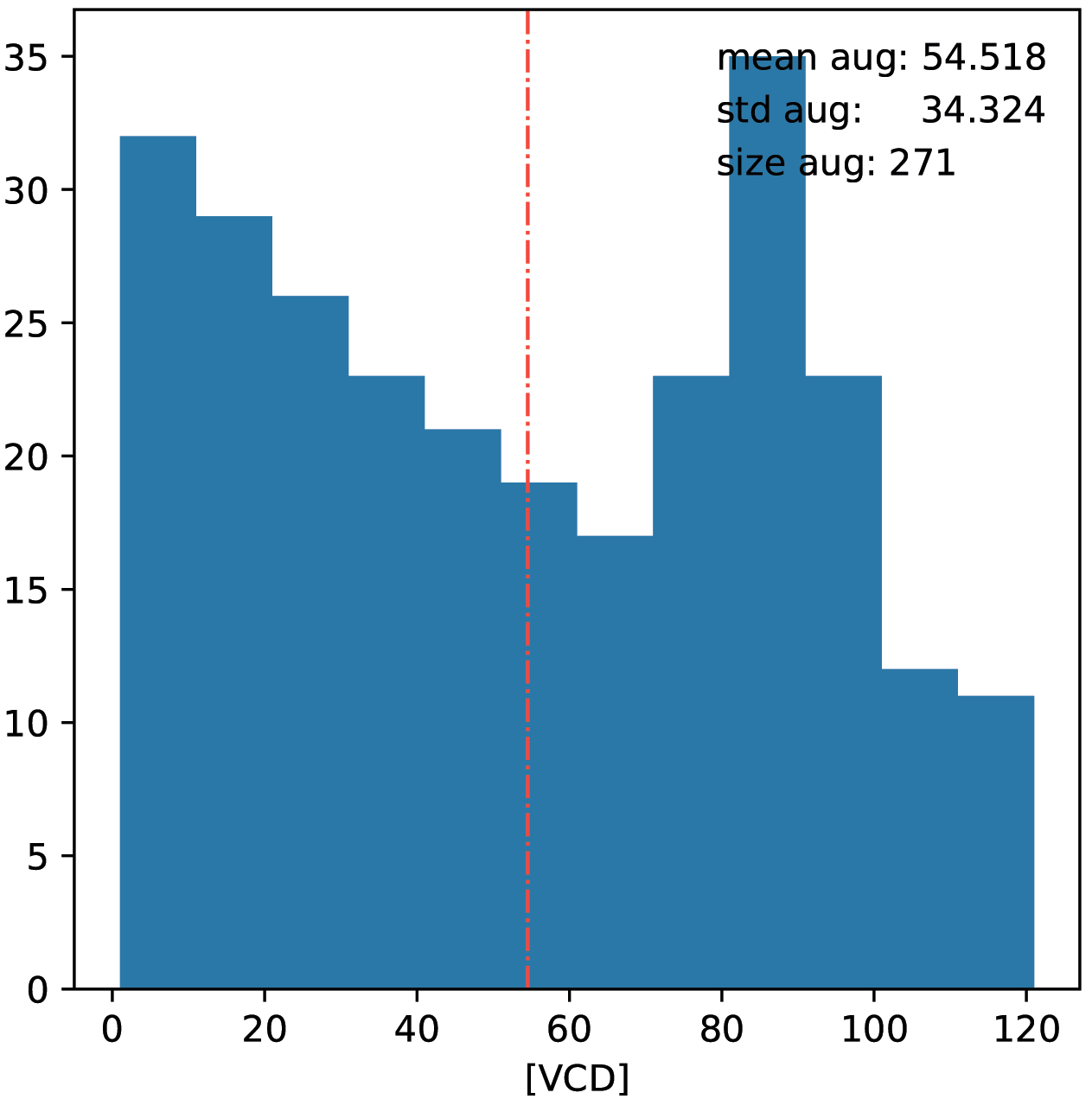}
  \end{minipage}
  \hfill
  \caption{Augmented distributions of Target Variables in DS2.}
  \label{img:dsTOCIresampled}
\end{figure}

The resulting distribution would have a similar size and shape as the desired distribution. The two described algorithms allowed to augment and resample the original distribution in a very flexible way, which means that finding the optimal scheme was just a matter of time. The exact implementation of the algorithm is shown in Appendix A. 3. Due to the time constrains of this project, the data sets were resampled using a distribution closest to the original distribution of the target variable: a Gaussian was used for Glucose and and exponentially decaying distributions were used for Titer, Lactate and VCD. The results of the augmentation are illustrated in Fig. \ref{img:ds13resampled} and Fig. \ref{img:dsTOCIresampled}.\\

\begin{figure}[h]
  \begin{minipage}[h]{\textwidth}
  \centering
    \includegraphics[width = 0.28\textwidth]{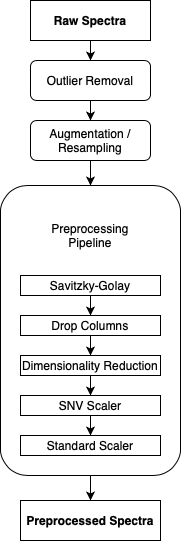}
  \end{minipage}
  \hfill
  \caption{Full pretreatment process flow of the analysed Raman spectra: outlier removal, augmentation/resampling and original preprocessing pipeline with dimensionality reduction}
  \label{img:pipeline}
\end{figure}

Fig. \ref{img:pipeline} shows a summary of all preprocessing steps described above.



\cleardoublepage

\chapter{Regression Models}

This section gives a brief introduction to regression models previously used in the raman project as well as recently introduced ones. The models can be divided into 2 families: ML models and DL models. The later can be subdivided into Feed-forward Neural Networks (NN), Convolutional Neural Networks (CNN) and Auto-encoders. The focus of this work are DL models.

\section{ML Models}
\begin{figure}[!h]
    \centering
    \includegraphics[width=\textwidth]{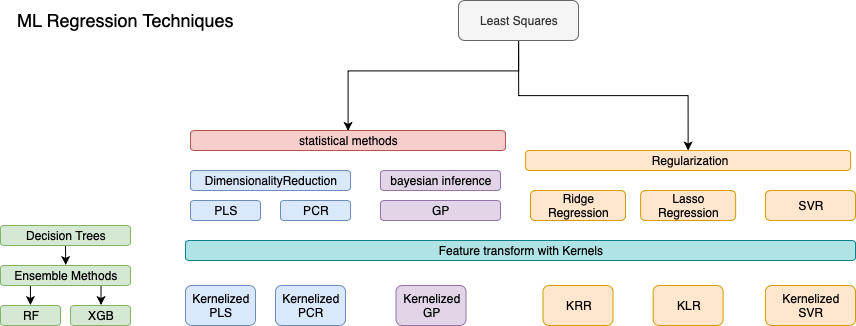}
    \caption{A taxonomy of some popular ML regression techniques: The methods are subdivided into Decision Tree-based methods, statistical methods and basic machine learning methods with respective kernelized extensions. The later will be explained in detail for SVR.}
    \label{fig:MLregressiontaxonomy}.
\end{figure}

All of the following ML models, except for XGB were already tested in previous experiments of the Raman regression project, using implementations from the scikit learn (\texttt{sklearn}) ML library \cite{scikit-learn}. These models were taken as-is without further parameter tuning (previously established optimal hyperparameters were used for all experiments).\cite{FeidlF19}
To simplify the understanding of the models described in this section, a taxomy is shown in Fig. \ref{fig:MLregressiontaxonomy}. The goal was to separate statistical approaches from tree based methods and Least Squares regression methods with explicit regularization and show how different methods are related to each other.

\subsection{Partial Least Squares (PLS) Regression}

PLS Regression is a statistical method which consists in projecting feature and target variable into a latent space and then fitting a linear regression model in this latent space. It is particularly suitable for regression problems where the number of features exceed the number of observations and when there is high multicollinearity between the features. The method was originally proposed for regression in chemometrics \cite{wold_sjeriksson_2001}, where it quickly gained popularity, but is also a known tool in domains like banking and finance and sports science \cite{doi:10.1111/jofi.12060, weaving_jones_ireton_whitehead_till_beggs_2019}.\\
PLS is a state of the art method for building Raman models and is used in commercial software. This is why it was chosen as the benchmark model for measuring the performance of other models developed in the course of this project. In this work, the implementation from \texttt{sklearn.cross\_decomposition.PLSRegression} was used, with references to \cite{Wegelin00asurvey}, with the following hyperparameter search space:

\begin{verbatim}
    parameters = {'plsregression__n_components': list(range(7, 34))}
\end{verbatim}

\subsection{Gaussian Processes (GP) Regression}

A random process is called a Gaussian Process if its random variables are distributed normally and if their joint distribution is also normal.
In the given problem, every sample of the train set and test set can be considered to be a collection of random variables. Gaussian distributions are closed under conditioning and marginalization, which means that the resulting distributions are also Gaussian. This is particularly useful, as it allows allows to use Bayesian Inference:

\begin{enumerate}
    \item First, a prior belief is formed about the distribution we want to predict: $P(\boldsymbol{X})$. For centered data (zero mean) this distribution is fully defined by the covariance, which is specified by a kernel function (a design-/hyperparameter to be optimized, $\theta$)
    \item Then, the GP regression algorithm uses training observations to come up with parameters $\boldsymbol{\theta}$ which are used to calculate the likelihood $P(\boldsymbol{\theta}|\boldsymbol{X})$ (see \cite{carledwardrasmussen2005} for the exact algorithm).
    \item Next, the likelihood is multiplied with the prior to give the posterior distribution $P(\boldsymbol{X}|\boldsymbol{\theta}) \propto P(\boldsymbol{\theta}|\boldsymbol{X}) \times P(\boldsymbol{X}) $, which is also a Gaussian, and where the mean is used for the prediction and the variance serves as a measure of uncertainty of the prediction.
    \item Finally the posterior becomes the new prior $P(\hat{\boldsymbol{X}})$ 
\end{enumerate}

The advantage of this algorithm is that it provides a confidence interval for the prediction and that it allows the use of custom kernels. In this project, the Radial Basis Functions (RBF) kernel was used, which is commonly used in non-linear problems. Its major disadvantage is the high computational cost for high dimensional data and large sample sets.
\\Here again, the implementation from \texttt{sklearn.gaussian\_process.GaussianProcessRegressor} was used, which is based on Algorithm 2.1 from \cite{carledwardrasmussen2005}, with the following search space for the RBF kernel's alpha :

\begin{verbatim}
    parameters = {'gaussianprocessregressor__alpha':
    [1e-10, 1e-6, 1e-4, 1e-3, 0.01, 0.1, 1, 1.5, 2, 3, 5]}
\end{verbatim}

\subsection{Random Forests (RF) Regression}

Random Forests are a supervised ensemble learning technique which makes use of multiple decision trees - which on their own are weak learners - to build one strong learner. Decision trees work by splitting up the feature space in every node by different True/False conditions. Finding these conditions is the subject of optimization. An example for a feature split would be for example the evaluation of the condition
\begin{math}
    (I(\lambda=1234) > 0.5 * 1e9)
\end{math}
which can evaluate to \texttt{True} or \texttt{False}.
This sums up the learning process of a decision tree. When it comes to making a prediction, a test sample is propagated through the tree and depending on the conditions it satisfies, it lands in one of the many leaf nodes of the tree which forms the prediction. An ensemble of such decision trees forms a decision \emph{forest} which forms the final predictions based on the average of the predictions of all trees (for a regression task; for a classification problem it would take the majority vote).
They are called \emph{random} because each decision tree makes decisions only on a random subset of the features when forming a prediction. The idea behind this is to form multiple diversified learners which specialize on specific features and use their joint 'knowledge' to produce a more robust final prediction. This procedure is called \emph{bagging} which stands for \emph{bootstrap aggregating}. Today, multiple versions of the algorithms exist, which can be found in \cite{ho_rf,Breiman2001}. Initially, this project made use of the \texttt{sklearn.ensemble.RandomForestRegressor} which is based on \cite{Breiman2001}, however, this model was dropped with the addition of the XGB Regressor described in the following subsection.

\subsection{Extreme Grandient Boosting (XGB) Regression}

\begin{figure}[h]
  \begin{minipage}[h]{0.48\textwidth}
    a)
    \includegraphics[width = \textwidth]{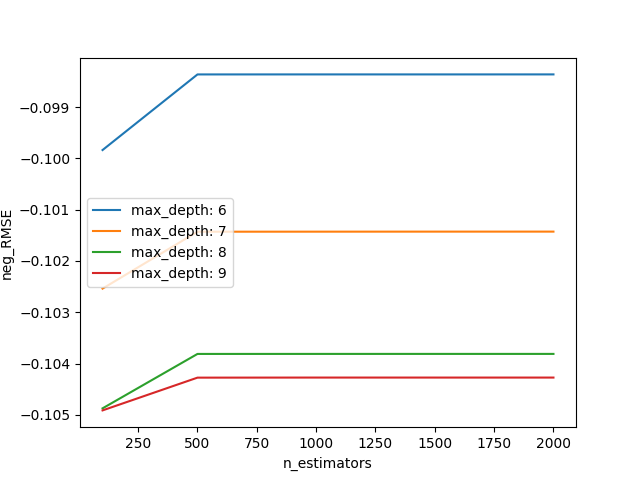}
  \end{minipage}
\begin{minipage}[h]{0.48\textwidth}
    \includegraphics[width = \textwidth]{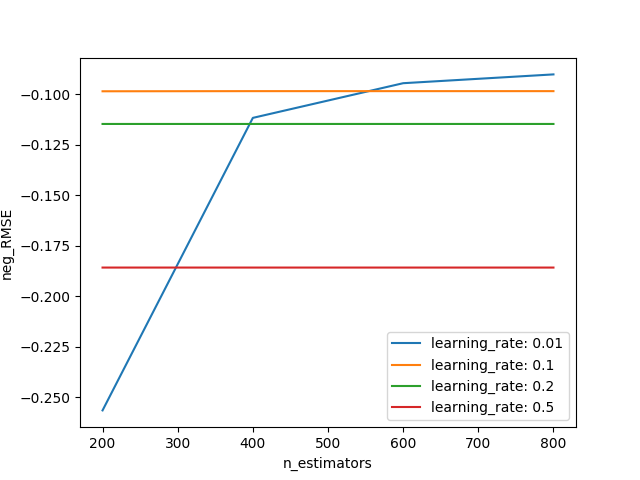}
    b)
 \end{minipage}
 
 \caption{Neg. squared error for different values of \texttt{n\_estimator} vs a) \texttt{max\_depth} and b) \texttt{learning\_rate} for the XGB Regressor.}
 \label{img:XGBhypersearch}
 \end{figure}

Similar to RF, this algorithm is also based on ensembles of decision trees. But instead of using bagging, where models (the trees) are trained in parallel, the trees are trained sequentially taking into account performance in previous iterations to minimize the objective function. In each iteration stronger learners gain weight. This technique is called boosting. There are different variants of boosting: In the \emph{AdaBoost} algorithm (Adaptive Boosting) \cite{Freund99ashort} samples with larger prediction error ("hard" examples) are given more importance. \emph{Gradient Boosting} \cite{friedman2000greedy}, on the other hand, uses the gradients of the errors of previous learners to create new learners that push the ensemble to the correct prediction. The optimized version of Gradient Boosting is called Extreme Gradient Boosting, which makes use of parallelization, shrinking of leafs,  and a new optimization algorithm. In this project, the python implementation \texttt{xgboost.XGBRegressor} with scikit-learn API is used, which is based on \cite{Chen2016XGBoostAS}. A hyperparameter search was performed in the very beginning of the project on the following space:

\begin{verbatim}
    parameters = {'xgbregressor__n_estimators': [400, 600, 800, 1000, 1400, 2000],
                  'xgbregressor__max_depth': [6,7,8,9],
                  'xgbregressor__learning_rate': [0.01, 0.1, 0.2, 0.5]}
\end{verbatim}

The results of the search are presented in Fig. \ref{img:XGBhypersearch}.
Improvements were marginal for parameters higher than:
\begin{verbatim}

    parameters = {'xgbregressor__n_estimators': 1000,
                  'xgbregressor__max_depth': 6,
                  'xgbregressor__learning_rate': 0.01}
                  
\end{verbatim}


without penalty on the error for, regardless of the target variable, so it was decided to fix these parameters for all experiments.
                  
One of the benefits of using Gradient Boosting is that it associates an importance measure to every tree, which in turn can be tracked to the importance of the input features.
The chosen (default) feature importance measure is \texttt{importance\_type : 'gain'} which is the average gain over all splits the feature is used in (the gain is the amount by which a feature split improves the performance).
The feature importance was recalculated multiple times (and with different hyper parameters) to ensure that this method yielded stable results.
Rankings were calculated for every target variable with the \texttt{RankFeatures.py} script (see Appendix A. 4. Its purpose is to generate a list of column \emph{indices} for the \texttt{DropColumnsTransformer} of the preprocessing pipeline, which drops all columns except the top N columns (N passed in the argument) and a list with the column \emph{names} to keep. It is important to notice that XGB operates on the preprocessed spectrum, which means all calculated indices are also related to the preprocessed spectrum and not the original one. This is why it is important to keep track of the column names (since now $\gamma \neq \text{index}$).

\subsection{Support Vector Regression (SVR)}

Support Vector Regression is a supervised ML Regression method based on Support Vector Machines. \cite{Cortes1995}.
Being an important step towards understanding the concepts of Neural Networks which will be introduced in the next section, it seems appropriate to briefly explain at this point some basic ML regression methods, how they are related to each other and the motivation behind SVR.
\bigbreak

In a regression problem we want to find $\vect{w}$ for :
\begin{equation}
    f(\vect{x}) = \vect{\hat{w}} \cdot \vect{x} + b = \vect{w} \cdot \vect{x}\\
    \text{where} f(\vect{x}) \approx y
    \label{regressioneq}
\end{equation}

Given a dataset $D = {(\vect{x}_1,y_1), (\vect{x}_2,y_2),...,(\vect{x}_n, y_n)}$ we want to find the optimal weight vector :

\begin{equation}
    \vect{w*} =  \argmin_{\vect{w}} \Sigma^{n}_{i=1}(y_i - \vect{w} \cdot \vect{x}_i)^2
\end{equation}

Which is called the least squares solution. This can be solved in closed form for linear problems.
However, there are cases where instead an iterative solution should be chosen, for example because of computational complexity or when we don't require an optimal solution, or when a problem does not admit a closed form solution.
Nevertheless, solving this equation leads to overfitting the test set. A possible way to counter this is to introduce regularization. Regularization aims to penalize the complexity of the model by forcing $\vect{w}$ to be sparse. Popular examples are :
\bigbreak
Rigde Regression, wich uses L2 norm as penalty for weights (with lambda controlling the strength of this penalty), forcing as many as possible of them to be close to zero :

\begin{equation}
\vect{w*} =  \argmin_{\vect{w}} \Sigma^{n}_{i=1}(y_i - \vect{w} \cdot \vect{x}_i)^2
+\lambda \norm{\vect{w}}_2^2
\end{equation}

Lasso (Least Absolute Schrinkage) Regression, which uses L1 norm as penalty, aiming to make as many componenets of $\vect{w}$ as possible exactly zero:

\begin{equation}
\vect{w*} =  \argmin_{\vect{w}} \Sigma^{n}_{i=1}(y_i - \vect{w} \cdot \vect{x}_i)^2
+ \lambda \norm{\vect{w}}_1
\end{equation}

There is another way to obtain sparse solutions, by introducing the concept of margin, noted $\epsilon$, which defines a tolerance bound for the residuals of \ref{regressioneq}. This margin can be chosen as "hard":

then solution is constrained by :
\begin{equation}
y_i - \vect{w} \cdot \vect{x}_i \leq \epsilon
y_i - \vect{w} \cdot \vect{x}_i \leq \epsilon
\end{equation}
and consists in minimizing : 
\begin{equation}
\min \frac{1}{2} \norm{\vect{w}}^2
\end{equation}

or "soft", for example when we can tolerate solutions with data points outside of the optimal margin. For this, the concept of slack variables $\xi_i^+$ and $\xi_i^-$ is introduced, defining the distance of each point from either side of the margin, outside of the margin.

this constrains the solution to:
\begin{equation}
y_i - \vect{w} \cdot \vect{x}_i \leq \epsilon + \xi_i^+ \\
\vect{w} \cdot \vect{x}_i - y_i \leq \epsilon + \xi_i^-
\end{equation}
and consists in minimizing : 
\begin{equation}
\min \frac{1}{2} \norm{\vect{w}}^2 + C \ \mathlarger{\Sigma}^{n}_{i=1} \left( \xi_i^+ +  \xi_i^- \right)
\end{equation}

where C is a parameter setting the penalty for slack variables.

The above solutions can solve linear tasks, i.e. where the solution can be defined with hyperplanes.
For data, where such a fit is not possible, a feature transformation can be used called \emph{kernel trick}. The original problem can be reformulated as :

\begin{equation}
    y = \mathlarger{\Sigma}^{n}_{i=1} \left(\alpha_i - \alpha_i^* \right)\vect{x}_i \cdot \vect{x}
\end{equation}

as the optimal $\vect{w}^*$ can be expressed in the span of the features:
\begin{equation}
    \vect{w}^* = \mathlarger{\Sigma}^{n}_{i=1} \left(\alpha_i y_i\right)\vect{x}_i
\end{equation}

and since $\vect{x}_i \cdot \vect{x}_i$ is a dot product, we can use kernel functions $K \left(\vect{x}_i, \vect{x}\right)$ to map the data to a higher dimensional space where a linear fit becomes possible with the previously described algorithms  :

\begin{equation}
    y = \mathlarger{\Sigma}^{n}_{i=1} \left(\alpha_i y_i \right) \phi(\vect{x}_i) \cdot \phi(\vect{x})
    y = \mathlarger{\Sigma}^{n}_{i=1} \left(\alpha_i y_i \right) K \left(\vect{x}_i, \vect{x}\right)
\end{equation}

Now, instead of looking for the optimal coefficients of $w^*$, the aim is to find the optimal $\alpha_i^*$, minimizing the same loss functions as defined earlier.
\bigbreak

There exist different types of kernels, for example linear, polynomial or RBF, and choosing a suitable kernel is often very hard, especialy for high dimensional problems. It essentially requires either very good domain knowledge on the input data or a brute force / heuristic search with cross validation. The later was chosen for this work, which uses the \texttt{sklearn.svm.SVR} implementation based on \texttt{libsvm}.\cite{libsvm}

The hyperparameter space was constrained by :

\begin{verbatim}
    parameters = {'svr__C': [0.1, 0.5, 1, 5, 10],
              'svr__epsilon': [0.01, 0.1, 0.2, 0.5],
              'svr__kernel': ['rbf', 'poly'],
              'svr__degree': [4.0, 5.0, 6.0],
              }
\end{verbatim}

\bigbreak
An alternative way to achieve non-linear mappings are Neural Networks, which will be discussed in the following section.

\newpage
\section{DL Models}

In this context, the word \emph{deep} describes the number of non-linearities, or layers in an Artificial Neural Network. In the following section one of its simplest variants is introduced: Feed-forward Neural Networks, also sometimes called Multi-layer Perceptron (MLP). \cite{reed_marks_1999}\\

\subsection{Feed Forward Neural Networks (NN)}

A Perceptron models the function of a neuron. It multiplies its input by a weight vector, adds a bias and feeds the result into a non linear activation function which evaluates to the output. Without this non-linearity, it would just correspond to a linear regression model.\\

Depending on the problem type, different activation functions can be used. An important requirement is differentiability, which is necessary for the optimization algorithm to work. Examples are:

\begin{description}
    \item[Sigmoid:] $f(x)= \frac{1}{1+e^-x}$. S shaped function, bounded between 0 and 1 and therefore often used for classification tasks (approximates the Heaviside step function). One of it's major drawbacks is the vanishing gradient at both ends, which inhibits the flow of gradients from later layers and makes training hard in some cases.
    \item[Hyperbolic Tangent:] $f(x)=tanh(x)$. similar to the previously described function, but bounded between -1 and +1, which makes it zero-centered. Suffers from the same vanishing gradient problem.
    \item[Rectified Linear Unit (ReLU):] $f(x)=max(0,x)$. Zero for negative input and linear for positive, it has one point were it is non differentiable, which can lead to weights getting "stuck" and neurons to "dying" at certain values. Nevertheless, ReLu's have been a popular choice for a long time, since they are computationaly inexpensive and can approximate any non-linear function.
    \item[Leaky ReLU):] Has a small slope on the negative, allowing gradients to flow even for negative input. There have been studies on how to make the network learn the value of this slope on its own \cite{DBLP:journals/corr/HeZR015}, however results were not always consistent.
    \item[Softplus:] $log(1+e^x)$. Despite being differentiable everywhere, it's advantages over ReLu and Leaky ReLu have yet to be discovered.
    \item[Exponential Linear Unit (ELU):] $x$ if $x > 0$ and $\alpha(e^x -1)$ else, Where $\alpha$ is a parameter which requires tuning.
    \item[Scaled ELU (SELU):] $f(x) = \lambda ELU(x, \alpha)$. As the name says, this scales the ELU activation by the factor $\lambda$, where $\alpha$ and $\lambda$ are chosen so that the mean and variance between layers are preserved. In order for this to work, it requires a special weights initialization scheme called \texttt{lecun\_normal}. As a result, the network is self-normalizing, which leads to faster convergence. In the corresponding paper, the vanishing and exploding gradient problems are proven to be impossible in this setting. \cite{DBLP:journals/corr/KlambauerUMH17}. During this project, experiments with such networks have indeed shown faster convergence, however exploding gradients were also observed, with a recovery to stable values after a few training epochs.
\end{description}

The NN models developped in this work rely essentially on Selu and Leaky Relu as main activation functions.

\newpage
When the output of one perceptron is fed as the input into another one, the overall model gains an additional non-linearity \emph{layer}, hence the name MLP. A layer that is not an input or an output is also called \emph{hidden}, \emph{fully connected} or \emph{dense} layer. Depending on the framework, activation functions can also be treated as layers. In general, a layer can be defined as an operation on an input resulting in an output.
Formally, a 3-layer network can be written as:
\begin{equation}
y = f_3 \left(\vect{W}_3 \cdot f_2\left(\vect{W}_2 \cdot f_1\left(\vect{W}_1 \cdot \vect{x} \right) \right) \right)
\end{equation}
Where $x$ is the input, $y$ the output, $f_i$ the non-linear activation functions and $W_i$ the weight matrices to be optimized (biases, included for simplicity of notation). Note that their size is (n x m), where m corresponds to the size of the input (1D in our case) and n corresponds to the number of outputs, also called \emph{neurons} or \emph{units} of the layer. Different layers can have a different number of neurons, which makes it an important design parameter. The \emph{Universal approximation theorem} \cite{cybenko_1989}, states that any function on a compact real subset can be approximated by a neural network with a single layer and a finite number of neurons, under mild assumptions on the activation function. This statement might convey optimism to someone out of the domain, however in practice, this is a quite weak statement, as often there is no information how large this "finite number of neurons" might be. This is why instead of using a single very dense layer, researchers have focused on developing deep models with multiple layers, introducing a hierarchy of layers and building up abstraction from the input to the output.
Similarly, building such a model requires either very good domain knowledge on the input data, transfer from similar problems, or a (semi-) automated search, which was also the procedure in this project.

\bigbreak
The \emph{training} of the network can be summarized as follows:

\begin{enumerate}
    \item First, the weights are initialised with an initialisation scheme which is often an important design parameter.
    \item A training sample is propagated through the network which results in a prediction with an error which is measured by a loss function (see Section \ref{tab:lossfunctions}).
    \item Then, an optimization algorithm tries to calculate new values for the weights so that the loss is minimized. This is done using the backward propagation of errors, or short \emph{backpropagation}, which uses the chain rule to calculate the gradients of the loss functions w.r.t. the weights in each layer. The original optimizer is called \emph{Gradient Descent Algorithm} and minimizes the loss by moving the weights in the direction of the steepest gradient (hence the name), however there exists a number of more advanced algorithms that can be chosen. The choice, as well as the learning rate of the optimizer (equivalent to the step size of the descent) are design parameters. This is why it is often convenient to use an optimizer with an adaptive learning rate, like \emph{Adam} for example, which is used in this work.\cite{KingmaB14}
\end{enumerate}

\newpage
In addition to dense layers, researchers have introduced other types of layers performing different operations, for example:

\begin{description}
    \item [Dropout:] Randomly ignores a fraction of layer inputs during each training update, which serves to prevent overfitting. \cite{JMLR:v15:srivastava14a}
    \item [Batch Normalization (BN) :] Applies a transformation that maintains the mean activation close to 0 and standard deviation close to 1. \cite{DBLP:journals/corr/IoffeS15}. Originally developped in order to reduce internal covariate shift during each training iteration and thus improve convergence. Depending on the type of activation, it should be either placed before (sigmoid, tanh) or after it (ReLU, Selu). It is recommended to not use Dropout with BN. \cite{DBLP:journals/corr/IoffeS15}
\end{description}

This whole variety of hyperparameters lead to a significant design problem which is difficult to solve manually. This is why an automated approach was chosen. Automated Learning has been gaining on demand recently, which can be noticed by the growth of multiple ML/AI vendors ranging from startups like H2O.ai to companies like Google with their Cloud AutoML solution. There were also major developments taking place in the open source community, for example called \texttt{keras-tuner}, which specializes on hyperparameter and architecture tuning of NN models built with the keras API\footnote{As of now, there are two pieces of software referred to by \texttt{keras}, the original implementation by F. Chollet and keras-team and the \texttt{tensorflow.keras} module incorporated into the \texttt{tensorflow} library developped and maintained by Google. In fact, the original keras is now an integral part of \texttt{tensorflow}. This is why the original raman project was ported to \texttt{tensorflow.keras} of the latest r2.1 release},
which was used to build the NN models used in this work.

\bigbreak
The library implements 4 main classes:

\begin{description}
    \item[HyperParameter:] Defines the type of the tunable hyperparameter. It can be a formulated as a Boolean, Int, Float, a choice from a list of predifined values, or it can be conditioned on the value of a parent HyperParameter. 
    \item[HyperModel:] Essentially consists of a build method used to construct the model based on the passed HyperParameter.
    \item[Tuner:] This defines the search algorithm (and instance) used for tuning. Predefined algorithms are: RandomSearch, BaysianOptimization and HyperBand, which is a Bandit-based approach based on \cite{hyperband}, especially suitable for very large search spaces. Multiple tuner instances can work simultaneously, which allows for a distributed search process.
    \item[Oracle:] An instance controlling the tuning process and telling the Tuner(s) which hyperparameters to try next.
\end{description}

\begin{figure}[!h]
    \centering
    \includegraphics[width=0.68\textwidth]{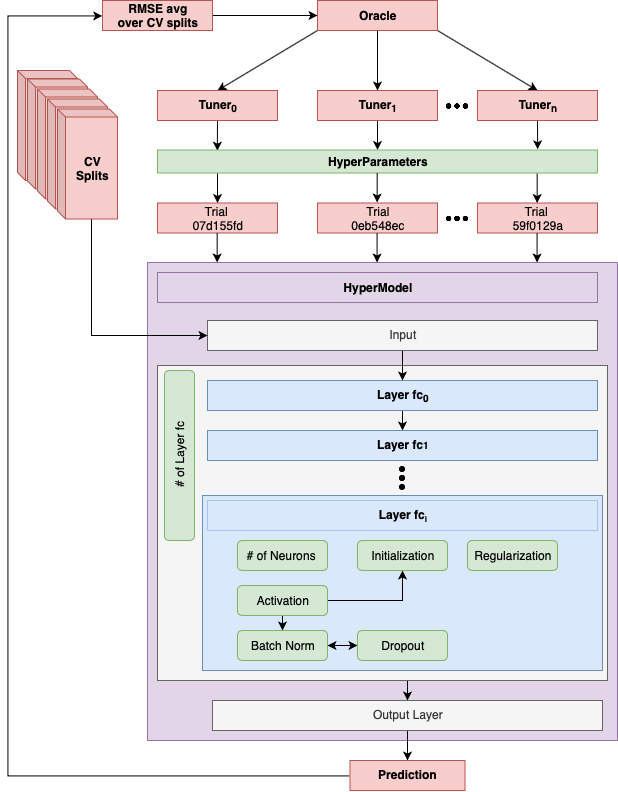}
    \caption{Flowchart of Hyperparameter tuning with \texttt{keras-tuner}. Instances related to the tuning \& scoring (via crossvalidation) algorithm are colored in red. The HyperModel, serving as a skeleton for all resulting models is colored in purple. The static, non-tunable part of the hypermodel is shaded in light grey. Hyperparameters are marked by round green boxes and their conditioning is marked by arrows. The $fc_i$ layers are coloured in blue, showing the dynamic part of the HyperModel's architecture (which depends on the number of layers).}
    \label{fig:KTflowchart}
\end{figure}

Unfortunately, none of the Algorithms supports crossvalidation by default, so a custom method had to be implemented. It also supports parallelized training, so that each cv-split can be run simultaneously. The exact algorithm can be found in Appendix A. 5. A flowchart of the Hyperparameter tuning process is shown in Fig. \ref{fig:KTflowchart}.

\newpage

At the top of the process flow, the oracle orchestrates $n$ Tuners, which can be distributed over one or multiple machines (by setting some environment variables referring to the address of the chief oracle), by instructing them to try a specific combination of hyperparameters, which form a so called \emph{trial}. Then, the tuner builds a model from these hyperparameters and scores it based on the average score over all crossvalidation splits (the base Tuner's class was subclassed with additional code to allow this custom scoring) and reports this score to the chief oracle, which determines a new combination of parameters for the next trial, based on the chosen search algorithm (taking into account past an ongoing trials).

\newpage
The search space was constrained as follows:

\begin{description}
    \item[\# of Dense Layers $fc_i$]: Between 1 and 3 in steps of 1
    \item[\# of Neurons in $fc_i$]: Between 64 and \texttt{input\_size} in steps of 64 (reduced to 32 in case of 128 input features)
    \item[Activations for $fc_i$]: Choice between \texttt{selu, relu, prelu, lrelu}
    \item[BN for $fc_i$]: Boolean, if True adds BN after Activation
    \item[Dropout for $fc_i$]: Boolean, if True and BN False, added after Activation
    \item[L1 kernel regularization for $fc_i$]: Choice between \texttt{0.0, 0.0001, 0.001, 0.01}
    \item[L2 kernel regularization for $fc_i$]: Choice between \texttt{0.0, 0.0001, 0.001, 0.01}
    \item[Kernel initializer for $fc_i$]: Conditioned on type of activation. If selu -- \texttt{lecun\_normal} (see \cite{DBLP:journals/corr/KlambauerUMH17}), else -- \texttt{he\_normal} (sometimes called \texttt{kaiming\_normal}), which is reported to work well with activations from the ReLU family
    \item[Dropout rate (global) :] Choice between \texttt{0.0, 0.2, 0.5}
\end{description}

The parameters above fully define all but the last layer, which consists of only one neuron (because the dimension of the output is 1) and no activation (linear).
Different search algorithms were tested, but a clear advantage of one versus the other could not be established.
One of the major drawbacks of the toolbox is that there is no central data structure which stores the scores of every trial, they are saved individually in their respective folder, which also contain the checkpoints for each model. This is very inconvenient, as for large search spaces their total size can grow very quickly (also because of increasing model complexity) into to the order of magnitude of TB. This is roughly the size of the scratch storage of the Euler cluster which powered the hyperparameter search. However, storing such volumes of data is not allowed in the long term (a number of results were lost that way) and is simply impractical. This was an unexpected problem which had to be solved. The rankings are stored in the tuner object itself, which loads them from the individual trial folders. An interruption of the search can break the retrieval of the scores by the tuner (can be corrected by deleting the unfinished trial folders). It was not straightforward to save the top n models of the search, which is why a short script was written for this purpose.\\
The above described models were tested on preprocessed data, where some of the valuable feature could have been unintentionally lost during smoothing. This is why it was decided to extend the Neural Network models with convolutional layers, which will be described in the following section.

\newpage
\subsection{Convolutional Neural Networks (CNN)}

ConvNets or CNNs are a type of Artificial Neural Networks widely used in Image Analysis, Recognition and Classification. Their architecture is inspired by the organization of the visual cortex.
Their main difference from previously described feed-forward networks is the use of convolutional layers. In these layers, the input is convolved with a filter -- the kernel -- whose values are learned by the network. Usually, a convolutional layer is composed of multiple channels to which the input is copied and a different filter operates on each channel. This enables the Network to learn different high-level features of the input: the number of learnable features grows with the number of channels. Their main advantage over normal neurons in dense layers is that they are able to learn the spatial dependencies of the input much more efficiently.
\\Another important distinction of ConvNets is the usage of pooling layers, which shrinks the input by retaining only the regions with the highest values. This operation can be thought of as downsampling of an image for example. Usually, this layer is added after a convolutional layer.
A deep CNN can be formed by chaining these layers in an alternating fashion. In order to keep the model's complexity (it's capacity to learn) constant, a Pooling layer (with shrinking factor 2x) is usually followed by convolutional layer 2x higher number of filters than in the previous layer.
As the size of the input decreases (approaches the size of the filter) and the number of filters increases, the layers are able to capture more specific, low-level features. This architecture allows the model to learn a representation of the input on multiple levels of abstraction simultaneously.

Depending on the task of the network, the output of the CNN sequence can be chained to a feed-forward network designed for a specific task (classification, or regression in our case) through a flattening layer. As a result, the convolutional layers act as feature extractors, often simplifying the regression task, as only the 'significant' features are passed to the input of the regressor.

Training this type of network is usually more difficult. Depending on the hardware, convolution operations (matrix multiplications) can be more difficult than dense operations (scalar products). Furthermore, efficient learning of these models requires more training examples than dense networks. There is a growing number of advanced CNN architectures that have shown success in the CV domain, namely in the ImageNet project, for example ResNet, GoogLeNet, Xception and others.\cite{DBLP:journals/corr/HeZRS15, DBLP:journals/corr/SzegedyLJSRAEVR14, DBLP:journals/corr/Chollet16a} However, these networks are usually \emph{very} deep (+100 layers) and operate on \emph{very} large ($+10^9$) high-dimensional (2D image x 3 channels) datasets  and a \emph{very} large number of outputs (1000 classes), and were therefore not considered for this project. Instead, the architecture of the VGG network was taken as basis, which is an older architecture that follows the basic CNN scheme described above.\cite{DBLP:journals/corr/SimonyanZ14a}

Unlike for NN models, the architecture of the developed CNN models was tuned manually, as it was considered that using basic guidelines for conventional CNN architectures, such as VGG, would give stable results. Another reason for ruling out an automated tuning approach was the much larger hyperparameter space in the case of CNNs. The designed models were based on the choice of the following global hyperparameters (with some variations):

\begin{description}
\item[A. Conv Layers:] Following the VGG architecture, convolutional layers were added in pairs, doubling the number of filters after every pair, starting from 4 filters, up to 32.
    \item[Kernel:] A kernel with size 3 was chosen, which is small enough to capture any edge. Choosing a larger kernel would lead to more smoothing, which can be also achieved by stacking multiple small kernels together, and therefore not necessary. Canonically, a stride of 1 was chosen. The \texttt{padding='same'} option was chosen to keep the output size consistent with the input. The kernel initializer was chosen in agreement with the activation functions, as it was described for NNs in the previous section.
    \item[Activation:] Normally, convolutional layers do necessarily require activation functions to perform their task, however for some variants of the CNN models it was decided to add a \texttt{selu} or \texttt{lrelu} for each convolution.
\item[B. Dense Layers:] For the regressor part of the network, some of the best architectures yielded by the NN hyperparameter search were selected manually.
\end{description}

\newpage
An example CNN model is shown in  Fig. \ref{fig:CNN_screen4}:

\begin{figure}[h]
    \centering
    \includegraphics[width=\textwidth]{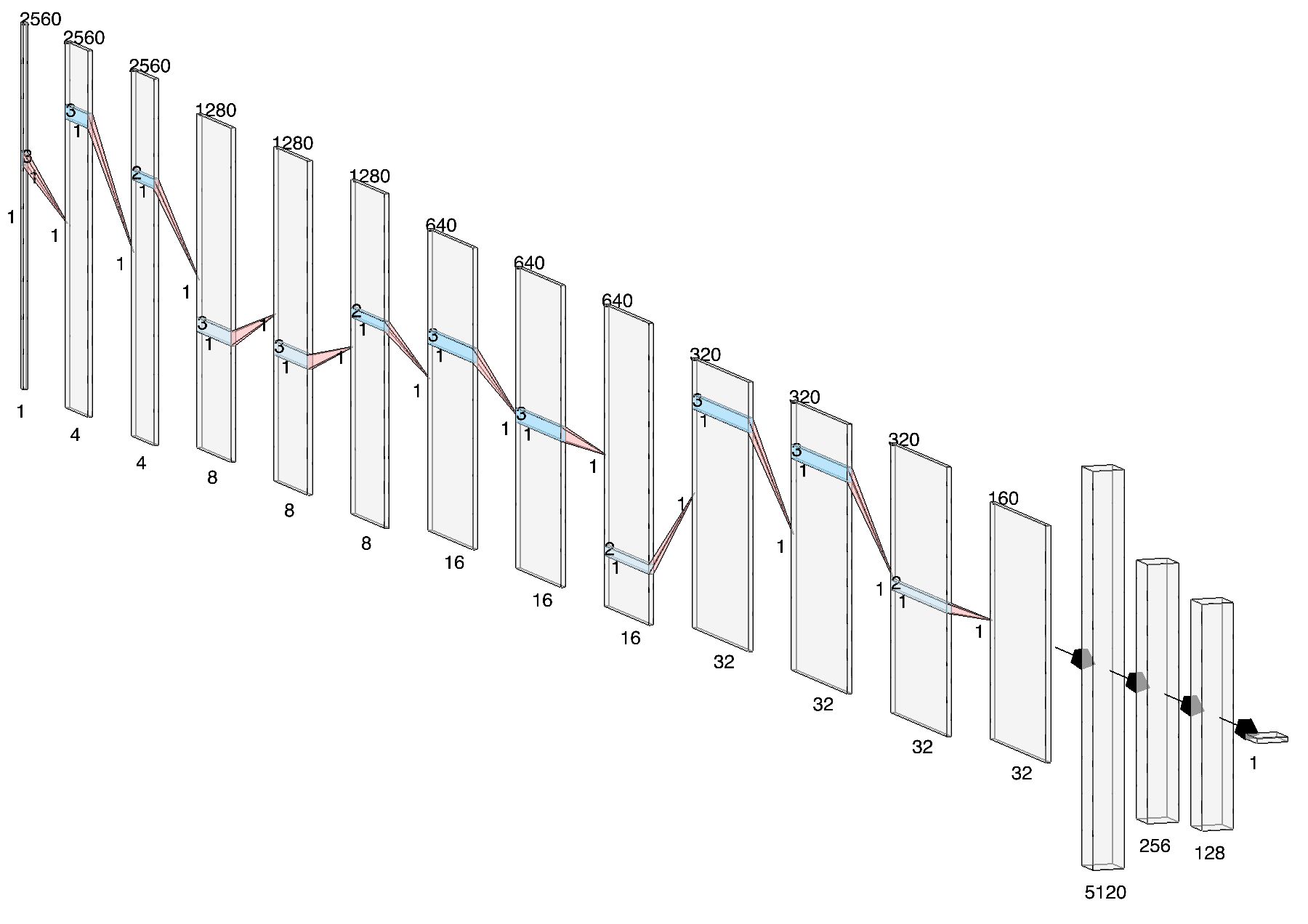}
    \caption{Outputs of CNN with 8 Conv layers (filter size=3). Note the pooling layers with filter size 2. The depth corresponds to the number of outputs (channels) of each layer. The transition from the last layer of the Conv sequence ($320\times32\times1$) to the input into the dense network ($5120\times1$) is done by a Reshape / Flatten layer}
    \label{fig:CNN_screen4}
\end{figure}

As can be seen from Fig. \ref{fig:CNN_screen4}, despite the 4 pooling operations, the input to the regressor has more features than the original input, which adds unwanted complexity to the problem. Correcting this would lead either to less convolutional layers (and therefore less capacity for feature extraction) or to a too deep network (with 3-4 more layers) which would be more difficult to train, given the size of the training set. A solution had to be found, which should allow keeping the capacity of the convolutional sequence, but alleviate the input complexity for the regressor.

In other words, the goal was to obtain a low-dimensional representation of the input, to simplify the regression problem.
The key to this turned out to be a convolutional encoder-decoder network, which will be described in the next section.

\newpage
\subsection{Autoencoders}

\begin{figure}[!h]
    \centering
    \includegraphics[width=\textwidth]{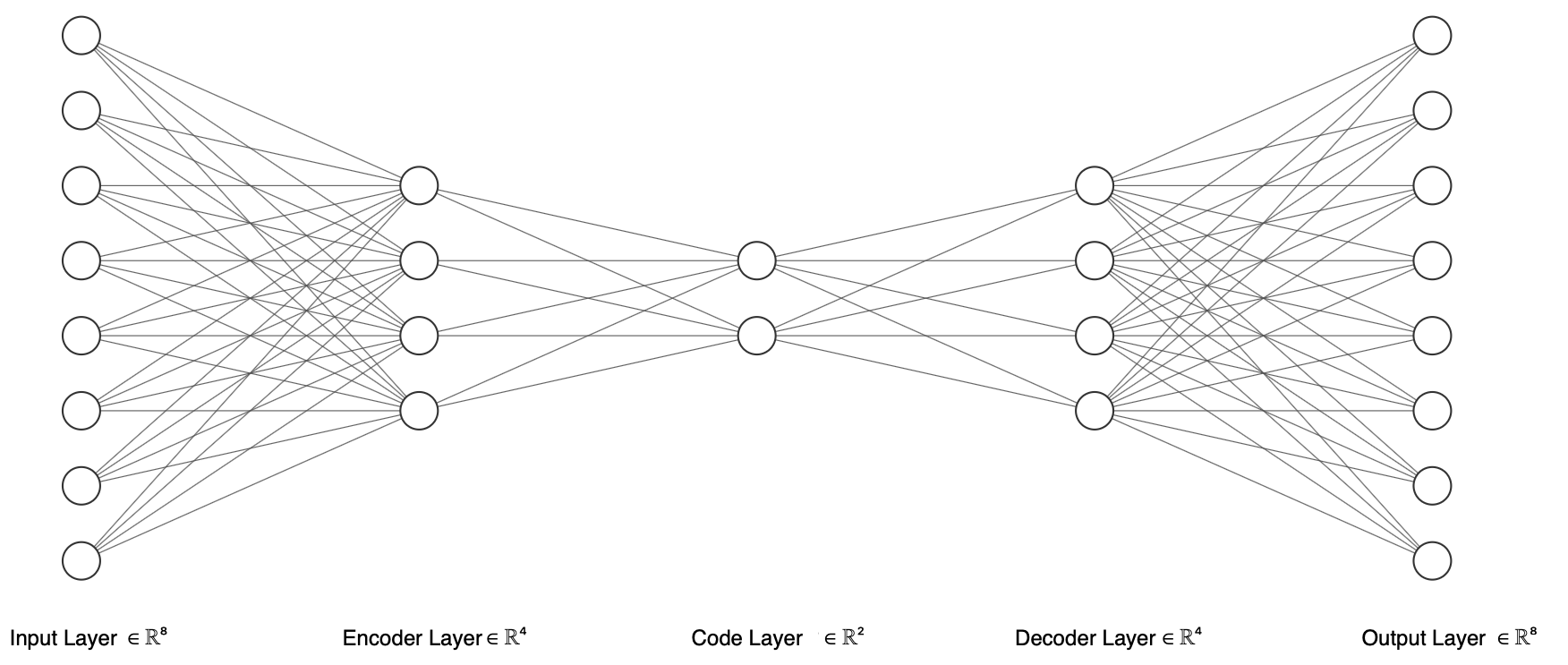}
    \caption{Example of a 5-layer stacked autoencoder}
    \label{fig:basicautoencoder}
\end{figure}

Autoencoders are neural networks that are able to learn a low dimensional representation of the input and reconstruct it from this low dimensional representation with minimal error. \cite{rumelhart_hinton_williams_1985}. The part responsible for learning the input, or encoding it, is called encoder, whereas the reconstruction, or decoding from the \emph{latent} (also called \emph{code}) layer, is carried out by the decoder.\\
An autoencoder is called \emph{stacked} when it consists of a sequence of encoding and decoding layers, where each layer operates on a different level of abstraction of the original input: the output of an encoder net shrinks with every layer, whereas in the decoder net it increases in size every layer. An example of such a network is shown in Fig. \ref{fig:basicautoencoder}

\bigbreak
If a noised image is fed into the input of the network while feeding a clean version into the loss function, such networks can be efficiently used for denoising. 
In autoencoders with linear activations, the nodes in the code layer directly correspond to the principal components from PCA. However, to the advantage of neural nets, typically notlinear activations are used, which enables the autoencoder to learn non-linear transformations.\\
Furthermore, these networks can be used for content generation: one can assume that sampling from the latent space and decoding it would generate an output similar to the data in the training set. Another option could be the interpolation between different training points in the latent space to produce a mixed output in the original space. However, this would only produce "realistic" results if the latent space is organized in a regular and continuous fashion: distinct features should be well separated from each other, with smooth transitions and without gaps. Otherwise, decoding a sampled point from the latent space would produce an "undefined" or nonsensical result, not representative of the original data.\\
The required constraints on the latent space can be achieved by a variational inference approach, which will be described in the following section.

\newpage
\subsection{Variational Autoencoders (VAE)}
\begin{figure}[!h]
    \centering
    \includegraphics[width=\textwidth]{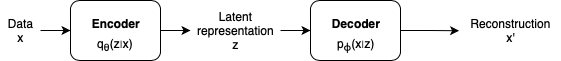}
    \caption{Encoder and Decoder formulateed as conditional distributions}
    \label{fig:vae1}
\end{figure}

In the previous section we have mentioned that it is not possible to achieve continuity in the latent space (and thus meaningful output when decoded) without training the model for an arbitrary input. One might think that this would require sampling from an infinitly large distribution of the input data (which is impossible), but there are more suitable ways to enforce a continuous latent space. 
\bigbreak
Until now, the latent space was defined as a vector with size equal to the number of neurons in the code layer. But instead of learning a vector representation of the latent space, the network could also encode it as a distribution by learning its parameters.\\
From a probabilistic perspective, the encoder and decoder networks can be reformulated as conditional distributions: $q_\theta (z|x)$ for the encoder and $p_\phi (x|z)$ for the decoder, where x represents the datapoint, z its latent representation. Then, $\theta$ describes the weights and biases of the encoder and $\phi$ describes the weights and biases of the decoder. Schematically, this setup is shown in Fig. \ref{fig:vae1}:

\bigbreak

Thus, a Variational Autoencoder tries to learn a specific probability model for the data $x$ and the latent variables $z$. The joint probability of this model can be written as $p(x,z) = p(x|z) p(z)$, where $p(z)$ is a prior for the latent variables and $p(x|z)$ the likelihood of generating $x$ given $z$. However, what we are trying to model with the encoder is the inference of $z$ given $x$, which can be written using Bayes' rule as :

\begin{equation}
    p(z|x) = \frac{p(x,z)}{p(x)} = \frac{p(x|z) p(z)}{p(x)}
\end{equation}

For the moment, let's assume we know $p(x|z)$. Unfortunately, $p(x)=\int p(x|z)p(z) dz$ takes exponential time to compute as it requires the calculation of the integral for all configurations of $z$. However, we can transform this problem into an optimization task and approximate the (true) posterior distribution $p(z|x)$ by a parametrized distribution $q_\theta (z|x)$, which corresponds to the encoder network. A common choice is to parametrize it as a family of Gaussians with $\mu(x_i)$ and $\sigma^2(x_i)$ as mean and variance for each data point $x_i$. In other words, the encoder network is trained to learn the (globally, for all $x_i$) optimal $\theta^*$ which would allow the network to yield the $\mu(x_i)$ $\sigma^2(x_i)$ that (locally, for given $x_i$) optimally approximate the true distribution of $p(z|x)$ for a given $x_i$.\\
A measure for the approximation quality of a distribution is the so called Kullback-Leibler Divergence, which is 0 for a perfectly approximated distribution and positive otherwise:

\begin{equation}
\mathbb{KL}(q_\theta (z|x) || p(z|x))
\label{eq:KLbasic}
\end{equation}

The goal of the optimization would be to find :

\begin{equation}
\begin{split}
\theta^* & = \argmin_{\theta} \mathbb{KL}(q_\theta (z|x) || p(z|x)) \\
& = \argmin_{\theta} \mathbb{E}_{z\sim q} \left[\log q_\theta (z|x)\right] - \mathbb{E}_{z\sim q} \left[\log \frac{p(x|z)p(z)}{p(x)} \right]\\
& = \argmin_{\theta} \mathbb{E}_{z\sim q} \left[\log  q_\theta (z|x)\right] - \mathbb{E}_{z\sim q} \left[\log  p(z)\right] - \mathbb{E}_{z\sim q} \left[\log p(x|z)\right] + \mathbb{E}_{z\sim q} \left[\log  p(x)\right] \\
& = \argmax_{\theta} \mathbb{E}_{z\sim q} \left[\log  p(x|z)\right] - \mathbb{KL}(q_\theta (z|x) || p(z))
\end{split}
\label{eq:vae_kl_opt}
\end{equation}
Note that to go from line 2 of Eq. \ref{eq:vae_kl_opt} to line 3 Bayes' rule and the product rule for logarithm to expand the joint likelihood was used. The last term in line 3 does not depend on z and therefore does not depend on $\theta$ so it has no importance for the optimization. The last line of Eq. \ref{eq:vae_kl_opt} is called Evidence Lower Bound -- ELBO.
This term corresponds to the negative of the loss funtion in VAEs: The first term, which can be interpreted as the negative log-likelihood is the reconstruction loss of the Autoencoder. The second term, which corresponds to the KL divergence is a regularization term, which ensures that the latent representations $z$ of $x$ follow the distribution of $p(x)$ (which will be discussed in more details soon).\\
For demonstration purposes, it was assumed previously that $p(x|z)$ which describes the decoding scheme, is known. However, the \emph{true} distribution is not known, but can be inferred from previously made assumptions on $p(z)$:\\
When choosing the prior that follows a gaussian distribution, the last KL divergence would ensure that the posterior $q_\theta (z|x)$ follows the same distribution. Since the product of two gaussians is a gaussian, it can be admitted that the likelihood $p(x|z)$ is also a gaussian.
Summarizing the above, the distributions of interest are:

\begin{equation}
    \begin{split}
        &\text{Prior} : &p(z) = \ & \mathcal{N}(0, I) \ & \text{a constrain we set on the latent space z}\\
        &\text{Likelihood}  : &p(x|z) = \ & \mathcal{N}(f(z), cI) \ & \text{where f is a function and c a parameter}\\
        &\text{Posterior}  : &q_\theta (z|x) = \ &  \mathcal{N}(\mu(x), \sigma(x)I) \ & \text{where $\mu$ and $\sigma$ are functions} 
    \end{split}
\end{equation}

Since neural networks are universal function approximators, they can be used to approximate the above distributions through the learned functions $\mu(x)$ and $\sigma(x)$ by the encoder, and through $f(z)$ by the decoder. Finally, rewriting Eq. \ref{eq:vae_kl_opt}, now with $p_\phi (x|z)$ (where $\phi$ are the weights and biases of the decoder network, now also part of the optimization objective) gives:

\begin{equation}
\begin{split}
(\theta^* , \phi^*) & = \argmax_{\theta, \phi} \mathbb{E}_{z\sim q} \left[\log  p_\phi (x|z)\right] - \mathbb{KL}(q_\theta (z|x) || p(z))\\
& =  \argmax_{\theta, \phi} \mathbb{E}_{z\sim q} \left[- \frac{\norm{x - f(z)}^2}{2c}  \right] - \mathbb{KL}(q_\theta (z|x) || p(z))
\end{split}
\label{eq:vae_kl_opt2}
\end{equation}

Where the transformation of the first term is possibile due to the fact that the likelihood is gaussian. In this form, it becomes even more apparent that this term describes the negative mean squared error between the original value of x and it's reconstruction, scaled by a factor of $1/2c$, which can be interpreted as the confidence of the reconstruction.\\

Lastly, one important aspect related to the training of the networks needs to be addressed. The sampling operation from the normal distribution $z \sim \mathcal{N}(0, I) $, as it is defined now, is not differentiable, hence training is not possible. This is why the sampling operation needs to be rewritten as:

\begin{equation}
    z = \mu(x) + \sigma(x)\epsilon \ \text{with} \ \epsilon \sim \mathcal{N}(0, I)
\end{equation}

When differentiating, this enables taking the derivative of the parameters of the random function instead of differentiating the random function itself.

\bigbreak

\begin{figure}
    \centering
    \includegraphics[width=\textwidth]{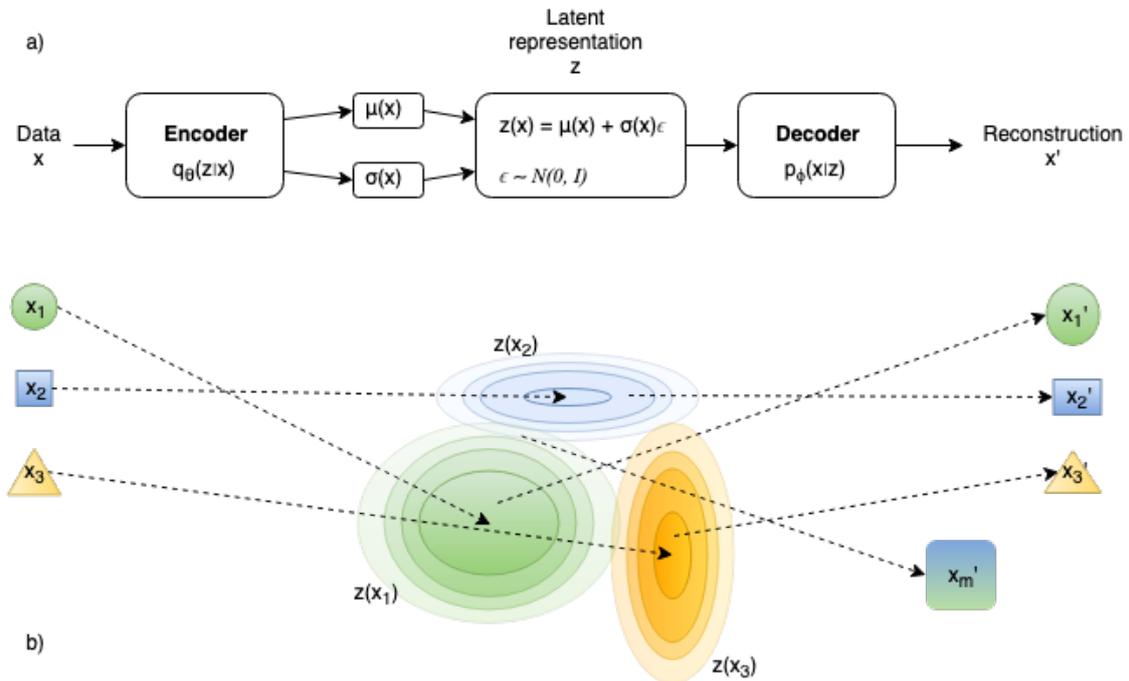}
    \caption{a)Schematic representation of VAE showing the reparameztrization of z. b) Toy example showing encoding-decoding process. Note the reconstruction from point between $z(x_1)$ and $z(x_2)$ results in something similar to 'interpolation'}
    \label{fig:vaelatenspace}
\end{figure}

Now that the that the main principles of the VAE have been clarified, a practical toy example of the encoding-decoding process can be analyzed, illustrated in Fig. \ref{fig:vaelatenspace} b): Assuming the network is trained, and the globally optimal $\theta^*$ and $\phi^*$ are found and given input data sample $x_1$, the decoder outputs values $\mu(x_1)$ and $\sigma(x_1)$ for a Gaussian distribution. In other words, the sample $x_1$ is encoded as $z \approx \mu(x_1)$, with some uncertainty $\sigma(x_1)$. Then the decoder draws a sample from this distribution $q_\theta(\mu(x_1),\sigma(x_1))$ which corresponds to $z'$, which can be different from the 'true' z. However, the shape of the distribution enforces the sampled point $z'$ to be 'close' to the 'true' z, allowing the decoder to reconstruct to x' which is as close as possible to the original x.\\ Now, if the network had to decode a sampled $z_m$ with $\mu$ between $\mu(x_1)$ and $\mu(x_2)$ and $\sigma$ between $\sigma(x_1)$ and $\sigma(x_2)$, due to continuity of the latent space, the reconstruction would result in an interpolation $x_m'$ between $x_1$ and $x_2$. Conversely, a distanced point $x_3$ would be encoded to $\mu(x_3) \not \approx \mu(x_1), \mu(x_2)$ and $\sigma(x_3) \not \approx \sigma(x_1), \sigma(x_2)$ and hence the sampled point would be distanced from the previous two and lead to a different reconstruction $x_3' \not \approx x_1', x_2'$.

Having acquired an intuition for the essential principles of VAE, it is now possible to discuss how to embed a regressor into this framework.

\newpage
\subsection{Convolutional VAEs for Regression}

In this work, the standard VAE model was extended in two ways:
Convolutional and pooling layers were added to the encoder network, which would act as feature detectors as discussed in the previous section. On the decoder side this structure was mirrored by adding the same convolutional layers in reverse order with upsampling layers, instead of the pooling layers.\\
Additionaly, a regressor network, similar to previously described architectures was added to the output of a copy of the encoder network. 
Then the original encoder-decoder network was trained (on the whole spectral dataset) until a good reconstruction error was achieved. The weights of the encoder were then loaded into the (encoder+regressor) and the corresponding layers where flagged as untrainable, which would ensure that further training would not alter the learned representation of the latent space.\\
Unfortunately, it was quickly realized that a regression on the learned latent space did not yield good results. The reason for this was that the latent space was optimized solely with the objective of minimizing the reconstruction error, without imposing the minimization of the regression error. It could be concluded, that the low-dimensional representation, which allowed for a very good reconstruction of the spectrum, did not capture the required information necessary for a good regression. This suggests that the majority of the learned details of the spectrum where not relevant for the prediction of the target concentration. \\
Therefore, it was decided to build a composite model, consisting of an encoder, decoder and regressor, where all layers would be jointly trained on the spectral data as well as the target concentrations. As a result, the loss function consisted of three terms:
\begin{description}
    \item[Reconstruction loss:] MSE(spectrum, reconstructed spectrum) -- \texttt{rec\_loss}
    \item[Regression loss:] MSE(target concentration, predicted concentration)  -- \texttt{reg\_loss}
    \item[KL Divergence:] regularization term constraining the latent space  -- \texttt{kl\_loss}
\end{description}

\begin{figure}[h!]
  \begin{minipage}[ht]{0.24\textwidth}a)
    \includegraphics[width =\textwidth]{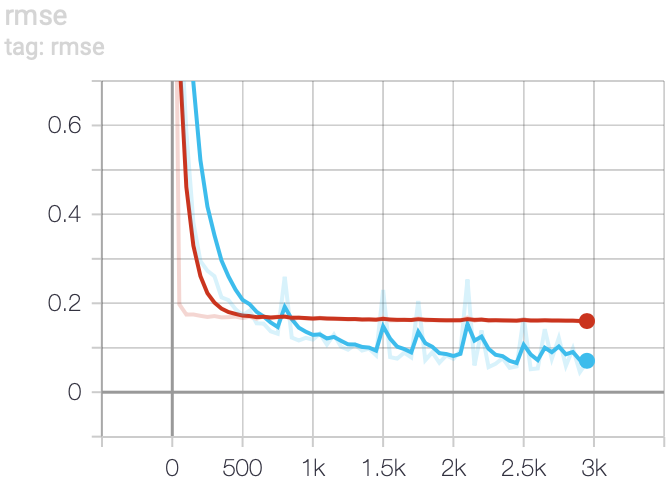}
  \end{minipage}
  \begin{minipage}[ht]{0.24\textwidth}
    \includegraphics[width = \textwidth]{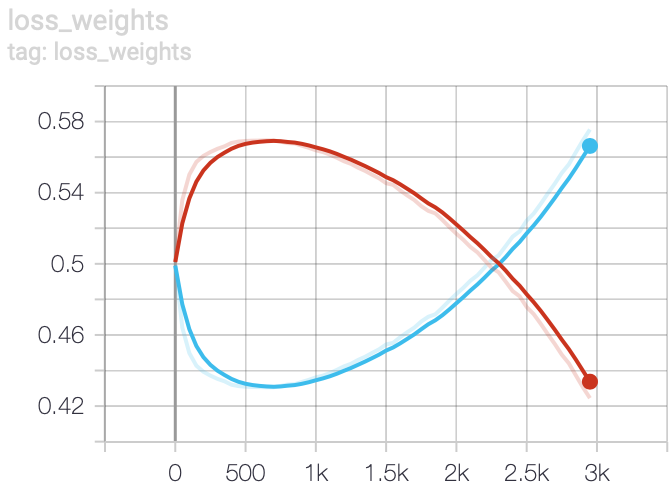}
  \end{minipage}
  \begin{minipage}[ht]{0.24\textwidth}b)
    \includegraphics[width = \textwidth]{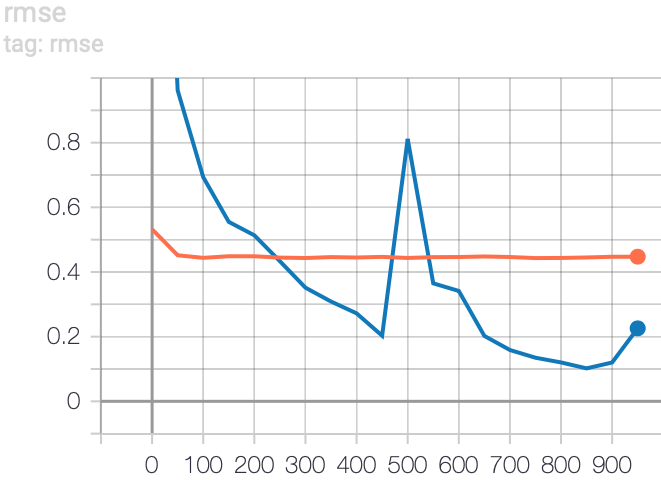}
  \end{minipage}
  \begin{minipage}[ht]{0.24\textwidth}
    \includegraphics[width = \textwidth]{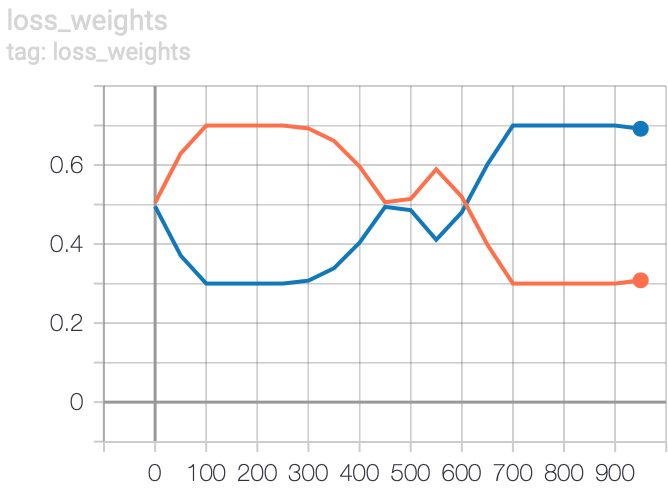}
  \end{minipage}
  \caption{Evolution of (unweighted) losses (RMSE) and weights learned for two different models a) and b). Light blue and dark blue lines show regression loss and red and orange lines show the reconstruction loss (both before multiplication by the weights). Note how in both cases, weights are at first higher for reconstruction loss and then shift towards the regression loss (after reconstruction loss stabilized). It is also interesting to see how the LossWeighter handles the sudden jump in regression loss for model b) between 400 and 500 epochs (might be explained by the too steep increase of the weight, which is corrected after some epochs)}
  \label{img:lossweighter}
\end{figure}

In the previous section the parameter $c$ defining the importance of the \texttt{rec\_loss} compared to the regularization was mentioned. The addition of the \texttt{reg\_loss} term required introducing an additional weighting factor. Intuitively, one would think that in order to perform a good regression, first the reconstruction should be on point. This suggests a time dependence (\# of training iterations) for the weights. The idea was to give high value to the \texttt{rec\_loss} in the beginning and neglect the \texttt{reg\_loss}, and after a certain number of training iterations gradually increase the weight of the \texttt{reg\_loss}. This procedure is sometimes called \emph{Weight Annealing}.\\
Unfortunately, this does not really simplify the problem, as the optimal annealing curve can only be found through trial and error, and would obviously change with model complexity and hyperparameters.\\
However, since the problem is already formulated in a learning framework, there is little difficulty to approximate the optimal annealing curve with a neural network. In order to do so, a custom layer was developped, called \texttt{LossWeighter}:

\begin{description}
    \item[LossWeighter:] Acceptes two inputs, corresponding to the two losses that need to be weighted. The output is then calculated as:
    \begin{equation}
        \texttt{tot\_loss} = w \times \texttt{reg\_loss} + (1 - w) \times \texttt{rec\_loss} \ \text{with : }\ w_{min} < w < w_{max}
    \end{equation}
\end{description}
The weight was initialized with $w = 0.5$ and constrained by $w_{min} = 0.2$ and $w_{max} = 0.8$. The LossWeighter layer was the final layer of the network and its output was registered as a custom loss, without specifying any other losses in the \texttt{.build()} method.\footnote{Usually, the standard loss functions have two inputs: the output of the network and the target input. However, in \texttt{keras} it is possible to define custom loss layers and custom loss functions, which do not have restrictions on the input. Doing so allows to use the optimizer algorithm to perform the minimization of an arbitrary quantity (use it like a numerical solver)}
An example of the learned annealing curves is shown in Fig. \ref{img:lossweighter}.

\newpage

\bigbreak

Including an additional loss function was not the only possible improvement. In the KL regularization term of \ref{eq:vae_kl_opt}, instead of regularizing the posterior with a normal prior, it can also be regularized with a \emph{target-specific} prior $p(z|c)$, where $c$ is the target concentration, as proposed by \cite{DBLP:journals/corr/abs-1904-05948}.

The resulting objective is then formulated by :

\begin{equation}
\begin{split}
(\theta^* , \phi^*) & = \argmax_{\theta, \phi} \log  p_\phi (c|x) +
\mathbb{E}_{z \sim q} \left[\log(p_\phi (x|z) \right]
- \mathbb{KL}(q_\theta (z|x) || p_\phi (z|c))
\end{split}
\label{eq:vaereg_kl}
\end{equation}

In this setting, the main difference to a traditional VAE is the formulation of $p(z|c)$, which is called \emph{latent generator} and serves to generate samples from the latent space, given the regression target. It is formulated as $p(z|c) \sim \mathcal{N}(u^T c, \sigma_{gen}I)$, where $u^T u = 1$.
$u$ is called \emph{disentangled dimension}\cite{DBLP:journals/corr/abs-1904-05948}. It's purpose is to map different $c$ to different $z$ (basically it's augmenting the dimensionality of $c$, producing a vector $z$ where for different $c$ the values of $z$ are separated in a 'meaningful' way). $\sigma_{gen}$ is the uncertainty of this mapping, currently ignored (commented out in code).
This distribution is also parametrized with a neural network which shares weights with the encoder as follows: the output of the regressor (yields c) is fed into a (1-layer) network with an output the size of the latent vector z. This single layer is equivalent to the previously introduced mapping $u^T c$.\\
Looking at the terms in Eq. \ref{eq:vaereg_kl}, in the first term the \texttt{reg\_loss} can be recognized, in the second the \texttt{rec\_loss} and in the last term the desired regularization.
In the actual code, the first two can be trivially formulated using the inputs and outputs of the network and the standard MSE.\footnote{Since \texttt{reg\_loss} and \texttt{rec\_loss} are defined as custom loss layers in \texttt{keras} and inner layers don't have access to the targets (y\_true) as only loss functions added via the \texttt{.build()} method do, the only way to fetch the targets was to pass them in the input layers.}\\
The regularization term can be expaneded using the KL divergence formula for gaussians:

\begin{equation}
\begin{split}
\mathbb{KL}(\mathcal{N}(\mu_1,\sigma_1), \mathcal{N}(\mu_2,\sigma_2)) = \log \sigma_2 - \log \sigma_1 + \frac{\sigma_1^2 + (\mu_1 - \mu_2)^2}{2\sigma_2^2}-\frac{1}{2}
\end{split}
\label{eq:klbasic}
\end{equation}

Where the layers of the network are named as follows:
\begin{equation}
\begin{split}
\texttt{z\_log\_var} & = \log \sigma_1 \quad  & \texttt{z\_mean} = & \mu_1 \quad \text{probabilistic encoder $q(z|x)$}\\
\texttt{pz\_log\_var} & = \log \sigma_2 \quad & \texttt{pz\_mean} = & \mu_2  \quad \text{latent generator $p(z|c)$} \\
\end{split}
\label{eq:klnames}
\end{equation}

The computation graph of the latent part of the network is shown in Fig. \ref{fig:vae_latent}:

\begin{figure}[!h]
    \centering
    \includegraphics[width=\textwidth]{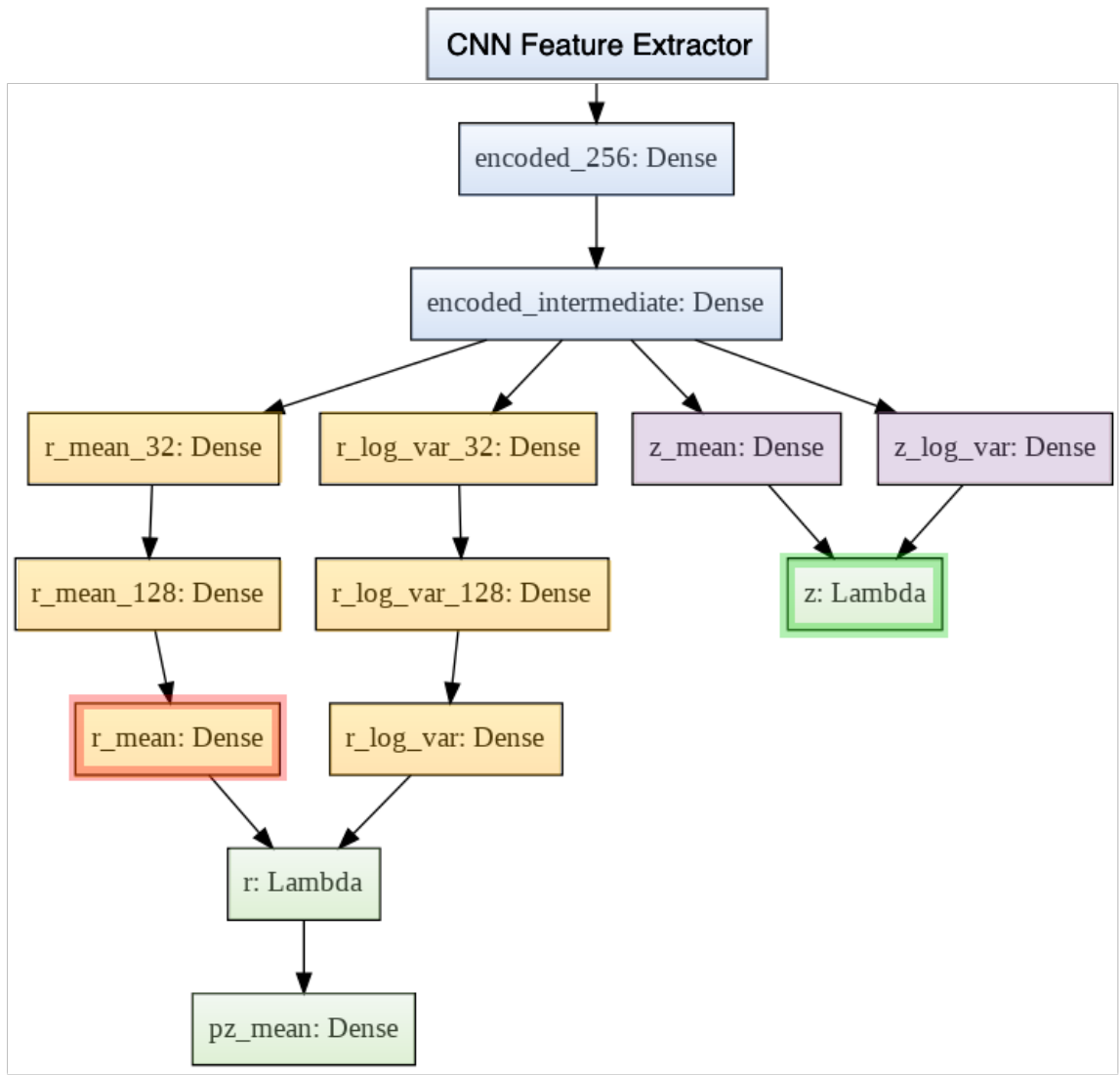}
    \caption{Computation graph of the latent (core) of VAE Rergressor network. The encoder is shaded in blue, including the CNN Feature extractor. Its last layer, named \texttt{encoded\_intermediate} has 32 nodes, whereas the latent layers\texttt{z}, \texttt{z\_mean} and \texttt{z\_log\_var} have 64 nodes. The regression net is shaded in orange and has (32,128,1) nodes, with the output framed with a red box. The generative part of the network is shaded in green: r and z originate from sampling the gaussian distributions parametriezed by the previous layers. The input to the decoder is framed by a green box.}
    \label{fig:vae_latent}
\end{figure}

\newpage
An important difference from earlier VAE regression models is that here the regressor shares layers with the encoder network, except for the latent layers \texttt{z\_mean} and \texttt{z\_log\_var}. This is because in this setting the regressor is used to impose the previously discussed regularization on z in form of \texttt{pz\_mean} (and \texttt{pz\_log\_var} which is not included in the current version).

\clearpage
\section{Model Fusion}

The idea behind model fusion is to use the predictions of multiple models to come up with a single, more accurate prediction. Ideally, a fused model should be more robust (produce errors with smaller variance compared to the individual models).\\
A fused model consists of the individual (base) models and a meta estimator operating on the output of the base models. This is sometimes called \emph{stacking}.\cite{DBLP:journals/corr/abs-0911-0460} There are different approaches to design this meta-estimator:

\subsection{Linear Stacking}
\label{subsec:linearstacking}

In the most simple form of stacking, a regression model is applied to the output vector containing the base model's predictions:

\begin{equation}
    y_{fused}(x) = \vect{w} \cdot \vect{y_{base}}(x)
\end{equation}

This results in a linear combination of the base models' predictions with no further dependence.

\subsection{Feature-Weighted Linear Stacking (FWLS)}

For the FWLS method the fusion weights are dependent on the input x.\cite{DBLP:journals/corr/abs-0911-0460} The task of the meta-estimator is to learn this feature dependence of the weights, which can be written as a linear combination, using the matrix $\vect{V}$:

\begin{equation}
\begin{split}
    y_{fused}(x) & = & \vect{w}(x) \cdot \vect{y_{base}}(x) \\
    & = & \vect{f}(x) \cdot \vect{V} \cdot \vect{y_{base}}(x)
\end{split}
\end{equation}

Once the feature function $\vect{f}(x)$ is chosen, this problem is equivalent to solving a linear system.

In the beginning of the project, the implementation idea of this method was dismissed, because it was not clear how to select the feature function. Passing the whole set of features as $\vect{f(x)} = \vect{x}$ would make "little sense since these are already used by the base estimators" [former team member, August 2019]

\begin{figure}[hb]
    \centering
    \includegraphics[width=0.36\textwidth]{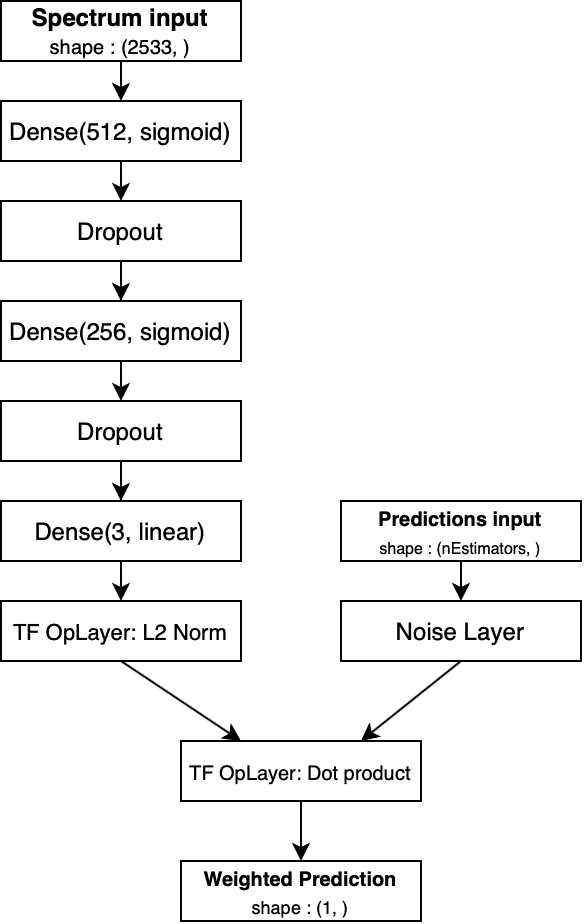}
    \caption{FWLS implementation scheme with a Neural Network. The network has two inputs: the input spectrum and the input predictions of the base estimators, which are estimated from the same input spectrum. The network learns to weight the predictions based on the features extracted from the input spectrum. A noise layer is active during the training phase of the network which is used to simulate 'test predictions' from the training predictions.}
    \label{fig:FWLS_NN}
\end{figure}

Later on, it was decided to use a NN as a feature detector to learn $\vect{f(x)}$. The solution is presented in Fig. \ref{fig:FWLS_NN}. A 3-layer network (\emph{feature detector}) learns to output normalized weights for the predictions vector based on the input spectrum. A series of experiments have shown that the normalization layer might not be necessary, however it was decided to keep this layer as it facilitates the visualization and interpretation of the results. Another considered solution, was to concatenate the predictions with the the learned features (may be of arbitrary size, but chosen to be same as the length of the prediction vector) and feed it into another dense layer for the final output. This idea was dropped, as experiments showed that in this case the network would completely neglect the spectral input for the weighting (constant output for the left side of the network). This behaviour might improve with an increased number of output features from the feature extractor, however, this hypothesis was not tested yet.\\
One distinctive feature of this design is the Gaussian noise layer applied to the input predictions vector. This addition was originally more a necessity, but turned out to be a significant improvement. It is only active during the training phase of the network and serves mainly as a regularizer, which makes the input predictions (in the current implementation coming from the training set, so they are assumed to be 'well fitted' to the training set) look more like test predictions (with a larger error compared to the train set). This is mainly a trick, required for the current implementation of the training and evaluation algorithm (will be discussed in the following chapter), which allows to reuse scarce training data for both the base estimators and the meta estimator.

\subsection{Best Linear Unbiased Estimator (BLUE) Method}
\label{subsec:blue}

\begin{figure}[hb]
\centering
    \includegraphics[width = 0.98\textwidth]{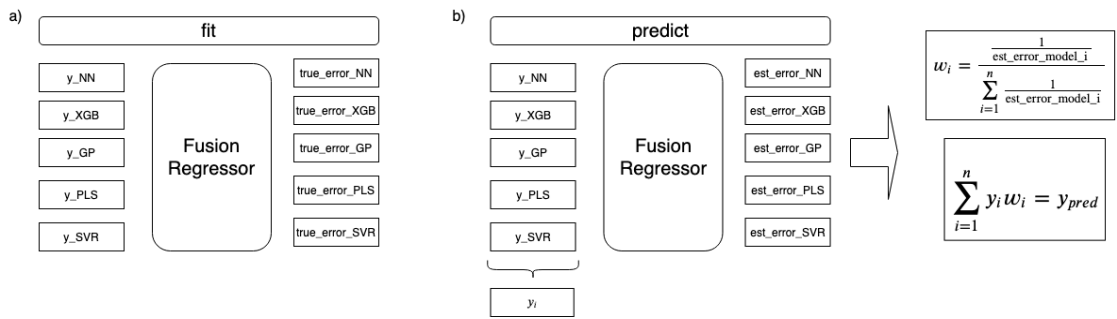}
    \caption{Implementation of BLUE using the estimated errors for weighting base models' predictions. a) The fusion regressor fits the vector of base models' predictions to the errors. b) The fusion regressor predicts the errors from the vector of base models' predictions and calculates the relative weights of each model based on the estimated errors. The resulting prediction is a linear combination of these weights and the individual predictions}
    \label{fig:fusionBLUE1}
\end{figure}

The name of this method might seem misleading: it is unbiased if the true uncertainties of the base estimators and their correlations are known, however this is not the normal case, as usually only estimates of these uncertainties are available. Hence, BLUE may exhibit bias. However, the authors claim that an iterative application of their algorithm may reduce the bias of the fused model.\cite{Lista_2017}\\
The fundamental idea of this method is to weight the base models' predictions with relative weights inversely proportional to the uncertainty of the prediction.
Same as in the original raman project, in this work, this uncertainty is quantified as the \emph{estimated error}: This \emph{estimated error} is predicted by a regression model (currently implemented are: Polynomial Regression, KRR, DT, NN and XGB, chosen as default for the evaluation), which fits the vector of the base model's predictions to the vector of errors of those predictions. The weight of each model corresponds to the inverse of the estimated error. Weights are then normalized. The prediction is the calculated as a linear combination of the relative weights and the individual model's predictions. A flowchart of the model is shown in Fig. \ref{fig:fusionBLUE1}.

\cleardoublepage

\chapter{Implementation Details}

\begin{figure}[h]
    \centering
    \includegraphics[width=0.98\textwidth]{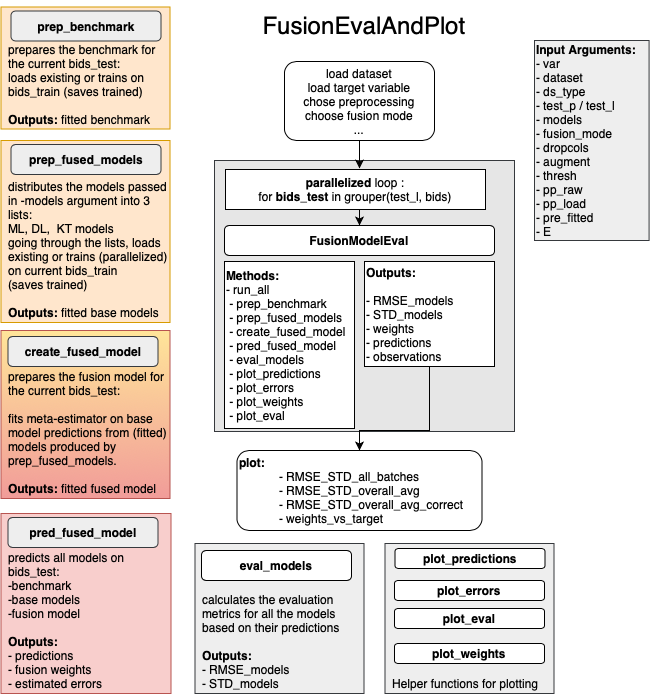}
    \caption{Flowchart of the training and evaluation script for the different models, including the fused model. Training steps shaded in orange, prediction steps shaded in red. The gradient indicates that in \texttt{create\_fused\_model} first the base models need to produce predictions before fitting the fusion model.}
    \label{fig:fusionevalplot1}
\end{figure}

This chapter briefly covers some of the implementation details that were not yet discussed, namely the training and evaluation sequence of the models and how the results were obtained. The flowchart in Fig. \ref{fig:fusionevalplot1} shows the structure of the script used to obtain the results. It is based on the FusionModelEval class, which contains the methods for training, predicting, evaluating and plotting.\\
First the user specifies the dataset, test ratio or length, target variables, preprocessing and fusion mode in the input arguments. Based on these arguments, train and test batch id's (bids) are generated, and fed into as arguments into a parallelized loop (this corresponds to the outer CV loop discussed in \ref{fig:nestedcv}). The parallelization is achieved using \texttt{joblib.Parallel}: this allows to run each iteration as a separate process (worker) on a dedicated core of the machine (default \# of workers : 12).\\
In each iteration, a FusionModelEval object is created. It defines methods for training, prediction, evaluation and plotting and keeps track of the training and model configuration (train/test bids, model names, augmentation etc...). A brief description of the methods is given in Fig. \ref{fig:fusionevalplot1}.\\

All newly developed regression models were added in form of new classes (except for XGB):

\begin{description}
\item[NNRegressor:] A base class for NN models, with 4 predefined architectures. Acts as a wrapper for Sklearn models (implements \texttt{.fit()} and \texttt{.predict()}) and implements a \texttt{.load\_model()} method which is help to load custom objects into the keras model (custom layers, custom lossfunctions, etc...).
\item[CNNRegressor:] Subclasses NNRegressor, with different predefined CNN architectures.
\item[VAERegressor:] Subclasses NNRegressor, with predefined VAE architecture and helper functions.
\item[VAERegressor:] Subclasses NNRegressor, with predefined VAE architecture and helper functions.
\item[KTRegressor:] loads and builds keras models obtained from \texttt{keras-tuner} optimization and wraps them as a Sklearn model.
\item[FusionModelRegressor:] Implements the metaestimator for fusion. Supported types of metaestimator: Polynomial Regression, KRR, Tree, XGB, NN. Supported fit targets: \texttt{true\_y} -- corresponds to basic linear stacking as in \ref{subsec:linearstacking} \texttt{errors} -- corresponds to BLUE as in \ref{subsec:blue}. 

\end{description}

\begin{figure}[h!]
    \centering
    \includegraphics[width=0.88\textwidth]{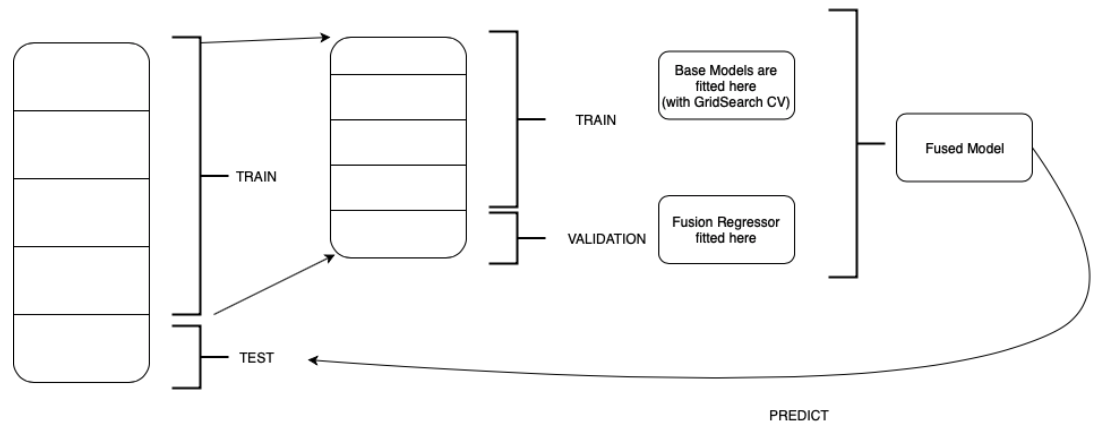}
    \caption{Flowchart depicting the training of the fusion regressor. The basemodels and the fusion regressor should be trained on different sets. After fitting the fusion regressor, the base models can be refitted on the train $+$ validation set for better prediction performance. However, this increases the overall training time of the model. Also it would be more difficult to setup a reloading scheme for pretrained models (because of the difficulty to keep track of seen/unseen data in different configurations).}
    \label{fig:fusionregressor1}
\end{figure}

\newpage
At this point, it is important to note that the current implementation relies on reloading previously trained models when possible: saved models encode the \texttt{test\_bids} in their name, so they are only reloaded for the current \texttt{test\_bids}. This makes the process of model fusion slightly complicated: Ideally, the predictions that are necessary to train the meta-estimator should come from previously unseen data, as shown in Fig. \ref{fig:fusionregressor1}. This would require an additional split of the (unseen) dataset which is already small. Furthermore, the base models should be retrained on full training set after fitting the meta-estimator in order to not waste training data. This idea was dropped in favor of an acceptable training for the models and the ability to reload pretrained models for a given test set.

\cleardoublepage

\chapter{Results and Discussion}\label{sec:resultsTiter}

In this chapter, the performance of the previously introduced models will be compared for different target variables. This chapter is divided into different sections for each target variable.
The effect of different proposed improvement methods will be highlighted in the first section (for Titer). The most effective methods will be then tested for the other target variables in the remaining sections.\\

\section{Titer}

In order to set a common baseline, each section starts with the comparison of the initial ML models DS1 with the standard preprocessing (previously developed for the raman project). Within each section, further subsections describe obtained results for different pretreatment settings and models, according to the relevance of the results. Main results for DS2 are mentioned at the end of each subsection for reference. The different settings are abbreviated as follows:

\begin{description}
    \item [STD:] Standard pretreatmnet (SG Filter + Drop Columns + SNV Scaler + Standart Scaler), also refers to the dataset with outliers in the tails removed (concentrations > 6.5 g/L for Titer)
    \item [NOPP:] Drop Columns + Standard Scaler
    \item [VIP:] Input variable selection according to variable importance, as in Fig. \ref{img:ds13varimportanceall}. 
    \item [RES:] Augmentation with resampling: according to Algorithm in Appendix A. \ref{pythonresampling}. 
    \item [AUG:] Augmentation without resampling (for ref. to RES): Dataset augmented by 100\% 
    \item [CLN:] Datapoints where the fit error (of an earlier KT model) was larger than 1 standard deviation were removed and models were refitted on the cleaned set (CLN set has 975/1081 data points).
\end{description}

The settings may be combined as follows:

\begin{figure}[h]
  \begin{minipage}[h]{0.48\textwidth}
    \begin{itemize}
        \item STD + RES
        \item STD + AUG
        \item STD + VIP 
        \item STD + VIP 
        \item NOPP + AUG
        \item NOPP + RES
        
    \end{itemize}
  \end{minipage}
  \begin{minipage}[h]{0.48\textwidth}
    \begin{itemize}
        \item CLN + RES
        \item CLN + AUG
        \item CLN + VIP 
        \item CLN + NOPP + AUG
        \item CLN + NOPP + RES 
    \end{itemize}
 \end{minipage}
 \label{lst:ppsettings}
 \end{figure}

In the following comparison tables, if not otherwise stated, only the last setting is indicated for combinations of settings and the base setting is presented in the first column.

\subsection{ML Models}

\newtoggle{inTableHeader}
\toggletrue{inTableHeader}
\newcommand*{\StartTableHeader}{\global\toggletrue{inTableHeader}}%
\newcommand*{\EndTableHeader}{\global\togglefalse{inTableHeader}}%

\let\OldTabular\tabular%
\let\OldEndTabular\endtabular%
\renewenvironment{tabular}{\StartTableHeader\OldTabular}{\OldEndTabular\StartTableHeader}%

\newcommand*{\MinNumber}{0.2}%
\newcommand*{\MidNumber}{0.27} %
\newcommand*{\MaxNumber}{0.35}%

\newcommand{\ApplyGradient}[1]{%
  \iftoggle{inTableHeader}{#1}{
    \ifdim #1 pt > \MidNumber pt
        \pgfmathsetmacro{\PercentColor}{max(min(100.0*(#1 - \MidNumber)/(\MaxNumber-\MidNumber),100.0),0.00)} %
        \pgfsetfillopacity{0.1}\colorbox{red!\PercentColor!yellow}{\pgfsetfillopacity{1}#1}
    \else
        \pgfmathsetmacro{\PercentColor}{max(min(100.0*(\MidNumber - #1)/(\MidNumber-\MinNumber),100.0),0.00)} %
        \pgfsetfillopacity{0.3}\colorbox{green!\PercentColor!yellow}{\pgfsetfillopacity{1}#1}
    \fi
  }}

\newcolumntype{R}{>{\collectcell\ApplyGradient}c<{\endcollectcell}}

Originally, the RES and AUG methods were intended to be used with DL models only, however, it can still be interesting to see the effect on ML models. Tab. \ref{tab:Titer_ML_STD_all} shows a comparison of the ML models in the STD, STD + AUG and STD + RES settings.
It can be seen that PLS outperforms the other ML models in every setting except for AUG, where XGB and SVR performed better. Fitting a larger resampled dataset showed only a minor benefit for SVR and XGB models, but only on some of the test batches, which translated into an overall better score. For GP and PLS models the used augmentation method degraded the performance on the majority of test batches, an thus on the overall score.

As PLS is considered to be a state of the art method for regression problems with high multicolinearity, it was chosen to be the benchmark model for all comparisons, unless otherwise indicated.

\begin{table}[h!]
\renewcommand{\arraystretch}{1.2}
    \centering
    \caption{ML models N RMSE for (STD, STD+RES, STD+AUG) on DS1 Titer}
    \resizebox{\columnwidth}{!}{%
    \begin{tabular}{l
    R@{\hskip -0.3\tabcolsep}R@{\hskip -0.3\tabcolsep}R
    R@{\hskip -0.3\tabcolsep}R@{\hskip -0.3\tabcolsep}R
    R@{\hskip -0.3\tabcolsep}R@{\hskip -0.3\tabcolsep}R
    R@{\hskip -0.3\tabcolsep}R@{\hskip -0.3\tabcolsep}R
    R@{\hskip -0.3\tabcolsep}R@{\hskip -0.3\tabcolsep}R}
    \toprule
    
    \multirow{3}{*}{TEST BATCHES} &
      \multicolumn{3}{c}{XGB} &
      \multicolumn{3}{c}{GP} &
      \multicolumn{3}{c}{SVR} &
      \multicolumn{3}{c}{PLS} &
      \multicolumn{3}{c}{Benchmark}\\
    & {STD} & {AUG} & {RES} & {STD} & {AUG} & {RES} & {STD} & {AUG} & {RES} & {STD} & {AUG} & {RES} & {STD} & {AUG} & {RES}\EndTableHeader\\
    \midrule 
     1,7,52,67,97,103,117,127 &         0.306 &  0.330 &  0.310 &        0.257 &  0.480 & 0.292 &   0.359 & 0.293 &  0.390 &    0.234 & 0.409 & 0.254 &       0.234 & 0.409 & 0.254  \\
     2,8,53,92,98,104,118     &         0.284 &  0.301 &  0.277 &        0.256 &  0.409 & 0.277 &   0.328 & 0.248 &  0.304 &    0.294 & 0.372 & 0.288 &       0.294 & 0.372 & 0.288  \\
     3,9,54,93,99,105,119     &         0.309 &  0.222 &  0.237 &        0.277 &  0.434 & 0.284 &   0.405 & 0.259 &  0.301 &    0.226 & 0.372 & 0.230 &       0.226 & 0.372 & 0.230  \\
     4,10,55,94,100,106,124   &         0.139 &  0.169 &  0.176 &        0.128 &  0.323 & 0.160 &   0.195 & 0.141 &  0.243 &    0.191 & 0.253 & 0.194 &       0.191 & 0.253 & 0.194  \\
     5,38,65,95,101,107,125   &         0.313 &  0.274 &  0.314 &        0.271 &  0.383 & 0.273 &   0.395 & 0.270 &  0.376 &    0.272 & 0.372 & 0.313 &       0.272 & 0.372 & 0.313  \\
     6,39,66,96,102,116,126   &         0.447 &  0.468 &  0.486 &        0.407 &  0.583 & 0.448 &   0.487 & 0.454 &  0.507 &    0.403 & 0.577 & 0.400 &       0.403 & 0.577 & 0.400  \\
     AVG BATCHES              &         0.300 &  0.294 &  0.300 &        0.266 &  0.435 & 0.289 &   0.361 & 0.278 &  0.353 &    0.270 & 0.393 & 0.280 &       0.270 & 0.393 & 0.280  \\
     AVG OVERALL              &         0.309 &  0.298 &  0.301 &        0.274 &  0.437 & 0.292 &   0.370 & 0.282 &  0.352 &    0.268 & 0.393 & 0.277 &       0.268 & 0.393 & 0.277  \\
    \bottomrule
    \end{tabular}%
    }
    \label{tab:Titer_ML_STD_all}
\end{table}

Evaluating the models on the CLN data showed significant improvement for XGB and GP, also combined with the AUG and RES settings (Tab. \ref{tab:Titer_ML_CLN_all}). SVR showed minor improvement for CLN and CLN + RES but not for CLN + AUG settings compared to the STD combinations. PLS did not benefit from CLN.

\begin{table}[h!]
\renewcommand{\arraystretch}{1.2}
    \centering
    \caption{ML models N RMSE for (CLN, CLN+RES, CLN+AUG) on DS1 Titer}
    \resizebox{\columnwidth}{!}{%
    \begin{tabular}{l
    R@{\hskip -0.3\tabcolsep}R@{\hskip -0.3\tabcolsep}R
    R@{\hskip -0.3\tabcolsep}R@{\hskip -0.3\tabcolsep}R
    R@{\hskip -0.3\tabcolsep}R@{\hskip -0.3\tabcolsep}R
    R@{\hskip -0.3\tabcolsep}R@{\hskip -0.3\tabcolsep}R
    R@{\hskip -0.3\tabcolsep}R@{\hskip -0.3\tabcolsep}R}
    \toprule
    
    \multirow{3}{*}{TEST BATCHES} &
      \multicolumn{3}{c}{XGB} &
      \multicolumn{3}{c}{GP} &
      \multicolumn{3}{c}{SVR} &
      \multicolumn{3}{c}{PLS} &
      \multicolumn{3}{c}{Benchmark}\\
    & {CLN} & {AUG} & {RES} & {CLN} & {AUG} & {RES} & {STD} & {AUG} & {RES} & {CLN} & {AUG} & {RES} & {CLN} & {AUG} & {RES}\EndTableHeader\\
    \midrule 
     1,6,38,55,93,98,103,119   &        0.179 & 0.171  & 0.179 &        0.248 & 0.210  & 0.250  &        0.323 & 0.281 & 0.324 &        0.292 & 0.283 & 0.281 &        0.292 & 0.283 & 0.281   \\
     2,7,39,65,94,99,104,124   &        0.299 & 0.289  & 0.265 &        0.224 & 0.296  & 0.223  &        0.297 & 0.273 & 0.298 &        0.250 & 0.313 & 0.235 &        0.250 & 0.313 & 0.235   \\
     3,8,52,66,95,100,105,125  &        0.212 & 0.208  & 0.223 &        0.307 & 0.252  & 0.343  &        0.411 & 0.376 & 0.414 &        0.256 & 0.232 & 0.210 &        0.256 & 0.232 & 0.210   \\
     4,9,53,67,96,101,106,126  &        0.339 & 0.355  & 0.403 &        0.333 & 0.444  & 0.341  &        0.391 & 0.365 & 0.382 &        0.390 & 0.453 & 0.384 &        0.390 & 0.453 & 0.384   \\
     5,10,54,92,97,102,107,127 &        0.230 & 0.234  & 0.251 &        0.193 & 0.212  & 0.206  &        0.292 & 0.250 & 0.279 &        0.295 & 0.261 & 0.240 &        0.295 & 0.261 & 0.240   \\
     AVG BATCHES               &        0.252 & 0.251  & 0.264 &        0.261 & 0.283  & 0.273  &        0.343 & 0.309 & 0.339 &        0.297 & 0.308 & 0.270 &        0.297 & 0.308 & 0.270   \\
     AVG OVERALL               &        0.257 & 0.258  & 0.276 &        0.267 & 0.295  & 0.281  &        0.348 & 0.315 & 0.345 &        0.303 & 0.317 & 0.277 &        0.303 & 0.317 & 0.277   \\
     \bottomrule
    \end{tabular}%
    }
    \label{tab:Titer_ML_CLN_all}
\end{table}


The variable selection (VIP) proposed in Sec. \ref{img:ds13varimportanceall} also had an effect on the ML models performance, as shown in Tab. \ref{tab:DS13_Titer_ML_VIP}. The number of input features was reduced to 512, 256 and 128. The dimensionality reduction improved the N RMSE of all ML models, except PLS. The optimal setting for XGB and GP was 512, while SVR showed best results for 128 input features.

\setlength{\tabcolsep}{2pt}
\begin{table}[h!]%
\caption{ML models N RMSE comparison for (STD, STD+512, STD+256, STD+128) on DS1 Titer}%
\renewcommand*{\MinNumber}{0.191}%
\renewcommand*{\MidNumber}{0.259}%
\renewcommand*{\MaxNumber}{0.37200000000000005}%
\centering%
\begin{adjustbox}{width=0.98\textwidth}%
\begin{tabular}{@{}lR@{\hskip-2.0\tabcolsep}R@{\hskip-2.0\tabcolsep}R@{\hskip-2.0\tabcolsep}RR@{\hskip-2.0\tabcolsep}R@{\hskip-2.0\tabcolsep}R@{\hskip-2.0\tabcolsep}RR@{\hskip-2.0\tabcolsep}R@{\hskip-2.0\tabcolsep}R@{\hskip-2.0\tabcolsep}RR@{\hskip-2.0\tabcolsep}R@{\hskip-2.0\tabcolsep}R@{\hskip-2.0\tabcolsep}RR@{\hskip-2.0\tabcolsep}R@{\hskip-2.0\tabcolsep}R@{\hskip-2.0\tabcolsep}R@{}}%
\toprule%
\multirow{2}{*}{TEST BATCHES}&\multicolumn{4}{c}{XGB}&\multicolumn{4}{c}{GP}&\multicolumn{4}{c}{SVR}&\multicolumn{4}{c}{PLS}&\multicolumn{4}{c}{Benchmark}\\%
&STD&512&256&128&STD&512&256&128&STD&512&256&128&STD&512&256&128&STD&512&256&128\\%
\midrule%
\EndTableHeader%
1,7,52,67,97,103,117,127&0.306&0.261&0.268&0.315&0.257&0.263&0.286&0.295&0.359&0.294&0.311&0.290&0.229&0.298&0.340&0.380&0.229&0.298&0.340&0.380\\%
2,8,53,92,98,104,118&0.284&0.237&0.227&0.322&0.256&0.220&0.239&0.294&0.328&0.232&0.248&0.286&0.294&0.298&0.339&0.399&0.294&0.298&0.339&0.399\\%
3,9,54,93,99,105,119&0.309&0.277&0.278&0.283&0.277&0.263&0.292&0.246&0.405&0.296&0.275&0.242&0.210&0.209&0.294&0.336&0.210&0.209&0.294&0.336\\%
4,10,55,94,100,106,124&0.139&0.160&0.186&0.214&0.128&0.175&0.196&0.236&0.195&0.205&0.200&0.199&0.191&0.250&0.320&0.355&0.191&0.250&0.320&0.355\\%
5,38,65,95,101,107,125&0.313&0.293&0.281&0.298&0.271&0.284&0.366&0.301&0.395&0.347&0.344&0.287&0.272&0.337&0.409&0.445&0.272&0.337&0.409&0.445\\%
6,39,66,96,102,116,126&0.447&0.443&0.437&0.445&0.407&0.383&0.403&0.408&0.487&0.422&0.424&0.402&0.372&0.487&0.523&0.566&0.372&0.487&0.523&0.566\\%
AVG BATCHES&0.300&0.279&0.279&0.313&0.266&0.265&0.297&0.297&0.361&0.299&0.300&0.284&0.261&0.313&0.371&0.414&0.261&0.313&0.371&0.414\\%
AVG OVERALL&0.309&0.285&0.285&0.315&0.274&0.268&0.298&0.294&0.370&0.302&0.301&0.284&0.259&0.311&0.366&0.407&0.259&0.311&0.366&0.407\\\bottomrule%
\end{tabular}%
\end{adjustbox}%
\label{tab:DS13_Titer_ML_VIP}
\end{table}

Table Tab. \ref{tab:TOCI_ML_Titer} shows the performance of the ML models in the STD setting on DS2 as a reference for the performance on DS1. As it can be seen, the performance is comparabel for PLS, SVR and GP, whereas XGB seems to do 30\% better on DS2.

\begin{table}[h!]%
\caption{ML models N RMSE comparison for (STD) on DS2 Titer}%
\renewcommand*{\MinNumber}{0.07}%
\renewcommand*{\MidNumber}{0.249}%
\renewcommand*{\MaxNumber}{0.41600000000000004}%
\centering%
\begin{adjustbox}{width=0.68\textwidth}%
\begin{tabular}{@{}lRRRRR@{}}%
\toprule%
\multirow{2}{*}{TEST BATCHES}&\multicolumn{1}{c}{XGB}&\multicolumn{1}{c}{GP}&\multicolumn{1}{c}{SVR}&\multicolumn{1}{c}{PLS}&\multicolumn{1}{c}{Benchmark}\\%
&STD&STD&STD&STD&STD\\%
\midrule%
\EndTableHeader%
97&0.170&0.287&0.366&0.416&0.416\\%
98&0.225&0.323&0.451&0.070&0.070\\%
99&0.165&0.149&0.154&0.156&0.156\\%
101&0.233&0.222&0.373&0.182&0.182\\%
152&0.223&0.251&0.170&0.410&0.410\\%
154&0.288&0.345&0.234&0.298&0.298\\%
AVG BATCHES&0.217&0.263&0.291&0.255&0.255\\%
AVG OVERALL&0.203&0.259&0.346&0.249&0.249\\\bottomrule%
\end{tabular}%
\end{adjustbox}%
\label{tab:TOCI_ML_Titer}
\end{table}

\newpage

\subsection{DL Models}

The comparison of DL Models is structured as follows: first, the performance of feedforward networks obtained via hyperparameter tuning with keras-tuner (KT models) for STD and CLN preprocessing and with the proposed AUG and RES augmentation method will be presented.
Then, the same will be done for KT models in the VIP setting (new KT models, obtained from new search space constrained by VIP). The effect of RES and AUG will also be analysed for these models.
Next, different CNN models will be compared with the addition of the NOPP setting. Similar comparison settings will be used for the VAE models.
Finally, results of the FWLS and BLUE fusion of some selected models will be demonstrated.

\subsection{DL Models - Feed Forward Networks}

In the STD setting, the search for KT models took significantly more time than in the different VIP settings, since the max number of units was capped by the size of the input (which is more than 80 \% smaller after input feature selection), the search produced models of significantly larger size. Despite the addition of a regularization parameter, the search favored models without regularization, which might lead to discrepancies betweem the CV scores during the hyperparameter trials and the CV scores during the actual testing.
In hindsight, defining the regularization rate as a hyperparameter in a search seemed to be harmful for models with better generalization than the top selected models without of with very little regularization.

Only the top 2/800 tested models were light enough (2.5 \& 3.1M hyperparameters) to be considered for evaluation, other models were discarded from the comparison as they showed lower performance in the trials, are significantly more complex (> 5M hyperparameters) and therefore harder to train. The results of the search are presented in Fig. \ref{tab:Titer_DL_KT_STD_top_2}. The top 2 models show a similar structure: same number layers, same number of nodes in the input layer, with an expansion in the further layers. Also, they both use activation functions from the ReLU family (same activation 'potentials' for positive inputs).

The performance of the two models is compared in Tab. \ref{tab:Titer_KT_2533_STD_res} and \ref{tab:Titer_KT_2533_CLN_all}.
In STD setting, the two models seem to slightly outperform the benchmark model, the difference is however minor.
In the CLN setting, simple augmentation seems to yield better results than resampling.
Overall, it can be suggested that there is room for improvement of DL models.

\begin{table}[h!]
\renewcommand{\arraystretch}{1.2}
    \centering
    \caption{Top 2/800 models from hyperparamter search in STD setting. Trial score is avg RMSE over first 3 test batch rotations (1-127, 2-118, 3-119)}
    \begin{tabular}{lrrr}
    \toprule
    Layer(HP)           & top 1 STD & top 2 STD\\
    \midrule        
     Layer 0            & 448       & 448      \\
     Activation 0       & LeakyReLu & PReLU     \\
     BN or DO 0         & BN        & -         \\
     Layer 1            & 1728      & 1280      \\
     Layer 2            & 704       & 640      \\
     Layer 3            & -         & -         \\
     Activation 1/2/3   & ReLu      & ReLU       \\
     BN or DO 1/2/3     & BN        & DO 0.2      \\
     L2 reg.            & 0         & 0.0001      \\
     Paramters          & 3,140,545 & 2,530,881    \\
     Trial Score        & 0.269     & 0.286    \\
    \bottomrule
    \end{tabular}
    \label{tab:Titer_DL_KT_STD_top_2}
\end{table}

\newpage
\begin{table}[h!]
\renewcommand{\arraystretch}{1.2}
    \centering
    \caption{KT STD models N RMSE comparison for (STD, STD+RES, STD+AUG) on DS1 Titer}
    \resizebox{\columnwidth}{!}{%
    \begin{tabular}{l
    R@{\hskip -0.3\tabcolsep}R@{\hskip -0.3\tabcolsep}R
    R@{\hskip -0.3\tabcolsep}R@{\hskip -0.3\tabcolsep}R
    R@{\hskip -0.3\tabcolsep}R@{\hskip -0.3\tabcolsep}R
    R@{\hskip -0.3\tabcolsep}R@{\hskip -0.3\tabcolsep}R
    R@{\hskip -0.3\tabcolsep}R@{\hskip -0.3\tabcolsep}R}
    \toprule
    
    \multirow{3}{*}{TEST BATCHES} &
      \multicolumn{3}{c}{2533 Top 1} &
      \multicolumn{3}{c}{2533 Top 2} &
      \multicolumn{3}{c}{Benchmark}\\
    & {STD} & {AUG} & {RES} & {STD} & {AUG} & {RES} & {STD} & {AUG} & {RES}\EndTableHeader\\
    \midrule 
     1,7,52,67,97,103,117,127 &         0.253 &  0.223 &  0.224 &     0.243 &  0.270 & 0.253 &     0.234 & 0.409 & 0.254  \\
     2,8,53,92,98,104,118     &         0.270 &  0.310 &  0.284 &     0.256 &  0.338 & 0.288 &     0.294 & 0.372 & 0.288  \\
     3,9,54,93,99,105,119     &         0.223 &  0.213 &  0.186 &     0.199 &  0.269 & 0.256 &     0.226 & 0.372 & 0.230  \\
     4,10,55,94,100,106,124   &         0.178 &  0.188 &  0.152 &     0.133 &  0.210 & 0.166 &     0.191 & 0.253 & 0.194  \\
     5,38,65,95,101,107,125   &         0.223 &  0.250 &  0.241 &     0.274 &  0.251 & 0.264 &     0.272 & 0.372 & 0.313  \\
     6,39,66,96,102,116,126   &         0.413 &  0.459 &  0.433 &     0.446 &  0.463 & 0.402 &     0.403 & 0.577 & 0.400  \\
     AVG BATCHES              &         0.260 &  0.274 &  0.253 &     0.259 &  0.300 & 0.271 &     0.270 & 0.393 & 0.280  \\
     AVG OVERALL              &         0.261 &  0.276 &  0.256 &     0.262 &  0.302 & 0.272 &     0.268 & 0.393 & 0.277  \\
    \bottomrule
    \end{tabular}%
    }
    \label{tab:Titer_KT_2533_STD_res}
\end{table}

\begin{table}[h!]
\renewcommand{\arraystretch}{1.2}
    \centering
    \caption{KT STD models N RMSE comparison for (CLN, CLN+RES, CLN+AUG) on DS1 Titer}
    \resizebox{\columnwidth}{!}{%
    \begin{tabular}{l
    R@{\hskip -0.3\tabcolsep}R@{\hskip -0.3\tabcolsep}R
    R@{\hskip -0.3\tabcolsep}R@{\hskip -0.3\tabcolsep}R
    R@{\hskip -0.3\tabcolsep}R@{\hskip -0.3\tabcolsep}R}
    \toprule
    
    \multirow{3}{*}{TEST BATCHES} &
      \multicolumn{3}{c}{2533 Top 1} &
      \multicolumn{3}{c}{2533 Top 2} &
      \multicolumn{3}{c}{Benchmark}\\
    & {CLN} & {AUG} & {RES} & {CLN} & {AUG} & {RES} & {CLN} & {AUG} & {RES}\EndTableHeader\\
    \midrule 
     1,6,38,55,93,98,103,119   &        0.253 & 0.219 & 0.249 &        0.254 & 0.253 & 0.289 &        0.292 & 0.283 & 0.281   \\
     2,7,39,65,94,99,104,124   &        0.294 & 0.209 & 0.228 &        0.275 & 0.270 & 0.258 &        0.250 & 0.313 & 0.235   \\
     3,8,52,66,95,100,105,125  &        0.221 & 0.170 & 0.214 &        0.180 & 0.222 & 0.228 &        0.256 & 0.232 & 0.210   \\
     4,9,53,67,96,101,106,126  &        0.364 & 0.371 & 0.331 &        0.358 & 0.348 & 0.351 &        0.390 & 0.453 & 0.384   \\
     5,10,54,92,97,102,107,127 &        0.198 & 0.205 & 0.263 &        0.214 & 0.199 & 0.211 &        0.295 & 0.261 & 0.240   \\
     AVG BATCHES               &        0.266 & 0.235 & 0.257 &        0.256 & 0.258 & 0.267 &        0.297 & 0.308 & 0.270   \\
     AVG OVERALL               &        0.270 & 0.246 & 0.261 &        0.261 & 0.261 & 0.271 &        0.303 & 0.317 & 0.277   \\
     \bottomrule
    \end{tabular}%
    }
    \label{tab:Titer_KT_2533_CLN_all}
\end{table}

More KT models were obtained from the following VIP settings: top 512, top 256 and top 128 input features.
The resulting architectures and their scores for VIP 512 are presented in Tab. \ref{tab:Titer_KT_512_top_5}. It can be noticed that the top 5/4000 tried models don't share common traits in their architecture, but have very similar trial scores, which are significantly better compared to the STD setting. This also suggests a better performance in the full 5-fold CV, which is shown in Tab. \ref{tab:Titer_KT_512_STD_res} and \ref{tab:Titer_KT_512_CLN_all}. Amongst all KT models, KT models in the VIP 512 setting outperformed all other KT models, but also ML and CNN models for the current target variable. The average N RMSE lies in the 0.23-0.24 range for the STD, STD + AUG and STD + RES settings, which translates to a 14\% improvement over the benchmark. For the CLN  no clear trend can be established. the best model performs more than 23\% better than the benchmark, the worst model performs similarly to the benchmark.

\begin{table}[h!]
\renewcommand{\arraystretch}{1.2}
    \centering
    \caption{Top 5/4000 models from hyperparamter search in VIP 512 setting. Trial score is avg RMSE on the first 3 test batch rotations}
    \resizebox{\columnwidth}{!}{%
    \begin{tabular}{lrrrrr}
    \toprule
    Layer(HP)           & KT 512 top 1 & KT 512 top 2 & KT 512 top 3 & KT 512 top 4 & KT 512 top 5\\
    \midrule                                                    
     Layer 0            & 448       & 320       & 448       & 256       & 192               \\
     Activation 0       & SeLU      & LeakyReLu & SeLU      & SeLU      & SeLU              \\
     BN or DO 0         & DO 0.5    & BN        & DO 0.2    & DO 0.5    & -                 \\
     Layer 1            & 128       & 384       & 128       & 320       & 256               \\
     Layer 2            & 192       & 512       & 64        & 128       & 448               \\
     Layer 3            & -         & 448       & -         & -        & -                 \\
     Activation 1/2/3   & ReLu      & ReLU      & LeakyReLu & SeLu      & PReLu             \\
     BN or DO 1/2/3     & -         & BN        & BN        & BN        & -                 \\
     L2 reg.            & 0         & 0         & 0         & 0         & 0                 \\
     Paramters          & 312,257   & 721,473   & 300,801   & 256,577   & 264,193           \\
     Trial Score        & 0.212     & 0.212     & 0.217     & 0.220     & 0.220             \\
    \bottomrule
    \end{tabular}%
    }
    \label{tab:Titer_KT_512_top_5}
\end{table}

\begin{table}[h!]
\renewcommand{\arraystretch}{1.2}
    \centering
    \caption{KT VIP 512 models N RMSE comparison for (STD,STD+RES,STD+AUG) on DS1 Titer}
    \resizebox{\columnwidth}{!}{%
    \begin{tabular}{l
    R@{\hskip -0.3\tabcolsep}R@{\hskip -0.3\tabcolsep}R
    R@{\hskip -0.3\tabcolsep}R@{\hskip -0.3\tabcolsep}R
    R@{\hskip -0.3\tabcolsep}R@{\hskip -0.3\tabcolsep}R
    R@{\hskip -0.3\tabcolsep}R@{\hskip -0.3\tabcolsep}R
    R@{\hskip -0.3\tabcolsep}R@{\hskip -0.3\tabcolsep}R
    R@{\hskip -0.3\tabcolsep}R@{\hskip -0.3\tabcolsep}R
    R@{\hskip -0.3\tabcolsep}R@{\hskip -0.3\tabcolsep}R}
    \toprule
    
    \multirow{3}{*}{TEST BATCHES} &
      \multicolumn{3}{c}{VIP 512 top 1} &
      \multicolumn{3}{c}{VIP 512 top 2} &
      \multicolumn{3}{c}{VIP 512 top 3} &
      \multicolumn{3}{c}{VIP 512 top 4} &
      \multicolumn{3}{c}{VIP 512 top 5} &
      \multicolumn{3}{c}{Benchmark}\\
    & {STD} & {AUG} & {RES} & {STD} & {AUG} & {RES} & {STD} & {AUG} & {RES} & {STD} & {AUG} & {RES} & {STD} & {AUG} & {RES} & {STD} & {AUG} & {RES}\EndTableHeader\\
    \midrule 
     1,7,52,67,97,103,117,127 &         0.232 &  0.263 &  0.222 &     0.264 &  0.244 &  0.237 &     0.225 &  0.232 &  0.229 &     0.233 &  0.223 &  0.249 &     0.261 &  0.260 &  0.249 &     0.234 & 0.409 & 0.254  \\
     2,8,53,92,98,104,118     &         0.178 &  0.174 &  0.190 &     0.204 &  0.190 &  0.213 &     0.190 &  0.203 &  0.232 &     0.184 &  0.165 &  0.171 &     0.204 &  0.178 &  0.200 &     0.294 & 0.372 & 0.288  \\
     3,9,54,93,99,105,119     &         0.147 &  0.143 &  0.154 &     0.191 &  0.173 &  0.170 &     0.181 &  0.171 &  0.136 &     0.161 &  0.172 &  0.163 &     0.152 &  0.138 &  0.139 &     0.226 & 0.372 & 0.230  \\
     4,10,55,94,100,106,124   &         0.125 &  0.152 &  0.121 &     0.159 &  0.141 &  0.135 &     0.156 &  0.169 &  0.175 &     0.146 &  0.164 &  0.180 &     0.136 &  0.168 &  0.126 &     0.191 & 0.253 & 0.194  \\
     5,38,65,95,101,107,125   &         0.235 &  0.234 &  0.261 &     0.233 &  0.217 &  0.228 &     0.243 &  0.221 &  0.223 &     0.230 &  0.224 &  0.238 &     0.242 &  0.224 &  0.229 &     0.272 & 0.372 & 0.313  \\
     6,39,66,96,102,116,126   &         0.425 &  0.435 &  0.437 &     0.403 &  0.390 &  0.362 &     0.385 &  0.406 &  0.395 &     0.363 &  0.419 &  0.388 &     0.377 &  0.410 &  0.438 &     0.403 & 0.577 & 0.400  \\
     AVG BATCHES              &         0.224 &  0.234 &  0.231 &     0.242 &  0.226 &  0.225 &     0.230 &  0.234 &  0.232 &     0.220 &  0.228 &  0.231 &     0.229 &  0.229 &  0.230 &     0.270 & 0.393 & 0.280  \\
     AVG OVERALL              &         0.230 &  0.240 &  0.236 &     0.244 &  0.229 &  0.226 &     0.231 &  0.235 &  0.234 &     0.221 &  0.231 &  0.233 &     0.232 &  0.234 &  0.238 &     0.268 & 0.393 & 0.277  \\
    \bottomrule
    \end{tabular}%
    }
    \label{tab:Titer_KT_512_STD_res}
\end{table}

\begin{table}[h!]
    \centering
    \caption{KT VIP 512 models N RMSE comparison for (CLN, CLN+RES, CLN+AUG) on DS1 Titer}
    \resizebox{\columnwidth}{!}{%
    \begin{tabular}{l
    R@{\hskip -0.3\tabcolsep}R@{\hskip -0.3\tabcolsep}R
    R@{\hskip -0.3\tabcolsep}R@{\hskip -0.3\tabcolsep}R
    R@{\hskip -0.3\tabcolsep}R@{\hskip -0.3\tabcolsep}R
    R@{\hskip -0.3\tabcolsep}R@{\hskip -0.3\tabcolsep}R
    R@{\hskip -0.3\tabcolsep}R@{\hskip -0.3\tabcolsep}R
    R@{\hskip -0.3\tabcolsep}R@{\hskip -0.3\tabcolsep}R
    R@{\hskip -0.3\tabcolsep}R@{\hskip -0.3\tabcolsep}R}
    \toprule
    
    \multirow{3}{*}{TEST BATCHES} &
      \multicolumn{3}{c}{VIP 512 Top 1} &
      \multicolumn{3}{c}{VIP 512 Top 2} &
      \multicolumn{3}{c}{VIP 512 Top 3} &
      \multicolumn{3}{c}{VIP 512 Top 4} &
      \multicolumn{3}{c}{VIP 512 Top 5} &
      \multicolumn{3}{c}{Benchmark}\\
    & {CLN} & {AUG} & {RES} & {CLN} & {AUG} & {RES} & {CLN} & {AUG} & {RES} & {CLN} & {AUG} & {RES} & {CLN} & {AUG} & {RES} & {CLN} & {AUG} & {RES}\EndTableHeader\\
    \midrule 
     1,6,38,55,93,98,103,119   &        0.159 & 0.150 & 0.175 &        0.137 & 0.150 & 0.177 &       0.196 & 0.173 & 0.162 &        0.178 & 0.159 & 0.162 &         0.130 & 0.137 & 0.144 &      0.292 & 0.283 & 0.281   \\
     2,7,39,65,94,99,104,124   &        0.246 & 0.268 & 0.273 &        0.245 & 0.218 & 0.233 &       0.218 & 0.218 & 0.210 &        0.235 & 0.216 & 0.216 &         0.243 & 0.228 & 0.220 &      0.250 & 0.313 & 0.235   \\
     3,8,52,66,95,100,105,125  &        0.217 & 0.196 & 0.196 &        0.193 & 0.167 & 0.207 &       0.207 & 0.185 & 0.196 &        0.219 & 0.204 & 0.209 &         0.224 & 0.222 & 0.219 &      0.256 & 0.232 & 0.210   \\
     4,9,53,67,96,101,106,126  &        0.369 & 0.330 & 0.324 &        0.341 & 0.338 & 0.348 &       0.369 & 0.332 & 0.288 &        0.379 & 0.340 & 0.302 &         0.313 & 0.337 & 0.318 &      0.390 & 0.453 & 0.384   \\
     5,10,54,92,97,102,107,127 &        0.202 & 0.149 & 0.209 &        0.192 & 0.166 & 0.234 &       0.178 & 0.198 & 0.178 &        0.219 & 0.217 & 0.255 &         0.205 & 0.188 & 0.199 &      0.295 & 0.261 & 0.240   \\
     AVG BATCHES               &        0.239 & 0.219 & 0.235 &        0.222 & 0.208 & 0.240 &       0.234 & 0.221 & 0.207 &        0.246 & 0.227 & 0.229 &         0.223 & 0.222 & 0.220 &      0.297 & 0.308 & 0.270   \\
     AVG OVERALL               &        0.249 & 0.227 & 0.240 &        0.232 & 0.219 & 0.248 &       0.244 & 0.229 & 0.211 &        0.257 & 0.237 & 0.236 &         0.231 & 0.232 & 0.228 &      0.303 & 0.317 & 0.277   \\
     \bottomrule
    \end{tabular}%
    }
    \label{tab:Titer_KT_512_CLN_all}
\end{table}

\clearpage
For reference purposes, the performance of the KT VIP 512 models was also analysed on DS2. The results are shown in table Tab. \ref{tab:TOCI_KT_512_Titer}. While model VIP 512 Top 1 performs badly, likely due to poor convergence during training, the other models seem to show very promising results. On multiple test batches the N RMSE is in the order of $10^{-2} - 10^{-3}$ for some models in the RES setting.
Such results suggest that the amount of training data as well as the quality of the reference measurements are crucial for fitting good model. 

\begin{table}[h!]%
\caption{KT models N RMSE comparison for (512, 512+RES) on DS2 Titer}%
\renewcommand*{\MinNumber}{0.07}%
\renewcommand*{\MidNumber}{0.249}%
\renewcommand*{\MaxNumber}{0.41600000000000004}%
\centering%
\begin{adjustbox}{width=0.98\textwidth}%
\begin{tabular}{@{}lR@{\hskip-1.0\tabcolsep}RR@{\hskip-1.0\tabcolsep}RR@{\hskip-1.0\tabcolsep}RR@{\hskip-1.0\tabcolsep}RR@{\hskip-1.0\tabcolsep}RR@{\hskip-1.0\tabcolsep}R@{}}%
\toprule%
\multirow{2}{*}{TEST BATCHES}&\multicolumn{2}{c}{top 1}&\multicolumn{2}{c}{top 2}&\multicolumn{2}{c}{top 3}&\multicolumn{2}{c}{top 4}&\multicolumn{2}{c}{top 5}&\multicolumn{2}{c}{Benchmark}\\%
&512&RES&512&RES&512&RES&512&RES&512&RES&STD&RES\\%
\midrule%
\EndTableHeader%
97&0.465&0.348&0.250&0.333&0.139&0.158&0.140&0.158&0.137&0.132&0.416&0.386\\%
98&0.408&0.329&0.133&0.098&0.071&0.160&0.247&0.089&0.118&0.161&0.070&0.066\\%
99&0.471&0.332&0.092&0.177&0.101&0.161&0.149&0.171&0.042&0.152&0.156&0.150\\%
101&0.422&0.348&0.123&0.143&0.083&0.159&0.231&0.183&0.058&0.144&0.182&0.191\\%
152&0.310&0.254&0.144&0.212&0.277&0.140&0.342&0.207&0.311&0.242&0.410&0.420\\%
154&0.481&0.322&0.152&0.529&0.170&0.187&0.336&0.200&0.482&0.378&0.298&0.324\\%
AVG BATCHES&0.426&0.322&0.149&0.249&0.140&0.161&0.241&0.168&0.191&0.201&0.255&0.256\\%
AVG OVERALL&0.427&0.327&0.159&0.228&0.113&0.156&0.210&0.152&0.150&0.165&0.249&0.241\\\bottomrule%
\end{tabular}%
\end{adjustbox}%
\label{tab:TOCI_KT_512_Titer}
\end{table}

 

\clearpage
The top 5/7000 models in the VIP 256 setting showed similar or lower performance compared to the previous setting. This can be seen from the trial scores in Tab. \ref{tab:Titer_KT_256_top_5} and the test scores in Tab. \ref{tab:Titer_KT_256_STD_res} and Tab. \ref{tab:Titer_KT_256_CLN_all}. In the STD setting, the best model showed a 10\% improvement vs the benchmark, whereas the worst showed similar performance. A similar statement can be made for STD + AUG and STD + RES setting, however, the best and worst models are different this time.

In the CLN, CLN+AUG, CLN+RES settings the best models were roughly 15\% better than the benchmark. These improvements are not consistent for the same models compared across the STD/CLN settings and their respective combination.

\begin{table}[h!]
\renewcommand{\arraystretch}{1.2}
    \centering
    \caption{Top 5/7000 models from hyperparamter search in VIP 256 setting. Trial score is RMSE}
    \resizebox{\columnwidth}{!}{%
    \begin{tabular}{lrrrrr}
    \toprule
    Layer(HP)           & KT 256 top 1 & KT 256 top 2 & KT 256 top 3 & KT 256 top 4 & KT 256 top 5\\
    \midrule                                                    
     Layer 0            & 192       & 128       & 192       & 128       & 128               \\
     Activation 0       & LeakyReLu & SeLU      & LeakyReLu & LeakyReLu & SeLU              \\
     BN or DO 0         & BN        & DO 0.5    & DO 0.2    & DO 0.5    & DO 0.2                 \\
     Layer 1            & 192       & 256       & 192       & 192       & 256               \\
     Layer 2            & 256       & 256       & 192       & 64        & 256               \\
     Layer 3            & -         & -         & -         & 192       & -                 \\
     Activation 1/2/3   & ReLu      & PReLU     & PReLu     & ReLu      & ReLu             \\
     BN or DO 1/2/3     & BN        & BN        & -         & BN        & BN                 \\
     L2 reg.            & 0         & 0         & 0         & 0         & 0                 \\
     Paramters          & 138,625   & 134,529   & 124,033   & 84,481    & 134,017           \\
     Trial Score        & 0.229     & 0.235     & 0.236     & 0.236     & 0.237             \\
    \bottomrule
    \end{tabular}%
    }
    \label{tab:Titer_KT_256_top_5}
\end{table}

\begin{table}[h!]
\renewcommand{\arraystretch}{1.2}
    \centering
    \caption{KT VIP 256 models N RMSE comparison for (STD, STD+RES, STD+AUG) on DS1 Titer)}
    \resizebox{\columnwidth}{!}{%
    \begin{tabular}{l
    R@{\hskip -0.3\tabcolsep}R@{\hskip -0.3\tabcolsep}R
    R@{\hskip -0.3\tabcolsep}R@{\hskip -0.3\tabcolsep}R
    R@{\hskip -0.3\tabcolsep}R@{\hskip -0.3\tabcolsep}R
    R@{\hskip -0.3\tabcolsep}R@{\hskip -0.3\tabcolsep}R
    R@{\hskip -0.3\tabcolsep}R@{\hskip -0.3\tabcolsep}R
    R@{\hskip -0.3\tabcolsep}R@{\hskip -0.3\tabcolsep}R
    R@{\hskip -0.3\tabcolsep}R@{\hskip -0.3\tabcolsep}R}
    \toprule
    
    \multirow{3}{*}{TEST BATCHES} &
      \multicolumn{3}{c}{VIP 256 top 1} &
      \multicolumn{3}{c}{VIP 256 top 2} &
      \multicolumn{3}{c}{VIP 256 top 3} &
      \multicolumn{3}{c}{VIP 256 top 4} &
      \multicolumn{3}{c}{VIP 256 top 5} &
      \multicolumn{3}{c}{Benchmark}\\
    & {STD} & {AUG} & {RES} & {STD} & {AUG} & {RES} & {STD} & {AUG} & {RES} & {STD} & {AUG} & {RES} & {STD} & {AUG} & {RES} & {STD} & {AUG} & {RES}\EndTableHeader\\
    \midrule 
     1,7,52,67,97,103,117,127 &         0.273 &  0.279 &  0.311 &     0.262 &  0.290 &  0.262 &     0.280 &  0.272 &  0.254 &     0.222 &  0.249 &  0.228 &     0.235 &  0.280 &  0.242 &     0.234 & 0.409 & 0.254  \\
     2,8,53,92,98,104,118     &         0.208 &  0.206 &  0.264 &     0.225 &  0.204 &  0.203 &     0.219 &  0.218 &  0.270 &     0.261 &  0.199 &  0.224 &     0.167 &  0.223 &  0.211 &     0.294 & 0.372 & 0.288  \\
     3,9,54,93,99,105,119     &         0.243 &  0.200 &  0.171 &     0.180 &  0.201 &  0.163 &     0.231 &  0.187 &  0.198 &     0.178 &  0.163 &  0.169 &     0.177 &  0.196 &  0.166 &     0.226 & 0.372 & 0.230  \\
     4,10,55,94,100,106,124   &         0.178 &  0.185 &  0.176 &     0.147 &  0.218 &  0.170 &     0.237 &  0.207 &  0.202 &     0.172 &  0.185 &  0.170 &     0.152 &  0.163 &  0.196 &     0.191 & 0.253 & 0.194  \\
     5,38,65,95,101,107,125   &         0.234 &  0.252 &  0.252 &     0.264 &  0.268 &  0.293 &     0.311 &  0.271 &  0.320 &     0.260 &  0.258 &  0.281 &     0.283 &  0.280 &  0.272 &     0.272 & 0.372 & 0.313  \\
     6,39,66,96,102,116,126   &         0.430 &  0.414 &  0.504 &     0.435 &  0.424 &  0.398 &     0.377 &  0.383 &  0.394 &     0.471 &  0.407 &  0.390 &     0.423 &  0.468 &  0.428 &     0.403 & 0.577 & 0.400  \\
     AVG BATCHES              &         0.261 &  0.256 &  0.280 &     0.252 &  0.268 &  0.248 &     0.276 &  0.256 &  0.273 &     0.261 &  0.244 &  0.244 &     0.239 &  0.268 &  0.253 &     0.270 & 0.393 & 0.280  \\
     AVG OVERALL              &         0.264 &  0.257 &  0.286 &     0.255 &  0.268 &  0.248 &     0.272 &  0.255 &  0.270 &     0.264 &  0.244 &  0.243 &     0.242 &  0.271 &  0.253 &     0.268 & 0.393 & 0.277  \\
    \bottomrule
    \end{tabular}%
    }
    \label{tab:Titer_KT_256_STD_res}
\end{table}

\begin{table}[h!]
\renewcommand{\arraystretch}{1.2}
    \centering
    \caption{KT VIP 256 models N RMSE comparison for (CLN, CLN+RES, CLN+AUG) on DS1 Titer}
    \resizebox{\columnwidth}{!}{%
    \begin{tabular}{l
    R@{\hskip -0.3\tabcolsep}R@{\hskip -0.3\tabcolsep}R
    R@{\hskip -0.3\tabcolsep}R@{\hskip -0.3\tabcolsep}R
    R@{\hskip -0.3\tabcolsep}R@{\hskip -0.3\tabcolsep}R
    R@{\hskip -0.3\tabcolsep}R@{\hskip -0.3\tabcolsep}R
    R@{\hskip -0.3\tabcolsep}R@{\hskip -0.3\tabcolsep}R
    R@{\hskip -0.3\tabcolsep}R@{\hskip -0.3\tabcolsep}R
    R@{\hskip -0.3\tabcolsep}R@{\hskip -0.3\tabcolsep}R}
    \toprule
    
    \multirow{3}{*}{TEST BATCHES} &
      \multicolumn{3}{c}{VIP 256 Top 1} &
      \multicolumn{3}{c}{VIP 256 Top 2} &
      \multicolumn{3}{c}{VIP 256 Top 3} &
      \multicolumn{3}{c}{VIP 256 Top 4} &
      \multicolumn{3}{c}{VIP 256 Top 5} &
      \multicolumn{3}{c}{Benchmark}\\
    & {CLN} & {AUG} & {RES} & {CLN} & {AUG} & {RES} & {CLN} & {AUG} & {RES} & {CLN} & {AUG} & {RES} & {CLN} & {AUG} & {RES} & {CLN} & {AUG} & {RES}\EndTableHeader\\
    \midrule 
     1,6,38,55,93,98,103,119   &        0.209 & 0.143 & 0.148 &        0.181 & 0.158 & 0.175 &       0.195 & 0.198 & 0.186 &        0.190 & 0.162 & 0.221 &         0.188 & 0.160 & 0.171 &      0.292 & 0.283 & 0.281   \\
     2,7,39,65,94,99,104,124   &        0.237 & 0.226 & 0.235 &        0.240 & 0.235 & 0.250 &       0.291 & 0.260 & 0.278 &        0.244 & 0.226 & 0.254 &         0.258 & 0.238 & 0.239 &      0.250 & 0.313 & 0.235   \\
     3,8,52,66,95,100,105,125  &        0.251 & 0.227 & 0.228 &        0.199 & 0.201 & 0.206 &       0.253 & 0.240 & 0.270 &        0.171 & 0.174 & 0.199 &         0.208 & 0.238 & 0.194 &      0.256 & 0.232 & 0.210   \\
     4,9,53,67,96,101,106,126  &        0.383 & 0.383 & 0.353 &        0.293 & 0.323 & 0.309 &       0.385 & 0.387 & 0.380 &        0.334 & 0.297 & 0.312 &         0.356 & 0.320 & 0.325 &      0.390 & 0.453 & 0.384   \\
     5,10,54,92,97,102,107,127 &        0.243 & 0.179 & 0.270 &        0.203 & 0.229 & 0.217 &       0.212 & 0.210 & 0.198 &        0.184 & 0.227 & 0.211 &         0.233 & 0.226 & 0.183 &      0.295 & 0.261 & 0.240   \\
     AVG BATCHES               &        0.265 & 0.231 & 0.247 &        0.223 & 0.229 & 0.231 &       0.267 & 0.259 & 0.262 &        0.224 & 0.217 & 0.239 &         0.249 & 0.236 & 0.223 &      0.297 & 0.308 & 0.270   \\
     AVG OVERALL               &        0.274 & 0.246 & 0.258 &        0.226 & 0.236 & 0.235 &       0.275 & 0.268 & 0.271 &        0.231 & 0.223 & 0.241 &         0.255 & 0.243 & 0.229 &      0.303 & 0.317 & 0.277   \\
     \bottomrule
    \end{tabular}%
    }
    \label{tab:Titer_KT_256_CLN_all}
\end{table}

\newpage

The results for the models obtained from the VIP 128 setting (Tab. \ref{tab:Titer_KT_128_top_5}) are shown in Tab. \ref{tab:Titer_KT_128_STD_res} and Tab. \ref{tab:Titer_KT_128_CLN_all}. From the trial RMSE and test N RMSE, it can be concluded that further decrease of the input size degrades the prediction performance of the models, leading to scores below the benchmark. Therefore, the VIP 256 and VIP 128 settings can be discarded from further comparisons.

\begin{table}[h!]
\renewcommand{\arraystretch}{1.2}
    \centering
    \caption{Top 5/1000 models from hyperparamter search in VIP 128 setting. Trial score is RMSE}
    \begin{tabular}{lrrrrr}
    \toprule
    Layer(HP)           & KT 128 top 1 & KT 128 top 2 & KT 128 top 3 & KT 128 top 4 & KT 128 top 5\\
    \midrule                                                    
     Layer 0            & 96        & 64        & 64        & 64        & 128               \\
     Activation 0       & ReLu      & SeLU      & PReLU     & ReLu      & SeLU              \\
     BN or DO 0         & -         & DO 0.2    & BN        & DO 0.5    & -                 \\
     Layer 1            & 128       & 64        & 64        & 64        & 64               \\
     Layer 2            & 96        & 64        & 96        & 128       & 64               \\
     Layer 3            & -         & -         & -         & 96        & 96                 \\
     Activation 1/2/3   & PReLu     & LeakyReLU & ReLU      & SeLu      & ReLu             \\
     BN or DO 1/2/3     & BN        & -         & BN        & BN        & -                 \\
     L2 reg.            & 0         & 0         & 0         & 0         & 0                 \\
     Paramters          & 38,401    & 16,641    & 19,713    & 34,369    & 35,265           \\
     Trial Score        & 0.275     & 0.294     & 0.295     & 0.296     & 0.296             \\
    \bottomrule
    \end{tabular}
    \label{tab:Titer_KT_128_top_5}
\end{table}

\begin{table}[h!]
\renewcommand{\arraystretch}{1.2}
    \centering
    \caption{KT VIP 128 models N RMSE comparison for (STD, STD+RES, STD+AUG) on DS1 Titer}
    \resizebox{\columnwidth}{!}{%
    \begin{tabular}{l
    R@{\hskip -0.3\tabcolsep}R@{\hskip -0.3\tabcolsep}R
    R@{\hskip -0.3\tabcolsep}R@{\hskip -0.3\tabcolsep}R
    R@{\hskip -0.3\tabcolsep}R@{\hskip -0.3\tabcolsep}R
    R@{\hskip -0.3\tabcolsep}R@{\hskip -0.3\tabcolsep}R
    R@{\hskip -0.3\tabcolsep}R@{\hskip -0.3\tabcolsep}R
    R@{\hskip -0.3\tabcolsep}R@{\hskip -0.3\tabcolsep}R
    R@{\hskip -0.3\tabcolsep}R@{\hskip -0.3\tabcolsep}R}
    \toprule
    
    \multirow{3}{*}{TEST BATCHES} &
      \multicolumn{3}{c}{VIP 128 top 1} &
      \multicolumn{3}{c}{VIP 128 top 2} &
      \multicolumn{3}{c}{VIP 128 top 3} &
      \multicolumn{3}{c}{VIP 128 top 4} &
      \multicolumn{3}{c}{VIP 128 top 5} &
      \multicolumn{3}{c}{Benchmark}\\
    & {STD} & {AUG} & {RES} & {STD} & {AUG} & {RES} & {STD} & {AUG} & {RES} & {STD} & {AUG} & {RES} & {STD} & {AUG} & {RES} & {STD} & {AUG} & {RES}\EndTableHeader\\
    \midrule 
     1,7,52,67,97,103,117,127 &         0.293 &  0.326 &  0.295 &     0.314 &  0.322 &  0.349 &     0.334 &  0.335 &  0.309 &     0.346 &  0.353 &  0.316 &     0.319 &  0.331 &  0.346 &     0.234 & 0.409 & 0.254  \\
     2,8,53,92,98,104,118     &         0.259 &  0.230 &  0.284 &     0.247 &  0.268 &  0.257 &     0.260 &  0.247 &  0.319 &     0.239 &  0.242 &  0.260 &     0.236 &  0.263 &  0.304 &     0.294 & 0.372 & 0.288  \\
     3,9,54,93,99,105,119     &         0.174 &  0.183 &  0.166 &     0.163 &  0.153 &  0.156 &     0.168 &  0.161 &  0.154 &     0.174 &  0.185 &  0.171 &     0.167 &  0.170 &  0.186 &     0.226 & 0.372 & 0.230  \\
     4,10,55,94,100,106,124   &         0.182 &  0.185 &  0.184 &     0.221 &  0.227 &  0.231 &     0.200 &  0.188 &  0.189 &     0.211 &  0.214 &  0.196 &     0.193 &  0.201 &  0.195 &     0.191 & 0.253 & 0.194  \\
     5,38,65,95,101,107,125   &         0.280 &  0.264 &  0.258 &     0.282 &  0.291 &  0.277 &     0.256 &  0.265 &  0.258 &     0.315 &  0.300 &  0.235 &     0.325 &  0.320 &  0.279 &     0.272 & 0.372 & 0.313  \\
     6,39,66,96,102,116,126   &         0.515 &  0.466 &  0.546 &     0.499 &  0.487 &  0.458 &     0.459 &  0.460 &  0.452 &     0.486 &  0.523 &  0.494 &     0.415 &  0.429 &  0.486 &     0.403 & 0.577 & 0.400  \\
     AVG BATCHES              &         0.284 &  0.276 &  0.289 &     0.288 &  0.291 &  0.288 &     0.280 &  0.276 &  0.280 &     0.295 &  0.303 &  0.279 &     0.276 &  0.286 &  0.299 &     0.270 & 0.393 & 0.280  \\
     AVG OVERALL              &         0.289 &  0.280 &  0.297 &     0.292 &  0.295 &  0.292 &     0.284 &  0.281 &  0.285 &     0.299 &  0.308 &  0.285 &     0.277 &  0.287 &  0.304 &     0.268 & 0.393 & 0.277  \\
    \bottomrule
    \end{tabular}%
    }
    \label{tab:Titer_KT_128_STD_res}
\end{table}

\begin{table}[h!]
\renewcommand{\arraystretch}{1.2}
    \centering
    \caption{KT VIP 128 models N RMSE comparison for (CLN, CLN+RES, CLN+AUG) on DS1 Titer}
    \resizebox{\columnwidth}{!}{%
    \begin{tabular}{l
    R@{\hskip -0.3\tabcolsep}R@{\hskip -0.3\tabcolsep}R
    R@{\hskip -0.3\tabcolsep}R@{\hskip -0.3\tabcolsep}R
    R@{\hskip -0.3\tabcolsep}R@{\hskip -0.3\tabcolsep}R
    R@{\hskip -0.3\tabcolsep}R@{\hskip -0.3\tabcolsep}R
    R@{\hskip -0.3\tabcolsep}R@{\hskip -0.3\tabcolsep}R
    R@{\hskip -0.3\tabcolsep}R@{\hskip -0.3\tabcolsep}R
    R@{\hskip -0.3\tabcolsep}R@{\hskip -0.3\tabcolsep}R}
    \toprule
    
    \multirow{3}{*}{TEST BATCHES} &
      \multicolumn{3}{c}{VIP 128 Top 1} &
      \multicolumn{3}{c}{VIP 128 Top 2} &
      \multicolumn{3}{c}{VIP 128 Top 3} &
      \multicolumn{3}{c}{VIP 128 Top 4} &
      \multicolumn{3}{c}{VIP 128 Top 5} &
      \multicolumn{3}{c}{Benchmark}\\
    & {CLN} & {AUG} & {RES} & {CLN} & {AUG} & {RES} & {CLN} & {AUG} & {RES} & {CLN} & {AUG} & {RES} & {CLN} & {AUG} & {RES} & {CLN} & {AUG} & {RES} \EndTableHeader\\
    \midrule 
     1,6,38,55,93,98,103,119   &        0.180 & 0.132 & 0.182 &        0.182 & 0.165 & 0.194 &       0.205 & 0.176 & 0.173 &        0.160 & 0.162 & 0.190 &         0.205 & 0.157 & 0.210 &      0.292 & 0.283 & 0.281   \\
     2,7,39,65,94,99,104,124   &        0.219 & 0.218 & 0.225 &        0.254 & 0.242 & 0.244 &       0.255 & 0.206 & 0.209 &        0.221 & 0.230 & 0.246 &         0.238 & 0.235 & 0.210 &      0.250 & 0.313 & 0.235   \\
     3,8,52,66,95,100,105,125  &        0.226 & 0.259 & 0.253 &        0.244 & 0.233 & 0.234 &       0.236 & 0.236 & 0.215 &        0.243 & 0.246 & 0.211 &         0.260 & 0.223 & 0.254 &      0.256 & 0.232 & 0.210   \\
     4,9,53,67,96,101,106,126  &        0.452 & 0.340 & 0.341 &        0.343 & 0.290 & 0.325 &       0.393 & 0.347 & 0.339 &        0.378 & 0.357 & 0.354 &         0.365 & 0.336 & 0.387 &      0.390 & 0.453 & 0.384   \\
     5,10,54,92,97,102,107,127 &        0.243 & 0.284 & 0.231 &        0.308 & 0.233 & 0.278 &       0.233 & 0.239 & 0.254 &        0.255 & 0.258 & 0.293 &         0.263 & 0.204 & 0.295 &      0.295 & 0.261 & 0.240   \\
     AVG BATCHES               &        0.264 & 0.247 & 0.246 &        0.266 & 0.232 & 0.255 &       0.264 & 0.241 & 0.238 &        0.251 & 0.251 & 0.259 &         0.266 & 0.231 & 0.271 &      0.297 & 0.308 & 0.270   \\
     AVG OVERALL               &        0.284 & 0.261 & 0.254 &        0.275 & 0.236 & 0.261 &       0.273 & 0.250 & 0.247 &        0.264 & 0.261 & 0.267 &         0.274 & 0.239 & 0.283 &      0.303 & 0.317 & 0.277   \\
     \bottomrule
    \end{tabular}%
    }
    \label{tab:Titer_KT_128_CLN_all}
\end{table}

\newpage
\subsection{DL Models - CNN}
5 different CNN architectures were analysed in the STD, STD+AUG, STD+RES (Tab. \ref{tab:Titer_CNN_STD_res}) and CLN, CLN+AUG, CLN+RES (\ref{tab:Titer_CNN_CLN_all}) settings. Additionally, this time the NOPP setting was tested in combination with STD+RES and CLN+RES, however, without success, and therefore omitted here.

\begin{table}[h!]
\renewcommand{\arraystretch}{1.4}
    \centering
    \caption{Selected CNN architectures}
    \resizebox{\columnwidth}{!}{%
    \begin{tabular}{llllll}
    \toprule
    Layer           & vanilla           & selu      & lrelu                 & selu sh       & lrelu sh \\
    \midrule    
    CONV FILTER     &                       &                   &                         &                   &                     \\
     Layer 0        & Conv1D(4)             & Conv1D(4,SeLU)    & Conv1D(4,LeakyRelu)     & Conv1D(4,SeLU)    & Conv1D(4,LeakyRelu) \\                       
     Layer 1        & Conv1D(4,LeakyRelu)   & Conv1D(4,SeLU)    & Conv1D(4,LeakyRelu)     & Conv1D(4,SeLU)    & Conv1D(4,LeakyRelu) \\                 
     Layer 2        & MaxPooling1D          & MaxPooling1D      & MaxPooling1D            & MaxPooling1D      & MaxPooling1D        \\                         
     Layer 4        & Conv1D(8)             & Conv1D(8,SeLU)    & Conv1D(8,LeakyRelu)     & Conv1D(8,SeLU)    & Conv1D(8,LeakyRelu) \\                         
     Layer 5        & Conv1D(8,LeakyRelu)   & Conv1D(8,SeLU)    & Conv1D(8,LeakyRelu)     & Conv1D(8,SeLU)    & Conv1D(8,LeakyRelu) \\                         
     Layer 7        & MaxPooling1D          & MaxPooling1D      & MaxPooling1D            & MaxPooling1D      & MaxPooling1D        \\                         
     Layer 8        & Conv1D(16)            & Conv1D(16,SeLU)   & Conv1D(16,LeakyRelu)    & Conv1D(16,SeLU)   & Conv1D(16,LeakyRelu) \\                         
     Layer 9        & Conv1D(16,LeakyRelu)  & Conv1D(16,SeLU)   & Conv1D(16,LeakyRelu)    & Conv1D(16,SeLU)   & Conv1D(16,LeakyRelu) \\                         
     Layer 11       & MaxPooling1D          & MaxPooling1D      & MaxPooling1D            & MaxPooling1D      & MaxPooling1D         \\                         
     Layer 12       & Conv1D(32)            &                   &                         & Conv1D(32,SeLU)   & Conv1D(32,LeakyRelu)  \\                         
     Layer 13       & Conv1D(32,LeakyRelu)  &                   &                         & Conv1D(32,SeLU)   & Conv1D(32,LeakyRelu)  \\                         
     Layer 15       & MaxPooling1D          &                   &                         & MaxPooling1D      & MaxPooling1D          \\
     \midrule 
     REGRESSOR      &                       &                   &                         &                   &                      \\   
     Layer 0        & Dense(60)             & Dense(256,SeLU)   & Dense(256,LeakyRelu)    & Dense(128,SeLU)   & Dense(128,LeakyRelu) \\                     
     Layer 1        & BN                    & Dense(256,SeLU)   & Dense(256,LeakyRelu)    & DO                & DO                   \\           
     Layer 2        & ReLU                  & Dense(64,SeLU)    & Dense(64,LeakyRelu)     & Dense(128,SeLU)   & Dense(128,LeakyRelu) \\           
     Layer 3        & DO                    &                   &                         & Dense(64,SeLU)    & Dense(64,LeakyRelu)  \\           
     Layer 4        & Dense(60)             &                   &                         &                   &                      \\                     
     Layer 5        & BN                    &                   &                         &                   &                      \\           
     Layer 6        & ReLU                  &                   &                         &                   &                      \\           
     Layer 7        & DO                    &                   &                         &                   &                      \\           
     Layer 8        & Dense(10)             &                   &                         &                   &                      \\                     
     Layer 9        & BN                    &                   &                         &                   &                      \\           
     Layer 10       & ReLU                  &                   &                         &                   &                      \\           
     Layer 11       & DO                    &                   &                         &                   &                      \\           
     Output Layer   & Dense(1)              & Dense(1)          & Dense(1)                & Dense(1)          & Dense(1)             \\                 
    \bottomrule
    \end{tabular}%
    }
    \label{tab:Titer_CNN_architectures}
\end{table}

In the standard setting, the CNN models did not outperform the benchmark. However, improvemetns were visible for the CLN, CLN+AUG and CLN+RES, the vanilla model showed best results in these settings. A major improvement compared to previous models is the above average performance on the 4th testsplit (batches 4,9,53,67,96,101,106,126) of the CLN set with RES setting (N RMSE of 0.253). The CNN models were then tested for the VIP 512 on the STD and the CLN sets. Only the shallow models and lrelu showed improvement, mainly for AUG and RES settings. This translates to a 15\% improvent vs the Benchmark (best performance for RES).
For reference, the performance of the CNN models on DS2 is shown in Tab.\ref{tab:TOCI_CNN_Titer}. The lrelu and shallow lrelu models showed above average performance for RES, with an average improvement of 40\% vs the bemchmark. Furthermore, the CNN models were also evaluated on the raw spectrum (NOPP). When trained on the RES set, the models performed significantly better compared to the original setting, yielding above average results. This highlights once more the importance of training the models on a sufficiently large and diverse training set.

\begin{table}[h!]
\renewcommand{\arraystretch}{1.2}
    \centering
    \caption{CNN models N RMSE comparison for (STD, STD+RES, STD+AUG) on DS1 Titer}
    \resizebox{\columnwidth}{!}{%
    \begin{tabular}{l
    R@{\hskip -0.3\tabcolsep}R@{\hskip -0.3\tabcolsep}R
    R@{\hskip -0.3\tabcolsep}R@{\hskip -0.3\tabcolsep}R
    R@{\hskip -0.3\tabcolsep}R@{\hskip -0.3\tabcolsep}R
    R@{\hskip -0.3\tabcolsep}R@{\hskip -0.3\tabcolsep}R
    R@{\hskip -0.3\tabcolsep}R@{\hskip -0.3\tabcolsep}R
    R@{\hskip -0.3\tabcolsep}R@{\hskip -0.3\tabcolsep}R
    R@{\hskip -0.3\tabcolsep}R@{\hskip -0.3\tabcolsep}R}
    \toprule
    
    \multirow{3}{*}{TEST BATCHES} &
      \multicolumn{3}{c}{vanilla} &
      \multicolumn{3}{c}{selu} &
      \multicolumn{3}{c}{selu sh} &
      \multicolumn{3}{c}{lrelu} &
      \multicolumn{3}{c}{lrelu sh} &
      \multicolumn{3}{c}{Benchmark}\\
    & {STD} & {AUG} & {RES} & {STD} & {AUG} & {RES} & {STD} & {AUG} & {RES} & {STD} & {AUG} & {RES} & {STD} & {AUG} & {RES} & {STD} & {AUG} & {RES} \EndTableHeader\\
    \midrule 
     1,7,52,67,97,103,117,127 &         0.310 &  0.229 &  0.267 &     0.341 &  0.250 &  0.253 &     0.371 &  0.279 &  0.272 &     0.358 &  0.327 &  0.295 &     0.467 &  0.275 &  0.291 &     0.234 & 0.409 & 0.254  \\
     2,8,53,92,98,104,118     &         0.295 &  0.269 &  0.274 &     0.321 &  0.357 &  0.303 &     0.364 &  0.381 &  0.301 &     0.243 &  0.290 &  0.250 &     0.285 &  0.317 &  0.241 &     0.294 & 0.372 & 0.288  \\
     3,9,54,93,99,105,119     &         0.261 &  0.258 &  0.191 &     0.212 &  0.263 &  0.188 &     0.280 &  0.280 &  0.203 &     0.237 &  0.281 &  0.180 &     0.341 &  0.232 &  0.189 &     0.226 & 0.372 & 0.230  \\
     4,10,55,94,100,106,124   &         0.214 &  0.247 &  0.169 &     0.265 &  0.149 &  0.149 &     0.264 &  0.151 &  0.144 &     0.190 &  0.186 &  0.154 &     0.223 &  0.202 &  0.159 &     0.191 & 0.253 & 0.194  \\
     5,38,65,95,101,107,125   &         0.261 &  0.228 &  0.263 &     0.296 &  0.282 &  0.258 &     0.258 &  0.275 &  0.271 &     0.353 &  0.284 &  0.296 &     0.359 &  0.297 &  0.275 &     0.272 & 0.372 & 0.313  \\
     6,39,66,96,102,116,126   &         0.473 &  0.426 &  0.402 &     0.538 &  0.425 &  0.389 &     0.539 &  0.392 &  0.418 &     0.544 &  0.411 &  0.413 &     0.600 &  0.393 &  0.407 &     0.403 & 0.577 & 0.400  \\
     AVG BATCHES              &         0.302 &  0.276 &  0.261 &     0.329 &  0.288 &  0.257 &     0.346 &  0.293 &  0.268 &     0.321 &  0.297 &  0.265 &     0.379 &  0.286 &  0.260 &     0.270 & 0.393 & 0.280  \\
     AVG OVERALL              &         0.303 &  0.276 &  0.261 &     0.330 &  0.291 &  0.259 &     0.350 &  0.298 &  0.271 &     0.325 &  0.298 &  0.267 &     0.387 &  0.284 &  0.262 &     0.268 & 0.393 & 0.277  \\
    \bottomrule
    \end{tabular}%
    }
    \label{tab:Titer_CNN_STD_res}
\end{table}

\begin{table}[h!]
\renewcommand{\arraystretch}{1.2}
    \centering
    \caption{CNN models N RMSE comparison for (CLN, CLN+RES, CLN+AUG) on DS1 Titer}
    \resizebox{\columnwidth}{!}{%
    \begin{tabular}{l
    R@{\hskip -0.3\tabcolsep}R@{\hskip -0.3\tabcolsep}R
    R@{\hskip -0.3\tabcolsep}R@{\hskip -0.3\tabcolsep}R
    R@{\hskip -0.3\tabcolsep}R@{\hskip -0.3\tabcolsep}R
    R@{\hskip -0.3\tabcolsep}R@{\hskip -0.3\tabcolsep}R
    R@{\hskip -0.3\tabcolsep}R@{\hskip -0.3\tabcolsep}R
    R@{\hskip -0.3\tabcolsep}R@{\hskip -0.3\tabcolsep}R
    R@{\hskip -0.3\tabcolsep}R@{\hskip -0.3\tabcolsep}R}
    \toprule
    
    \multirow{3}{*}{TEST BATCHES} &
      \multicolumn{3}{c}{vanilla} &
      \multicolumn{3}{c}{selu} &
      \multicolumn{3}{c}{selu sh} &
      \multicolumn{3}{c}{lrelu} &
      \multicolumn{3}{c}{lrelu sh} &
      \multicolumn{3}{c}{Benchmark}\\
    & {CLN} & {AUG} & {RES} & {CLN} & {AUG} & {RES} & {CLN} & {AUG} & {RES} & {CLN} & {AUG} & {RES} & {CLN} & {AUG} & {RES} & {CLN} & {AUG} & {RES} \EndTableHeader\\
    \midrule 
     1,6,38,55,93,98,103,119   &        0.187 & 0.176 & 0.211 &        0.248 & 0.242 & 0.230 &       0.216 & 0.242 & 0.207 &        0.150 & 0.246 & 0.177 &         0.169 & 0.268 & 0.144 &      0.292 & 0.283 & 0.281   \\
     2,7,39,65,94,99,104,124   &        0.338 & 0.217 & 0.232 &        0.311 & 0.234 & 0.222 &       0.273 & 0.294 & 0.281 &        0.320 & 0.263 & 0.254 &         0.335 & 0.256 & 0.288 &      0.250 & 0.313 & 0.235   \\
     3,8,52,66,95,100,105,125  &        0.265 & 0.260 & 0.254 &        0.280 & 0.228 & 0.207 &       0.246 & 0.242 & 0.226 &        0.255 & 0.287 & 0.308 &         0.312 & 0.265 & 0.265 &      0.256 & 0.232 & 0.210   \\
     4,9,53,67,96,101,106,126  &        0.349 & 0.280 & 0.253 &        0.363 & 0.297 & 0.308 &       0.420 & 0.310 & 0.318 &        0.374 & 0.338 & 0.323 &         0.439 & 0.359 & 0.312 &      0.390 & 0.453 & 0.384   \\
     5,10,54,92,97,102,107,127 &        0.253 & 0.191 & 0.208 &        0.297 & 0.219 & 0.212 &       0.257 & 0.192 & 0.229 &        0.254 & 0.221 & 0.247 &         0.281 & 0.215 & 0.230 &      0.295 & 0.261 & 0.240   \\
     AVG BATCHES               &        0.278 & 0.225 & 0.232 &        0.300 & 0.244 & 0.236 &       0.283 & 0.256 & 0.252 &        0.271 & 0.271 & 0.262 &         0.307 & 0.272 & 0.248 &      0.297 & 0.308 & 0.270   \\
     AVG OVERALL               &        0.282 & 0.229 & 0.232 &        0.302 & 0.245 & 0.239 &       0.292 & 0.256 & 0.254 &        0.280 & 0.274 & 0.269 &         0.320 & 0.276 & 0.254 &      0.303 & 0.317 & 0.277   \\
     \bottomrule
    \end{tabular}%
    }
    \label{tab:Titer_CNN_CLN_all}
\end{table}

\begin{table}[h!]
\renewcommand{\arraystretch}{1.2}
    \centering
    \caption{CNN VIP 512 models N RMSE comparison for (STD, STD+RES, STD+AUG) on DS1 Titer}
    \resizebox{\columnwidth}{!}{%
    \begin{tabular}{l
    R@{\hskip -0.3\tabcolsep}R@{\hskip -0.3\tabcolsep}R
    R@{\hskip -0.3\tabcolsep}R@{\hskip -0.3\tabcolsep}R
    R@{\hskip -0.3\tabcolsep}R@{\hskip -0.3\tabcolsep}R
    R@{\hskip -0.3\tabcolsep}R@{\hskip -0.3\tabcolsep}R
    R@{\hskip -0.3\tabcolsep}R@{\hskip -0.3\tabcolsep}R
    R@{\hskip -0.3\tabcolsep}R@{\hskip -0.3\tabcolsep}R
    R@{\hskip -0.3\tabcolsep}R@{\hskip -0.3\tabcolsep}R}
    \toprule
    
    \multirow{3}{*}{TEST BATCHES} &
      \multicolumn{3}{c}{vanilla} &
      \multicolumn{3}{c}{selu} &
      \multicolumn{3}{c}{selu sh} &
      \multicolumn{3}{c}{lrelu} &
      \multicolumn{3}{c}{lrelu sh} &
      \multicolumn{3}{c}{Benchmark}\\
    & {STD} & {AUG} & {RES} & {STD} & {AUG} & {RES} & {STD} & {AUG} & {RES} & {STD} & {AUG} & {RES} & {STD} & {AUG} & {RES} & {STD} & {AUG} & {RES} \EndTableHeader\\
    \midrule 
     1,7,52,67,97,103,117,127 &         0.360 &  0.279 &  0.247 &     0.335 &  0.299 &  0.291 &     0.350 &  0.254 &  0.312 &     0.331 &  0.321 &  0.304 &     0.365 &  0.284 &  0.318 &     0.234 & 0.409 & 0.254  \\
     2,8,53,92,98,104,118     &         0.313 &  0.307 &  0.231 &     0.278 &  0.331 &  0.231 &     0.263 &  0.227 &  0.228 &     0.266 &  0.218 &  0.209 &     0.230 &  0.210 &  0.223 &     0.294 & 0.372 & 0.288  \\
     3,9,54,93,99,105,119     &         0.291 &  0.238 &  0.222 &     0.266 &  0.218 &  0.183 &     0.203 &  0.196 &  0.186 &     0.224 &  0.174 &  0.239 &     0.220 &  0.201 &  0.208 &     0.226 & 0.372 & 0.230  \\
     4,10,55,94,100,106,124   &         0.268 &  0.205 &  0.224 &     0.253 &  0.167 &  0.155 &     0.163 &  0.146 &  0.161 &     0.238 &  0.156 &  0.167 &     0.219 &  0.226 &  0.190 &     0.191 & 0.253 & 0.194  \\
     5,38,65,95,101,107,125   &         0.267 &  0.330 &  0.315 &     0.430 &  0.263 &  0.311 &     0.373 &  0.375 &  0.287 &     0.339 &  0.255 &  0.261 &     0.351 &  0.239 &  0.275 &     0.272 & 0.372 & 0.313  \\
     6,39,66,96,102,116,126   &         0.466 &  0.408 &  0.403 &     0.497 &  0.554 &  0.389 &     0.499 &  0.423 &  0.427 &     0.443 &  0.450 &  0.436 &     0.567 &  0.390 &  0.444 &     0.403 & 0.577 & 0.400  \\
     AVG BATCHES              &         0.327 &  0.295 &  0.274 &     0.343 &  0.306 &  0.260 &     0.308 &  0.270 &  0.267 &     0.307 &  0.262 &  0.269 &     0.325 &  0.258 &  0.276 &     0.270 & 0.393 & 0.280  \\
     AVG OVERALL              &         0.328 &  0.292 &  0.270 &     0.340 &  0.314 &  0.261 &     0.313 &  0.271 &  0.269 &     0.305 &  0.269 &  0.272 &     0.330 &  0.258 &  0.278 &     0.268 & 0.393 & 0.277  \\
    \bottomrule
    \end{tabular}%
    }
    \label{tab:Titer_CNN_512_STD_res}
\end{table}

\begin{table}[h!]
\renewcommand{\arraystretch}{1.2}
    \centering
    \caption{CNN VIP 512 models N RMSE comparison for (CLN, CLN+RES, CLN+AUG) on DS1 Titer}
    \resizebox{\columnwidth}{!}{%
    \begin{tabular}{l
    R@{\hskip -0.3\tabcolsep}R@{\hskip -0.3\tabcolsep}R
    R@{\hskip -0.3\tabcolsep}R@{\hskip -0.3\tabcolsep}R
    R@{\hskip -0.3\tabcolsep}R@{\hskip -0.3\tabcolsep}R
    R@{\hskip -0.3\tabcolsep}R@{\hskip -0.3\tabcolsep}R
    R@{\hskip -0.3\tabcolsep}R@{\hskip -0.3\tabcolsep}R
    R@{\hskip -0.3\tabcolsep}R@{\hskip -0.3\tabcolsep}R
    R@{\hskip -0.3\tabcolsep}R@{\hskip -0.3\tabcolsep}R}
    \toprule
    
    \multirow{3}{*}{TEST BATCHES} &
      \multicolumn{3}{c}{vanilla} &
      \multicolumn{3}{c}{selu} &
      \multicolumn{3}{c}{selu sh} &
      \multicolumn{3}{c}{lrelu} &
      \multicolumn{3}{c}{lrelu sh} &
      \multicolumn{3}{c}{Benchmark}\\
    & {CLN} & {AUG} & {RES} & {CLN} & {AUG} & {RES} & {CLN} & {AUG} & {RES} & {CLN} & {AUG} & {RES} & {CLN} & {AUG} & {RES} & {CLN} & {AUG} & {RES} \EndTableHeader\\
    \midrule 
     1,6,38,55,93,98,103,119   &        0.240 & 0.257 & 0.190 &        0.209 & 0.198 & 0.207 &       0.225 & 0.187 & 0.171 &        0.173 & 0.182 & 0.154 &         0.164 & 0.182 & 0.167 &      0.292 & 0.283 & 0.281   \\
     2,7,39,65,94,99,104,124   &        0.285 & 0.272 & 0.254 &        0.276 & 0.226 & 0.244 &       0.292 & 0.207 & 0.254 &        0.307 & 0.289 & 0.251 &         0.263 & 0.273 & 0.224 &      0.250 & 0.313 & 0.235   \\
     3,8,52,66,95,100,105,125  &        0.269 & 0.240 & 0.232 &        0.314 & 0.221 & 0.225 &       0.257 & 0.266 & 0.214 &        0.254 & 0.282 & 0.216 &         0.259 & 0.310 & 0.250 &      0.256 & 0.232 & 0.210   \\
     4,9,53,67,96,101,106,126  &        0.342 & 0.338 & 0.312 &        0.431 & 0.335 & 0.330 &       0.435 & 0.327 & 0.284 &        0.292 & 0.311 & 0.313 &         0.428 & 0.318 & 0.299 &      0.390 & 0.453 & 0.384   \\
     5,10,54,92,97,102,107,127 &        0.299 & 0.199 & 0.211 &        0.335 & 0.244 & 0.217 &       0.326 & 0.267 & 0.226 &        0.284 & 0.240 & 0.224 &         0.255 & 0.202 & 0.244 &      0.295 & 0.261 & 0.240   \\
     AVG BATCHES               &        0.287 & 0.261 & 0.240 &        0.313 & 0.245 & 0.244 &       0.307 & 0.251 & 0.230 &        0.262 & 0.261 & 0.231 &         0.274 & 0.257 & 0.237 &      0.297 & 0.308 & 0.270   \\
     AVG OVERALL               &        0.290 & 0.263 & 0.243 &        0.325 & 0.251 & 0.248 &       0.318 & 0.259 & 0.233 &        0.266 & 0.264 & 0.237 &         0.289 & 0.263 & 0.243 &      0.303 & 0.317 & 0.277   \\
     \bottomrule
    \end{tabular}%
    }
    \label{tab:Titer_CNN_512_CLN_all}
\end{table}

\begin{table}[h!]%
\caption{CNN models N RMSE comparison for (STD, STD+RES) on DS2 Titer}%
\renewcommand*{\MinNumber}{0.07}%
\renewcommand*{\MidNumber}{0.249}%
\renewcommand*{\MaxNumber}{0.41600000000000004}%
\centering%
\begin{adjustbox}{width=0.98\textwidth}%
\begin{tabular}{@{}lR@{\hskip-1.0\tabcolsep}RR@{\hskip-1.0\tabcolsep}RR@{\hskip-1.0\tabcolsep}RR@{\hskip-1.0\tabcolsep}RR@{\hskip-1.0\tabcolsep}RR@{\hskip-1.0\tabcolsep}RR@{\hskip-1.0\tabcolsep}RR@{\hskip-1.0\tabcolsep}RR@{\hskip-1.0\tabcolsep}RR@{\hskip-1.0\tabcolsep}RR@{\hskip-1.0\tabcolsep}R@{}}%
\toprule%
\multirow{2}{*}{TEST BATCHES}&\multicolumn{2}{c}{ vanilla}&\multicolumn{2}{c}{ selu}&\multicolumn{2}{c}{ selu sh}&\multicolumn{2}{c}{ lrelu}&\multicolumn{2}{c}{ lrelu sh}&\multicolumn{2}{c}{ vanilla }&\multicolumn{2}{c}{ selu }&\multicolumn{2}{c}{ selu sh }&\multicolumn{2}{c}{ lrelu }&\multicolumn{2}{c}{ lrelu sh }&\multicolumn{2}{c}{Benchmark}\\%
&STD&RES&STD&RES&STD&RES&STD&RES&STD&RES&NOPP&RES&NOPP&RES&NOPP&RES&NOPP&RES&NOPP&RES&STD&RES\\%
\midrule%
\EndTableHeader%
97&0.215&0.227&0.158&0.149&0.310&0.136&0.413&0.134&0.284&0.119&0.240&0.256&0.237&0.159&0.201&0.172&0.181&0.162&0.220&0.135&0.416&0.386\\%
98&0.459&0.420&0.199&0.130&0.382&0.087&0.233&0.099&0.155&0.097&0.296&0.155&0.277&0.262&0.292&0.287&0.173&0.110&0.253&0.147&0.070&0.066\\%
99&0.215&0.176&0.265&0.197&0.233&0.187&0.177&0.172&0.233&0.178&0.279&0.192&0.182&0.175&0.184&0.196&0.222&0.165&0.192&0.165&0.156&0.150\\%
101&0.237&0.149&0.249&0.197&0.320&0.222&0.242&0.185&0.231&0.183&0.580&0.535&0.529&0.375&0.596&0.363&0.659&0.329&0.720&0.449&0.182&0.191\\%
152&0.444&0.149&0.339&0.119&0.326&0.122&0.370&0.378&0.467&0.252&0.794&0.219&0.326&0.132&0.545&0.128&0.446&0.167&0.616&0.128&0.410&0.420\\%
154&0.285&0.254&0.544&0.469&0.667&0.553&0.143&0.111&0.117&0.131&0.682&0.229&0.447&0.314&0.436&0.205&0.845&0.199&0.486&0.196&0.298&0.324\\%
AVG BATCHES&0.309&0.229&0.292&0.210&0.373&0.218&0.263&0.180&0.248&0.160&0.478&0.264&0.333&0.236&0.376&0.225&0.421&0.188&0.415&0.203&0.255&0.256\\%
AVG OVERALL&0.307&0.265&0.242&0.186&0.337&0.192&0.277&0.156&0.231&0.146&0.404&0.305&0.333&0.252&0.363&0.255&0.394&0.199&0.411&0.245&0.249&0.241\\\bottomrule%
\end{tabular}%
\end{adjustbox}%
\label{tab:TOCI_CNN_Titer}
\end{table}

\clearpage
\subsection{DL Models - VAE}

VAE models were particularly hard to train as they were composed from different submodels, requiring separate optimization objectives. This section focuses mainly on the network design results and findings rather than the N RMSE performance. The focus of the following experiments was the application of VAEs on the raw spectrum (NOPP), results for the STD spectrum will be shown briefly at the end of the subsection.

\begin{table}[h!]
\renewcommand{\arraystretch}{1.4}
    \centering
    \caption{Selected VAE architectures}
    \resizebox{\columnwidth}{!}{%
    \begin{tabular}{lllll}
    \toprule
    Layer           & VAE vanilla      & VAE convT           & VAE fc dec           & VAE fc  \\
    \midrule                    
    ENCODER         &                  &                     &                      &                   \\
    \midrule  
     Layer 0        & Conv1D(4,SeLU)   & Conv1D(4,SeLU)      & Conv1D(4,SeLU)       & MaxPooling1D      \\                     
     Layer 1        & MaxPooling1D     & MaxPooling1D        & MaxPooling1D         & Dense(1280,SeLU)  \\                   
     Layer 2        & Conv1D(16,SeLU)  & Conv1D(16,SeLU)     & Conv1D(16,SeLU)      & DO                \\                
     Layer 4        & MaxPooling1D     & MaxPooling1D        & MaxPooling1D         & Dense(512,SeLU)   \\ 
     Layer 5        & Conv1D(8,SeLU)   & Conv1D(8,SeLU)      & Conv1D(8,SeLU)       & DO                \\          
     Layer 7        & MaxPooling1D     & MaxPooling1D        & MaxPooling1D         & Dense(256,SeLU)   \\                             
     Layer 8        & Dense(256,SeLU)  & Dense(256,SeLU)     & Dense(256,SeLU)      &                   \\ 
     Encoder Output & Dense(64,SeLU)   & Dense(64,SeLU)      & Dense(64,SeLU)       & Dense(64,SeLU)    \\       
    \midrule
    LATENT LAYERS   &                               \\
    \midrule  
    z mean          & Dense(128)                    \\
    z log var       & Dense(128)                    \\             
    z               & Sampling(z mean, z log var) \\
    \midrule
    DECODER         &                    &                    &                      &                 \\  
    \midrule  
     Layer 0        & Dense(64,SeLU)     & Dense(64,SeLU)     & Dense(64,SeLU)       & Dense(64,SeLU)  \\  
     Layer 1        & Dense(256,SeLU)    & Dense(256,SeLU)    & Dense(128,SeLU)      & Dense(256,SeLU) \\
     Layer 2        & Dense(1280,SeLU)   & Dense(1280,SeLU)   &                      & Dense(1280,SeLU)\\                  
     Layer 3        & Reshape(160,8)     & Reshape(160,8)     &                      &                 \\  
     Layer 4        & Conv1D(8,SeLU)     & Upsampling1D       &                      &                 \\  
     Layer 5        & Upsampling1D       & Conv1DT(8,SeLU)    &                      &                 \\  
     Layer 6        & Conv1D(16,SeLU)    & Upsampling1D       &                      &                 \\  
     Layer 7        & Upsampling1D       & Conv1DT(16,SeLU)   &                      &                 \\  
     Layer 8        & Conv1D(8,SeLU)     & Upsampling1D       &                      &                 \\  
     Layer 9        & Upsampling1D       & Conv1DT(8,SeLU)    &                      &                 \\  
     Layer 10       & Conv1D(4,SeLU)     & Upsampling1D       &                      &                 \\  
     Layer 11       & Conv1D(1,SeLU)     & Conv1DT(4,SeLU)    &                      &                 \\  
     Decoder Output & Dense(input shape) & Conv1DT(1,SeLU)    & Dense(input shape)   & Upsampling1D    \\
     \midrule 
     REGRESSOR          &                               \\ 
     \midrule  
     Layer 0            & Dense(32,SeLU)                \\                     
     Layer 1            & Dense(128,SeLU)               \\           
     Regressor Output   & Dense(1)                      \\
    \bottomrule
    \end{tabular}%
    }
    \label{tab:VAE_architectures}
\end{table}

\clearpage
The main objective was to separate the concentration as a variable in the latent space, while keeping a low reconstruction error. It was quickly noticed that an Autoencoder with the best reconstruction performance will not automatically produce good regression results from its latent space. In other words, the latent space required for regression is quite different from the latent space found by an Autoencoder focused on reconstruction, even when its regularized by the variational constraint.

\begin{figure}[h!]
  \begin{minipage}[h]{0.98\textwidth}
    a)\\
    \includegraphics[width = \textwidth]{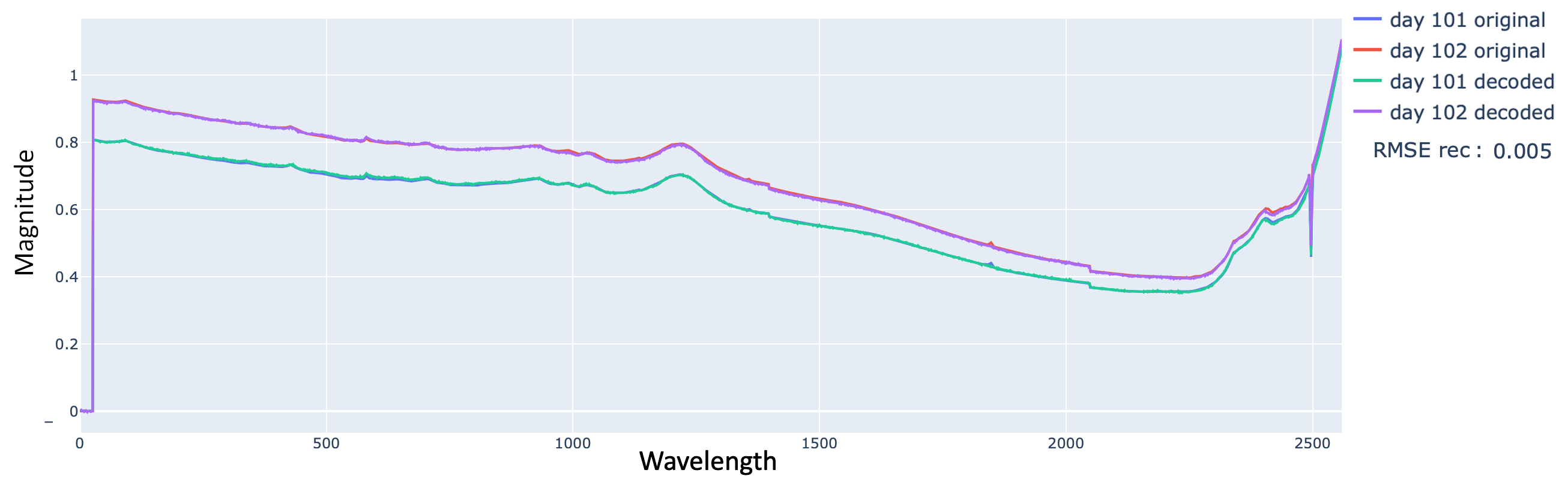}
  \end{minipage}
  \\
  \begin{minipage}[h]{0.5\textwidth}
    b)\\
    \includegraphics[width = \textwidth]{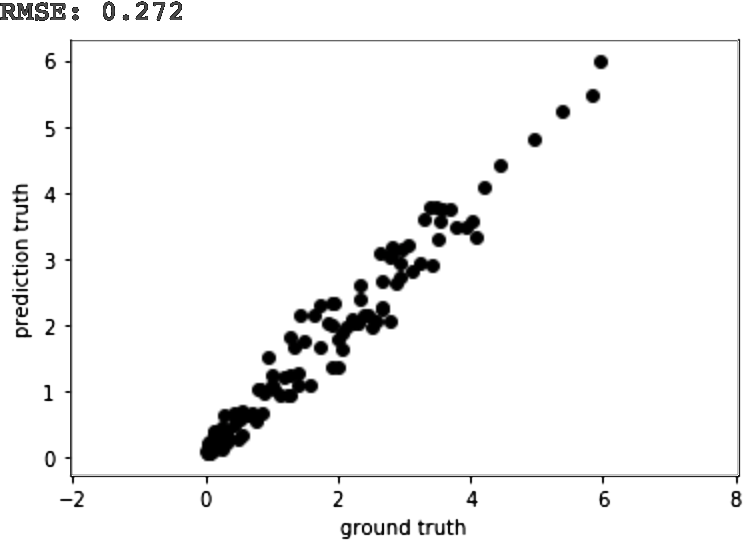}
  \end{minipage}
  \begin{minipage}[h]{0.46\textwidth}
    c)\\
    \includegraphics[width = \textwidth]{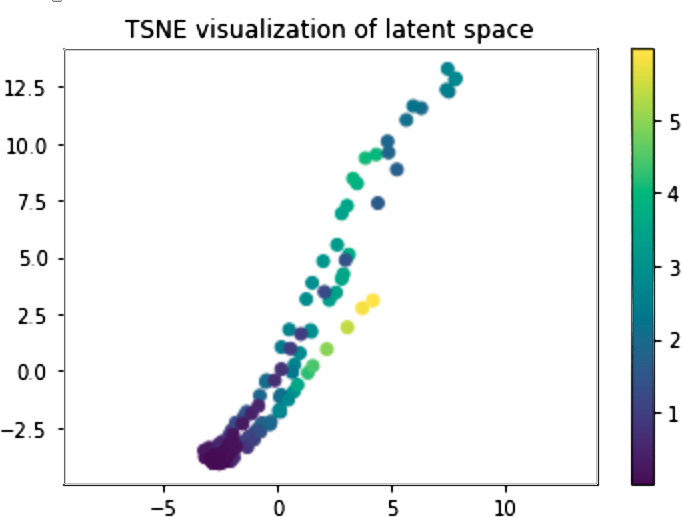}
  \end{minipage}
 \caption{Reconstruction a) and prediction b) performance (validation) of an earlier VAE fc model on DS1 batches [39, 65, 119, 103]. RMSE is absolute. While the reconstruction error is low, the regression error suggests a necessary improvement of the model's architecture. c) predictions representation in the latent space (z mean), axes represent most dissimilar directions}
 \label{fig:Titer_VAE_fc_raw_spect_tsne_rmse}
 \end{figure}

The results for a VAE which performed exceptionally well on the reconstruction of the NOPP spectrum are shown in Fig. \ref{fig:Titer_VAE_fc_raw_spect_tsne_rmse}. While the quality of the reconstructed raw spectrum in a) seems to be quite high (absolute RMSE in the order of $10^-3$) the N RMSE b) of the regression is not particularly low. Further analysis of the latent space structure c) shows that the concentration could not be fully decoupled from the other latent variables. The t-SNE plot (t-Distributed Stochastic Neighbor Embedding \cite{vanDerMaaten2008}) in c) shows a 2D representation of the latent space mapped on the two most dissimilar directions in the latent space. The input datapoints are mapped to the latent space via the encoder and colored by the true target concentration. While there seems to be an overall separation between high and low concentrations, it does not have a unique direction (three branches), which means that the concentration is not a fully independent latent variable. Also, points with low concentration seem to be scattered across the space, which proves a bad separation. The reason for such a behaviour is that in this VAE model, the decoder part has a significantly higher complexety (3.4M parameters) compared to the encoder (500k parameters) and regressor (100k parameters), as it's last layer is a dense layer of the size of the input.

\newpage

\begin{figure}[h!]
  \begin{minipage}[h]{0.98\textwidth}
    a)
    \includegraphics[width = \textwidth]{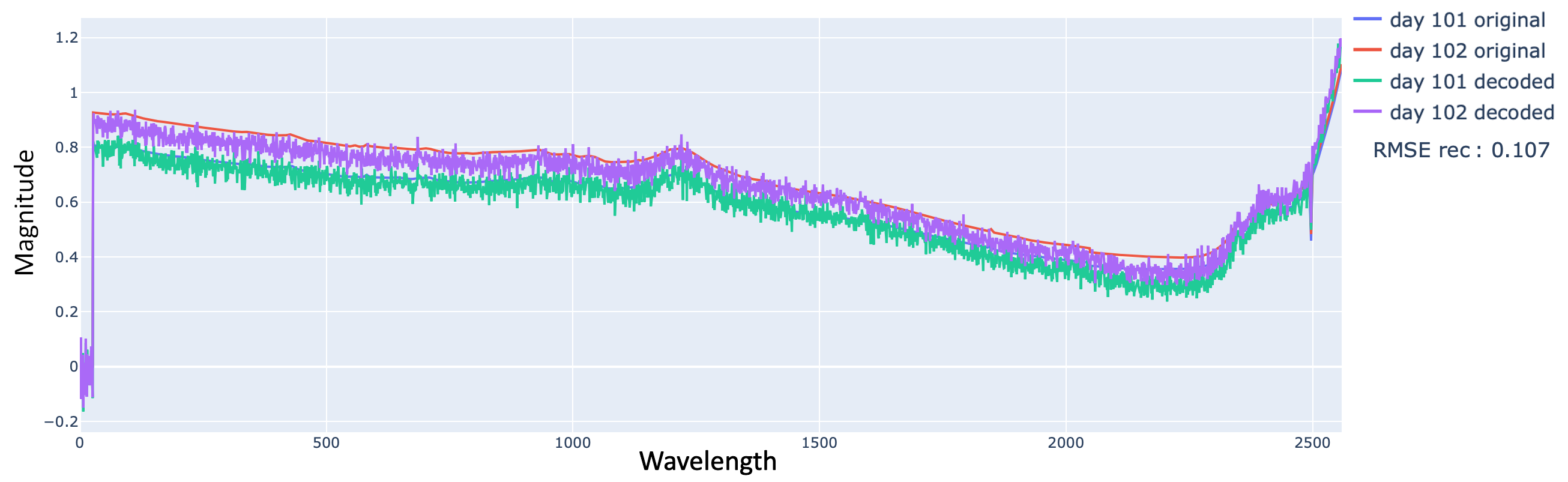}
  \end{minipage}
  \\
  \begin{minipage}[h]{0.5\textwidth}
    b)
    \includegraphics[width = \textwidth]{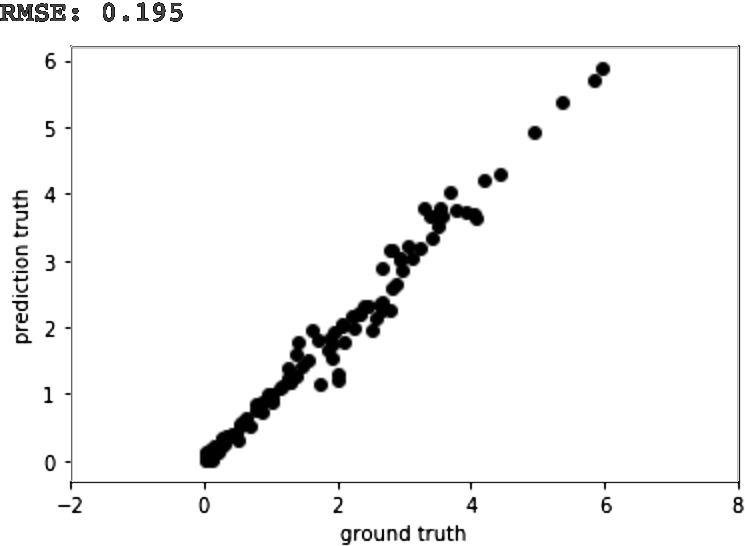}
  \end{minipage}
  \begin{minipage}[h]{0.46\textwidth}
    c)
    \includegraphics[width = \textwidth]{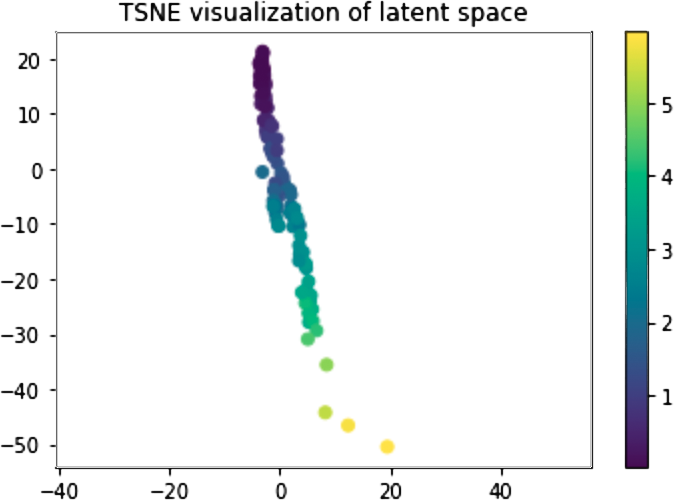}
  \end{minipage}
 
 \caption{Reconstruction and prediction performance (validation) of a vanilla VAE model on DS1 batches [39, 65, 119, 103]. RMSE is absolute.}
 \label{fig:Titer_VAE_conv_raw_spect_tsne_rmse}
 \end{figure}

Fig. \ref{fig:Titer_VAE_conv_raw_spect_tsne_rmse}, on the other hand, shows a VAE model which produced a noisy spectrum a) during reconstruction, but has a significantly (30\%) higher regression accuracy b), which also seemed to surpass the benchmark model on the STD data (not shown here, as it is not the focus of this comparison) and previous models on the NOPP data. Furthermore, the t-SNE plot c) shows a clear separation of the concentration from other target variables, except for higher concentrations above 5.5 g/L. It can be also noticed that almost no branching of the concentrations is present, the separation appears in a linear, continuous fashion. The difference of this model compared to the previous is that it is designed in a symmetrical, more balanced way, where the encoder and decoder have similar complexity. As a result, it is much more prone to information loss, manifested in the lower reconstruction quality.  However, the overall lower complexity allows to shape the latent space in a way that is beneficial for the regressor.

\begin{figure}[h]
\centering
\begin{minipage}{.98\textwidth}
    \centering
    \includegraphics[width=.58\textwidth]{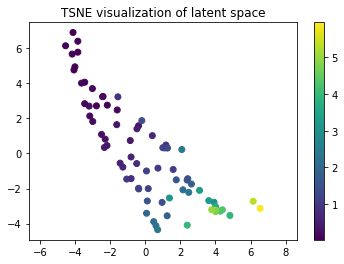}
    \captionof{figure}{t-SNE representation of latent space z mean obtained by adding non-linearity to the latent part of the VAE network. This adds complexity to the latent representation, which is not beneficial for the separation of points with different concentrations.}
    \label{fig:latentspacenonlin}
\end{minipage}%
\end{figure}

\bigbreak
Originally, the latent layers were implemented as \texttt{Dense(latent\_dim, linear)} layers (only one layer between the encoder output and latent output and linear activation). Adding non-linearities to the latent layer (\texttt{z mean = Dense(latent\_dim,SeLU)(Dense(256,SeLU)(Dense(128,SeLU))))}) severely affected the separation of points with different concentration, as shown in Fig. \ref{fig:latentspacenonlin}. Therefore the latent layer should be kept linear.

\begin{figure}[h]
\centering
\begin{minipage}{.98\textwidth}
  \centering
  \includegraphics[width=.98\textwidth]{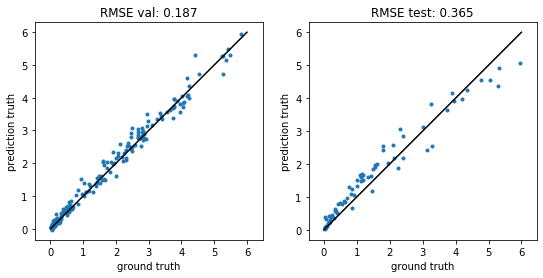}
  \captionof{figure}{RMSE test on [39, 65, 119, 103] (right) and RMSE validation randomly chosen from 20\% of remaining data, excluding training data and before augmentation (left)}
  \label{fig:VAE_titer_val_test}
\end{minipage}
\end{figure}

It shall be noted that the performance was significantly higher (50\%-100\%) when validating against batches [39, 65, 119, 103] (different from previous experiments, selected manually as 'difficult' test batches) than when using them for testing only and validating on a different set (random 20\% of remaining data, excluding training data and before augmentation). This behaviour is shown in Fig. \ref{fig:VAE_titer_val_test}. Both sets are unseen data for the model, the first is used to select the best training weights, while the second is used to verify the performance. The large spread in error suggests that the val-test data was highly inhomogeneous and the sets were disproportionate. Normally, a random split should be sufficient, but since the test data is not randomized (only batch-wise, but not intra-batch), the validation data should be chosen accordingly. This should be considered for future experiments, by using the equal-sized consecutive splits obtained from the shuffling algorithm defined in Appendix A. 1: for example, when testing against batches [1,7,52,67,97,103,117,127] a good choice for validation would be [2,8,53,92,98,104,118].

Unfortunately, the performance during cross-validation was not as exceptional as during the architecture selection. This is due to a number of reasons:\\
CV was performed on the Euler cluster (with different hardware, and different tensorflow implementations), while the above model selection was done on different hardware, using distribution strategies. The main difficulty was to find a suitable number of required training epochs (\# of passes through the entire dataset), without over fitting. This number depends on the chosen batchsize (\# of training samples fed through the network before calculating the gradients, not to be confused with the batches of the dataset, describing sequences of consecutive measurements). When using distributed training, the batch is split across the computation units and the computed gradients are then aggregated. This allows to use higher batchsizes, which results in 'faster' training without reducing generalization performance (contrary to normal large batch training \cite{DBLP:journals/corr/KeskarMNST16}). This might lead to different optima than the ones found during training on a single computation node, as in the case of cross-validation on the Euler cluster, where distribution is only used for training multiple models simultaneously and for multiple CV splits on different cores, however, the gradient descent is not distributed.\\
During gradient descent distribution, batchsize was set to 128 and the optimal number of epochs varied between 1000-2000 epochs, depending on the model. For cross-validation on Euler, the batchsize was 16, therefore the optimal number of epochs was assumed to scale accordingly to 125-250 epochs. However, not all models converged for this setting, therefore individual settings need to be found for each model. It was attempted to solve this problem by implementing a delayed early stopping method, which tracks the decrease of the loss starting after a given number of epochs (150) and stops the training if the loss did not improve for a certain number of epochs (50).
Furthermore, the performance could be improve by using more homogeneous validation-test splits, as mentioned earlier.\\
The the N RMSE performance of different models is presented in Tab. \ref{tab:Titer_VAE_STD_res} and \ref{tab:Titer_VAE_CLN_res} for DS1. In the STD setting, the fc VIP 512 model showed above average performance. It may be worth mentioning that the architecture of the VIP 512 model is different from the fc model (from fully-connected) in that it has lower number of layers (first three and last two layers removed) due to the different size of the input. In the NOPP setting, none of the models showed satisfactory results, most probably because the number of training epochs was not adjusted well enough.\\
Tab. \ref{tab:TOCI_Titer_VAE} shows the performance of the VAE models on DS2 for reference. It can be seen that the results are above average for the STD data for all models and average for NOPP data for the convT and fc dec models.\\
Having evaluated the performance of VAEs on just one targte variable, it is yet too early to draw conclusions about its overall performance. Therefore, it will be further analysed for the remaining target variables in the upcoming sections.

\begin{table}[h!]
    \centering
    \caption{VAE models N RMSE comparison for (STD, STD+RES, STD+AUG) on DS1 Titer }
    \resizebox{\columnwidth}{!}{%
    \begin{tabular}{l
    R@{\hskip -0.3\tabcolsep}R
    R@{\hskip -0.3\tabcolsep}R
    R@{\hskip -0.3\tabcolsep}R
    R@{\hskip -0.3\tabcolsep}R
    R@{\hskip -0.3\tabcolsep}R
    R@{\hskip -0.3\tabcolsep}R
    R@{\hskip -0.3\tabcolsep}R
    R@{\hskip -0.3\tabcolsep}R
    R@{\hskip -0.3\tabcolsep}R
    R@{\hskip -0.3\tabcolsep}R}
    \toprule
    
    \multirow{3}{*}{TEST BATCHES} &
      \multicolumn{2}{c}{vanilla} &
      \multicolumn{2}{c}{convT} &
      \multicolumn{2}{c}{fc dec} &
      \multicolumn{2}{c}{fc} &
      \multicolumn{2}{c}{fc VIP 512} &
      \multicolumn{2}{c}{vanilla} &
      \multicolumn{2}{c}{convT} &
      \multicolumn{2}{c}{fc dec} &
      \multicolumn{2}{c}{fc} &
      \multicolumn{2}{c}{Benchmark}\\
    & {STD} & {RES}     & {STD} & {RES}     & {STD} & {RES}   & {STD} & {RES}   & {STD} & {RES}     & {NOPP} & {RES}     & {NOPP} & {RES}    & {NOPP} & {RES}    & {NOPP} & {RES}    & {STD} & {RES} \EndTableHeader\\
    \midrule 
     1,7,52,67,97,103,117,127 &  0.312 &  0.215 &  0.299 &  0.298 &   0.318 &  0.292 &     0.241 &  0.209 &    0.232 &  0.223 &   0.312 &  0.279 &     0.353 &  0.328 &   0.322 &  0.305 &    0.457 &  0.529 &   0.234 & 0.254  \\
     2,8,53,92,98,104,118     &  0.292 &  0.335 &  0.315 &  0.277 &   0.310 &  0.285 &     0.306 &  0.324 &    0.217 &  0.237 &   0.417 &  0.372 &     0.369 &  0.362 &   0.429 &  0.325 &    0.726 &  0.570 &   0.294 & 0.288  \\
     3,9,54,93,99,105,119     &  0.199 &  0.218 &  0.196 &  0.196 &   0.194 &  0.219 &     0.170 &  0.190 &    0.170 &  0.179 &   0.323 &  0.299 &     0.308 &  0.324 &   0.359 &  0.287 &    0.485 &  0.568 &   0.226 & 0.230  \\
     4,10,55,94,100,106,124   &  0.192 &  0.122 &  0.179 &  0.177 &   0.193 &  0.155 &     0.174 &  0.141 &    0.159 &  0.135 &   0.221 &  0.226 &     0.199 &  0.234 &   0.238 &  0.199 &    0.440 &  0.607 &   0.191 & 0.194  \\
     5,38,65,95,101,107,125   &  0.271 &  0.246 &  0.265 &  0.267 &   0.288 &  0.290 &     0.251 &  0.268 &    0.260 &  0.253 &   0.339 &  0.376 &     0.295 &  0.334 &   0.341 &  0.342 &    0.554 &  0.626 &   0.272 & 0.313  \\
     6,39,66,96,102,116,126   &  0.437 &  0.441 &  0.411 &  0.446 &   0.405 &  0.409 &     0.429 &  0.442 &    0.358 &  0.377 &   0.718 &  0.611 &     0.701 &  0.672 &   0.694 &  0.639 &    0.692 &  0.755 &   0.403 & 0.400  \\
     AVG BATCHES              &  0.284 &  0.263 &  0.277 &  0.277 &   0.284 &  0.275 &     0.262 &  0.262 &    0.233 &  0.234 &   0.389 &  0.361 &     0.371 &  0.376 &   0.397 &  0.350 &    0.559 &  0.609 &   0.270 & 0.280  \\
     AVG OVERALL              &  0.285 &  0.270 &  0.279 &  0.279 &   0.285 &  0.276 &     0.264 &  0.267 &    0.232 &  0.235 &   0.398 &  0.362 &     0.383 &  0.381 &   0.404 &  0.355 &    0.560 &  0.601 &   0.268 & 0.277  \\
    \bottomrule
    \end{tabular}%
    }
    \label{tab:Titer_VAE_STD_res}
    \vspace{-5mm}
\end{table}

\begin{table}[h]
    \centering
    \caption{VAE models N RMSE comparison for (CLN, CLN+RES, CLN+AUG) on DS1 Titer}
    \resizebox{\columnwidth}{!}{%
    \begin{tabular}{l
    R@{\hskip -0.3\tabcolsep}R
    R@{\hskip -0.3\tabcolsep}R
    R@{\hskip -0.3\tabcolsep}R
    R@{\hskip -0.3\tabcolsep}R
    R@{\hskip -0.3\tabcolsep}R
    R@{\hskip -0.3\tabcolsep}R
    R@{\hskip -0.3\tabcolsep}R
    R@{\hskip -0.3\tabcolsep}R
    R@{\hskip -0.3\tabcolsep}R
    R@{\hskip -0.3\tabcolsep}R}
    \toprule
    
    \multirow{3}{*}{TEST BATCHES} &
      \multicolumn{2}{c}{vanilla} &
      \multicolumn{2}{c}{convT} &
      \multicolumn{2}{c}{fc dec} &
      \multicolumn{2}{c}{fc} &
      \multicolumn{2}{c}{fc VIP 512} &
      \multicolumn{2}{c}{vanilla} &
      \multicolumn{2}{c}{convT} &
      \multicolumn{2}{c}{fc dec} &
      \multicolumn{2}{c}{fc} &
      \multicolumn{2}{c}{Benchmark}\\
    & {CLN} & {RES}     & {CLN} & {RES}     & {CLN} & {RES}   & {CLN} & {RES}   & {STD} & {RES}     & {NOPP} & {RES}     & {NOPP} & {RES}    & {NOPP} & {RES}    & {NOPP} & {RES}    & {STD} & {RES} \EndTableHeader\\
    \midrule 
     1,6,38,55,93,98,103,119   &  0.239 &  0.208 &  0.244 &  0.206 &   0.253 &  0.221 &     0.228 &  0.230 &    000 &  000 &   0.275 &  0.343 &     0.251 &  0.275 &   0.267 &  0.252 &    0.549 &  0.513 &   0.292 & 0.281  \\
     2,7,39,65,94,99,104,124   &  0.218 &  0.248 &  0.221 &  0.247 &   0.235 &  0.254 &     0.240 &  0.226 &    000 &  000 &   0.605 &  0.354 &     0.337 &  0.323 &   0.391 &  0.339 &    0.618 &  0.617 &   0.250 & 0.235  \\
     3,8,52,66,95,100,105,125  &  0.209 &  0.217 &  0.221 &  0.206 &   0.230 &  0.211 &     0.207 &  0.171 &    000 &  000 &   0.331 &  0.303 &     0.298 &  0.242 &   0.229 &  0.279 &    0.598 &  0.583 &   0.256 & 0.210  \\
     4,9,53,67,96,101,106,126  &  0.375 &  0.409 &  0.357 &  0.335 &   0.356 &  0.338 &     0.414 &  0.346 &    000 &  000 &   0.459 &  0.432 &     0.480 &  0.471 &   0.509 &  0.518 &    0.615 &  0.773 &   0.390 & 0.384  \\
     5,10,54,92,97,102,107,127 &  0.257 &  0.230 &  0.240 &  0.255 &   0.226 &  0.259 &     0.244 &  0.223 &    000 &  000 &   0.298 &  0.269 &     0.246 &  0.254 &   0.297 &  0.282 &    0.405 &  0.671 &   0.295 & 0.240  \\
     AVG BATCHES               &  0.260 &  0.262 &  0.257 &  0.250 &   0.260 &  0.257 &     0.267 &  0.239 &    000 &  000 &   0.394 &  0.340 &     0.322 &  0.313 &   0.339 &  0.334 &    0.557 &  0.631 &   0.297 & 0.270  \\
     AVG OVERALL               &  0.268 &  0.274 &  0.263 &  0.255 &   0.265 &  0.261 &     0.278 &  0.246 &    000 &  000 &   0.399 &  0.342 &     0.333 &  0.323 &   0.351 &  0.348 &    0.557 &  0.641 &   0.303 & 0.277  \\
    \bottomrule
    \end{tabular}%
    }
    \label{tab:Titer_VAE_CLN_res}
    \vspace{-5mm}
\end{table}

\begin{table}[h!]%
\caption{VAE models N RMSE comparison for (STD, STD+RES) on DS2 Titer}%
\renewcommand*{\MinNumber}{0.07}%
\renewcommand*{\MidNumber}{0.249}%
\renewcommand*{\MaxNumber}{0.41600000000000004}%
\centering%
\begin{adjustbox}{width=0.98\textwidth}%
\begin{tabular}{@{}lR@{\hskip-1.0\tabcolsep}RR@{\hskip-1.0\tabcolsep}RR@{\hskip-1.0\tabcolsep}RR@{\hskip-1.0\tabcolsep}RR@{\hskip-1.0\tabcolsep}RR@{\hskip-1.0\tabcolsep}RR@{\hskip-1.0\tabcolsep}RR@{\hskip-1.0\tabcolsep}RR@{\hskip-1.0\tabcolsep}R@{}}%
\toprule%
\multirow{2}{*}{TEST BATCHES}&\multicolumn{2}{c}{ vanilla}&\multicolumn{2}{c}{ convT}&\multicolumn{2}{c}{ fc dec}&\multicolumn{2}{c}{ fc}&\multicolumn{2}{c}{ vanilla }&\multicolumn{2}{c}{ convT }&\multicolumn{2}{c}{ fc dec }&\multicolumn{2}{c}{ fc }&\multicolumn{2}{c}{Benchmark}\\%
&STD&RES&STD&RES&STD&RES&STD&RES&NOPP&RES&NOPP&RES&NOPP&RES&NOPP&RES&STD&RES\\%
\midrule%
\EndTableHeader%
97&0.298&0.315&0.265&0.172&0.293&0.230&0.126&0.164&0.274&0.264&0.160&0.149&0.150&0.131&0.230&0.244&0.416&0.386\\%
98&0.250&0.099&0.197&0.164&0.146&0.160&0.300&0.292&0.454&0.357&0.184&0.129&0.221&0.119&0.603&0.242&0.070&0.066\\%
99&0.165&0.160&0.166&0.158&0.167&0.159&0.155&0.156&0.189&0.210&0.162&0.180&0.161&0.174&0.298&0.207&0.156&0.150\\%
101&0.228&0.206&0.218&0.216&0.207&0.215&0.265&0.202&0.879&0.768&0.492&0.471&0.519&0.471&0.623&0.729&0.182&0.191\\%
152&0.235&0.139&0.117&0.170&0.133&0.117&0.502&0.268&0.381&0.379&0.159&0.139&0.198&0.139&1.244&0.183&0.410&0.420\\%
154&0.074&0.080&0.154&0.136&0.183&0.171&0.132&0.401&0.384&0.474&0.162&0.143&0.169&0.161&0.207&1.064&0.298&0.324\\%
AVG BATCHES&0.208&0.167&0.186&0.169&0.188&0.175&0.247&0.247&0.427&0.409&0.220&0.202&0.236&0.199&0.534&0.445&0.255&0.256\\%
AVG OVERALL&0.233&0.200&0.206&0.172&0.203&0.186&0.235&0.224&0.503&0.445&0.270&0.253&0.287&0.249&0.507&0.449&0.249&0.241\\\bottomrule%
\end{tabular}%
\end{adjustbox}%
\label{tab:TOCI_Titer_VAE}
\end{table}

\clearpage
\subsection{Fused Models}

Tab. \ref{tab:fusionDS13Titer} presents the results of fused VAE fc VIP 512, KT VIP 512 Top 1 and KT VIP 512 Top 2 models using the FWLS and BLUE methods. Both yield comparable performance, resulting in an overall lower N RMSE (10\% improvement vs the base model with the lowest score, here KT VIP 512 Top 2 and close to 20\% improvement vs the benchmark in the STD setting).

\begin{table}[h!]%
\caption{N RMSE comparison for FWLS and BLUE fusion of VAE fc VIP 512, KT VIP 512 Top 1 and KT VIP 512 Top 2 on DS1 Titer}%
\renewcommand*{\MinNumber}{0.191}%
\renewcommand*{\MidNumber}{0.259}%
\renewcommand*{\MaxNumber}{0.37200000000000005}%
\centering%
\begin{adjustbox}{width=0.98\textwidth}%
\begin{tabular}{lRRRRRR}
\toprule
\multirow{2}{*}{} &   VAE fc &   KT Top 1 &  KT Top 2 & FWLS & BLUE & Benchmark\\
TEST BATCHES & VIP 512 & VIP 512 & VIP 512 & VIP 512 & VIP 512 & STD \\
\midrule
\EndTableHeader
 1,7,52,67,97,103,117,127 &    0.232 &         0.232 &         0.264 &    0.225 &   0.217 &    0.242 \\
 2,8,53,92,98,104,118     &    0.217 &         0.178 &         0.204 &    0.169 &   0.168 &    0.318 \\
 3,9,54,93,99,105,119     &    0.170 &         0.147 &         0.191 &    0.152 &   0.133 &    0.229 \\
 4,10,55,94,100,106,124   &    0.159 &         0.125 &         0.159 &    0.129 &   0.131 &    0.207 \\
 5,38,65,95,101,107,125   &    0.260 &         0.235 &         0.233 &    0.233 &   0.224 &    0.305 \\
 6,39,66,96,102,116,126   &    0.358 &         0.425 &         0.403 &    0.384 &   0.412 &    0.421 \\
 AVG BATCHES              &    0.233 &         0.224 &         0.242 &    0.215 &   0.214 &    0.287 \\
 AVG OVERALL              &    0.232 &         0.230 &         0.244 &    0.218 &   0.220 &    0.285 \\
\bottomrule
\end{tabular}
\end{adjustbox}%
\label{tab:fusionDS13Titer}
\end{table}

\bigbreak

Overall, the lowest N RMSE on DS1 Titer STD was achieved with the FWLS fusion model for VIP 512 variable selection, beating the STD PLS benchmark by almost 20\%, followed by KT VIP 512  models and VAE fc VIP 512.

\clearpage
\section{Lactate}\label{sec:resultsLactate}

In the following sections, the lists of test batches are abbreviated as shown in Tab. \ref{tab:bids_rotations}:
\vspace{-2mm}
\begin{table}[h!]
\renewcommand{\arraystretch}{1.2}
    \centering
    \caption{Abreviations for TEST BATCHES for Lactate, Glucose and VCD }
    \begin{tabular}{lp{.5\textwidth}}
    \toprule
     ROTATION &   TEST BATCHES \\
    \midrule
    test split 1     &1,6,11,16,21,26,31,36,41,46,51,56,
                    61,66,71,76,81,86,91,96,101,106,111,116
                    121,126,131,136,141,146,151,156,161,166,171,176
                    181,186,191,196,201\\
    test split 2     &2,7,12,17,22,27,32,37,42,47,52,57                
                    62,67,72,77,82,87,92,97,102,107,112,117
                    122,127,132,137,142,147,152,157,162,167,172,177
                    182,187,192,197,202\\
    test split 3     &3,8,13,18,23,28,33,38,43,48,53,58                
                    63,68,73,78,83,88,93,98,103,108,113,118
                    123,128,133,138,143,148,153,158,163,168,173,178
                    183,188,193,198,203\\
    test split 4     &4,9,14,19,24,29,34,39,44,49,54,59                
                    64,69,74,79,84,89,94,99,104,109,114,119
                    124,129,134,139,144,149,154,159,164,169,174,179
                    184,189,194,199,204\\
    test split 5     &5,10,15,20,25,30,35,40,45,50,55,60               
                    65,70,75,80,85,90,95,100,105,110,115,120
                    125,130,135,140,145,150,155,160,165,170,175,180
                    185,190,195,200,205\\

    \bottomrule
    \end{tabular}
    \label{tab:bids_rotations}
    \vspace{-5mm}
\end{table}

\clearpage
\vspace{-2mm}
\subsection{ML Models}

Overall amongst the ML models, GP has shown the best performance on both DS1 and DS2. Augmentation (Tab. \ref{tab:DS13_ML_Lactate_AUG}) and dimensionality reduction (Tab. \ref{tab:DS13_ML_Lactate_VIP}) did not improve the avergage N RMSE (except for some cases with SVR).
\vspace{-2mm}


\begin{table}[h]%
\caption{ML models N RMSE comparison for (STD, STD+AUG, STD+RES) on DS1 Lactate}%
\renewcommand*{\MinNumber}{0.16}%
\renewcommand*{\MidNumber}{0.174}%
\renewcommand*{\MaxNumber}{0.205}%
\renewcommand{\arraystretch}{1.0}
\centering%
\begin{adjustbox}{width=0.98\textwidth}%
\begin{tabular}{@{}lR@{\hskip-1.5\tabcolsep}R@{\hskip-1.5\tabcolsep}RR@{\hskip-1.5\tabcolsep}R@{\hskip-1.5\tabcolsep}RR@{\hskip-1.5\tabcolsep}R@{\hskip-1.5\tabcolsep}RR@{\hskip-1.5\tabcolsep}R@{\hskip-1.5\tabcolsep}RR@{\hskip-1.5\tabcolsep}R@{\hskip-1.5\tabcolsep}R@{}}%
\toprule%
\multirow{2}{*}{TEST BATCHES}&\multicolumn{3}{c}{XGB}&\multicolumn{3}{c}{GP}&\multicolumn{3}{c}{SVR}&\multicolumn{3}{c}{PLS}&\multicolumn{3}{c}{Benchmark}\\%
&STD&AUG&RES&STD&AUG&RES&STD&AUG&RES&STD&AUG&RES&STD&AUG&RES\\%
\midrule%
\EndTableHeader%
test split 1&0.145&0.175&0.162&0.145&0.161&0.156&0.162&0.157&0.168&0.163&0.172&0.163&0.163&0.172&0.163\\%
test split 2&0.192&0.198&0.206&0.161&0.173&0.175&0.216&0.223&0.226&0.171&0.196&0.174&0.171&0.196&0.174\\%
test split 3&0.173&0.196&0.206&0.170&0.174&0.170&0.271&0.252&0.222&0.173&0.207&0.181&0.173&0.207&0.181\\%
test split 4&0.178&0.196&0.226&0.156&0.163&0.187&0.190&0.151&0.166&0.160&0.198&0.161&0.160&0.198&0.161\\%
test split 5&0.289&0.256&0.272&0.175&0.196&0.173&0.297&0.256&0.252&0.205&0.209&0.209&0.205&0.209&0.209\\%
AVG BATCHES&0.195&0.204&0.214&0.161&0.173&0.172&0.227&0.208&0.207&0.174&0.196&0.178&0.174&0.196&0.178\\%
AVG OVERALL&0.199&0.205&0.216&0.162&0.173&0.172&0.234&0.214&0.210&0.174&0.197&0.178&0.174&0.197&0.178\\\bottomrule%
\end{tabular}%
\end{adjustbox}%
\vspace{-5mm}
\label{tab:DS13_ML_Lactate_AUG}
\end{table}
\begin{table}[h]%
\caption{ML models N RMSE comparison for (STD, STD+512, STD+256, STD+128) on DS1 Lactate}%
\renewcommand{\arraystretch}{1.2}
\renewcommand*{\MinNumber}{0.16}%
\renewcommand*{\MidNumber}{0.174}%
\renewcommand*{\MaxNumber}{0.205}%
\centering%
\begin{adjustbox}{width=0.98\textwidth}%
\begin{tabular}{@{}lR@{\hskip-2.0\tabcolsep}R@{\hskip-2.0\tabcolsep}R@{\hskip-2.0\tabcolsep}RR@{\hskip-2.0\tabcolsep}R@{\hskip-2.0\tabcolsep}R@{\hskip-2.0\tabcolsep}RR@{\hskip-2.0\tabcolsep}R@{\hskip-2.0\tabcolsep}R@{\hskip-2.0\tabcolsep}RR@{\hskip-2.0\tabcolsep}R@{\hskip-2.0\tabcolsep}R@{\hskip-2.0\tabcolsep}RR@{\hskip-2.0\tabcolsep}R@{\hskip-2.0\tabcolsep}R@{\hskip-2.0\tabcolsep}R@{}}%
\toprule%
\multirow{2}{*}{TEST BATCHES}&\multicolumn{4}{c}{XGB}&\multicolumn{4}{c}{GP}&\multicolumn{4}{c}{SVR}&\multicolumn{4}{c}{PLS}&\multicolumn{4}{c}{Benchmark}\\%
&STD&512&256&128&STD&512&256&128&STD&512&256&128&STD&512&256&128&STD&512&256&128\\%
\midrule%
\EndTableHeader%
test split 1&0.145&0.155&0.165&0.178&0.145&0.141&0.148&0.178&0.162&0.150&0.154&0.160&0.163&0.169&0.182&0.207&0.163&0.169&0.182&0.207\\%
test split 2&0.192&0.199&0.204&0.203&0.161&0.157&0.166&0.185&0.216&0.183&0.202&0.208&0.171&0.239&0.219&0.234&0.171&0.239&0.219&0.234\\%
test split 3&0.173&0.180&0.172&0.250&0.170&0.178&0.207&0.222&0.271&0.203&0.221&0.243&0.173&0.215&0.244&0.238&0.173&0.215&0.244&0.238\\%
test split 4&0.178&0.213&0.221&0.179&0.156&0.219&0.162&0.204&0.190&0.177&0.165&0.151&0.160&0.224&0.217&0.235&0.160&0.224&0.217&0.235\\%
test split 5&0.289&0.264&0.286&0.292&0.175&0.185&0.195&0.215&0.297&0.220&0.213&0.210&0.205&0.257&0.257&0.270&0.205&0.257&0.257&0.270\\%
AVG BATCHES&0.195&0.202&0.209&0.220&0.161&0.176&0.176&0.201&0.227&0.187&0.191&0.194&0.174&0.221&0.224&0.237&0.174&0.221&0.224&0.237\\%
AVG OVERALL&0.199&0.204&0.211&0.226&0.162&0.177&0.178&0.202&0.234&0.188&0.194&0.200&0.174&0.222&0.226&0.237&0.174&0.222&0.226&0.237\\\bottomrule%
\end{tabular}%
\end{adjustbox}%
\vspace{-5mm}
\label{tab:DS13_ML_Lactate_VIP}
\end{table}
\begin{table}[h]%
\caption{ML models N RMSE comparison for (STD) on DS2 Lactate}%
\renewcommand{\arraystretch}{1.0}
\renewcommand*{\MinNumber}{0.231}%
\renewcommand*{\MidNumber}{0.37799999999999995}%
\renewcommand*{\MaxNumber}{0.71}%
\centering%
\begin{adjustbox}{width=0.48\textwidth}%
\begin{tabular}{@{}lRRRRR@{}}%
\toprule%
\multirow{2}{*}{TEST BATCHES}&\multicolumn{1}{c}{XGB}&\multicolumn{1}{c}{GP}&\multicolumn{1}{c}{SVR}&\multicolumn{1}{c}{PLS}&\multicolumn{1}{c}{Benchmark}\\%
&STD&STD&STD&STD&STD\\%
\midrule%
\EndTableHeader%
97&0.548&0.263&0.552&0.327&0.327\\%
98&0.765&0.469&0.302&0.710&0.710\\%
99&1.038&0.609&1.162&0.340&0.340\\%
101&0.606&0.406&0.507&0.466&0.466\\%
152&0.746&0.256&0.685&0.231&0.231\\%
154&0.301&0.156&0.505&0.231&0.231\\%
AVG BATCHES&0.667&0.360&0.619&0.384&0.384\\%
AVG OVERALL&0.614&0.324&0.561&0.378&0.378\\\bottomrule%
\end{tabular}%
\end{adjustbox}%
\vspace{-5mm}
\label{tab:TOCI_ML_Lactate}
\end{table}

\clearpage
\subsection{DL Models - KT}
On DS1, KT 512 models (obtained from the same KT optimisation as for Titer) showed below average results, compared to the benchmark in the STD setting (Tab. \ref{tab:DS13_KT_Lactate}). Training on the RES set allowed to show similar performance as the benchmark in the STD setting (see KT VIP 512 Top 3 and  KT VIP 512 Top 5).\\
On DS2 (Tab. \ref{tab:TOCI_KT_Lactate}), KT 512 models did not beat the benchmark model.
\begin{table}[h]%
\caption{KT models N RMSE comparison for (512, 512+RES) on DS1 Lactate}%
\renewcommand*{\MinNumber}{0.168}%
\renewcommand*{\MidNumber}{0.225}%
\renewcommand*{\MaxNumber}{0.258}%
\centering%
\begin{adjustbox}{width=0.98\textwidth}%
\begin{tabular}{@{}lR@{\hskip-1.0\tabcolsep}RR@{\hskip-1.0\tabcolsep}RR@{\hskip-1.0\tabcolsep}RR@{\hskip-1.0\tabcolsep}RR@{\hskip-1.0\tabcolsep}RR@{\hskip-1.0\tabcolsep}R@{}}%
\toprule%
\multirow{2}{*}{TEST BATCHES}&\multicolumn{2}{c}{top 1}&\multicolumn{2}{c}{top 2}&\multicolumn{2}{c}{top 3}&\multicolumn{2}{c}{top 4}&\multicolumn{2}{c}{top 5}&\multicolumn{2}{c}{Benchmark}\\%
&512&RES&512&RES&512&RES&512&RES&512&RES&512&RES\\%
\midrule%
\EndTableHeader%
test split 1&0.378&0.372&0.167&0.163&0.222&0.150&0.372&0.186&0.180&0.146&0.168&0.164\\%
test split 2&0.393&0.345&0.283&0.177&0.249&0.184&0.672&0.196&0.222&0.187&0.239&0.174\\%
test split 3&0.461&0.408&0.243&0.206&0.222&0.211&0.367&0.218&0.233&0.191&0.219&0.186\\%
test split 4&0.343&0.408&0.183&0.201&0.194&0.142&0.193&0.154&0.189&0.179&0.237&0.170\\%
test split 5&0.411&0.363&0.190&0.170&0.245&0.169&0.542&0.193&0.245&0.161&0.258&0.209\\%
AVG BATCHES&0.397&0.379&0.213&0.183&0.227&0.171&0.429&0.189&0.214&0.173&0.224&0.181\\%
AVG OVERALL&0.402&0.380&0.218&0.184&0.227&0.174&0.458&0.192&0.216&0.174&0.225&0.182\\\bottomrule%
\end{tabular}%
\end{adjustbox}%
\vspace{-5mm}
\label{tab:DS13_KT_Lactate}
\end{table}
\begin{table}[h]%
\caption{KT models N RMSE comparison for (512, 512+RES) on DS2 Lactate}%
\renewcommand*{\MinNumber}{0.231}%
\renewcommand*{\MidNumber}{0.37799999999999995}%
\renewcommand*{\MaxNumber}{0.71}%
\centering%
\begin{adjustbox}{width=0.98\textwidth}%
\begin{tabular}{@{}lR@{\hskip-1.0\tabcolsep}RR@{\hskip-1.0\tabcolsep}RR@{\hskip-1.0\tabcolsep}RR@{\hskip-1.0\tabcolsep}RR@{\hskip-1.0\tabcolsep}RR@{\hskip-1.0\tabcolsep}R@{}}%
\toprule%
\multirow{2}{*}{TEST BATCHES}&\multicolumn{2}{c}{top 1}&\multicolumn{2}{c}{top 2}&\multicolumn{2}{c}{top 3}&\multicolumn{2}{c}{top 4}&\multicolumn{2}{c}{top 5}&\multicolumn{2}{c}{Benchmark}\\%
&512&RES&512&RES&512&RES&512&RES&512&RES&STD&RES\\%
\midrule%
\EndTableHeader%
97&0.590&0.623&0.562&0.271&0.239&0.258&0.239&0.305&0.208&0.513&0.327&0.352\\%
98&0.433&0.529&0.278&0.277&0.204&0.344&0.357&0.349&0.106&0.231&0.710&0.609\\%
99&0.930&1.326&0.219&0.850&0.121&0.776&0.282&0.926&0.218&1.461&0.340&0.366\\%
101&0.768&0.598&0.148&0.855&0.083&0.720&0.181&0.345&0.353&1.340&0.466&0.432\\%
152&0.356&0.328&0.636&0.540&0.558&0.394&0.633&0.434&0.518&0.510&0.231&0.208\\%
154&0.548&0.519&0.534&0.442&0.636&0.229&0.488&0.245&0.624&0.249&0.231&0.132\\%
AVG BATCHES&0.604&0.654&0.396&0.539&0.307&0.454&0.363&0.434&0.338&0.717&0.384&0.350\\%
AVG OVERALL&0.523&0.572&0.435&0.483&0.364&0.403&0.386&0.391&0.359&0.668&0.378&0.342\\\bottomrule%
\end{tabular}%
\end{adjustbox}%
\vspace{-5mm}
\label{tab:TOCI_KT_Lactate}
\end{table}
\subsection{DL Models - CNN}

CNN models have shown poor results on both datasets for Lactate (Tab. \ref{tab:DS13_CNN_Lactate} and \ref{tab:TOCI_CNN_Lactate}).
\begin{table}[h]%
\caption{CNN models N RMSE comparison for (STD, STD+RES) on DS1 Lactate}%
\renewcommand*{\MinNumber}{0.16399999999999998}%
\renewcommand*{\MidNumber}{0.17800000000000002}%
\renewcommand*{\MaxNumber}{0.205}%
\centering%
\begin{adjustbox}{width=0.98\textwidth}%
\begin{tabular}{@{}lR@{\hskip-1.0\tabcolsep}RR@{\hskip-1.0\tabcolsep}RR@{\hskip-1.0\tabcolsep}RR@{\hskip-1.0\tabcolsep}RR@{\hskip-1.0\tabcolsep}RR@{\hskip-1.0\tabcolsep}RR@{\hskip-1.0\tabcolsep}RR@{\hskip-1.0\tabcolsep}RR@{\hskip-1.0\tabcolsep}RR@{\hskip-1.0\tabcolsep}RR@{\hskip-1.0\tabcolsep}R@{}}%
\toprule%
\multirow{2}{*}{TEST BATCHES}&\multicolumn{2}{c}{ vanilla}&\multicolumn{2}{c}{ selu}&\multicolumn{2}{c}{ selu sh}&\multicolumn{2}{c}{ lrelu}&\multicolumn{2}{c}{ lrelu sh}&\multicolumn{2}{c}{ vanilla }&\multicolumn{2}{c}{ selu }&\multicolumn{2}{c}{ selu sh }&\multicolumn{2}{c}{ lrelu }&\multicolumn{2}{c}{ lrelu sh }&\multicolumn{2}{c}{Benchmark}\\%
&STD&RES&STD&RES&STD&RES&STD&RES&STD&RES&STD&RES&NOPP&RES&NOPP&RES&NOPP&RES&NOPP&RES&STD&RES\\%
\midrule%
\EndTableHeader%
test split 1&0.200&0.185&0.171&0.150&0.198&0.151&0.199&0.151&0.158&0.150&0.645&0.253&0.242&0.224&0.289&0.179&0.776&0.707&0.803&0.367&0.164&0.164\\%
test split 2&0.268&0.296&0.206&0.179&0.209&0.199&0.228&0.236&0.225&0.241&0.953&0.274&0.446&0.203&1.159&0.275&0.923&0.313&0.902&0.308&0.170&0.173\\%
test split 3&0.231&0.211&0.230&0.202&0.235&0.183&0.257&0.192&0.226&0.192&1.018&0.240&0.335&0.319&0.377&0.269&1.026&0.849&0.634&0.451&0.177&0.187\\%
test split 4&0.179&0.221&0.186&0.186&0.201&0.215&0.163&0.162&0.143&0.206&0.917&0.535&0.259&0.213&0.314&0.240&0.758&0.242&0.800&0.514&0.169&0.170\\%
test split 5&0.258&0.190&0.233&2.375&0.265&0.179&0.230&0.228&0.200&0.195&0.565&0.224&0.294&0.255&0.257&0.230&0.762&0.230&0.450&0.507&0.205&0.210\\%
AVG BATCHES&0.227&0.220&0.205&0.619&0.221&0.185&0.215&0.194&0.190&0.197&0.820&0.305&0.315&0.243&0.479&0.239&0.849&0.468&0.718&0.429&0.177&0.181\\%
AVG OVERALL&0.230&0.222&0.207&1.076&0.224&0.185&0.220&0.197&0.195&0.197&0.839&0.316&0.323&0.250&0.578&0.241&0.862&0.555&0.728&0.435&0.178&0.182\\\bottomrule%
\end{tabular}%
\end{adjustbox}%
\vspace{-5mm}
\label{tab:DS13_CNN_Lactate}
\end{table}
\begin{table}[h!]%
\caption{CNN models N RMSE comparison for (STD, STD+RES) on DS2 Lactate}%
\renewcommand*{\MinNumber}{0.231}%
\renewcommand*{\MidNumber}{0.37799999999999995}%
\renewcommand*{\MaxNumber}{0.71}%
\centering%
\begin{adjustbox}{width=0.98\textwidth}%
\begin{tabular}{@{}lR@{\hskip-1.0\tabcolsep}RR@{\hskip-1.0\tabcolsep}RR@{\hskip-1.0\tabcolsep}RR@{\hskip-1.0\tabcolsep}RR@{\hskip-1.0\tabcolsep}RR@{\hskip-1.0\tabcolsep}RR@{\hskip-1.0\tabcolsep}RR@{\hskip-1.0\tabcolsep}RR@{\hskip-1.0\tabcolsep}RR@{\hskip-1.0\tabcolsep}RR@{\hskip-1.0\tabcolsep}R@{}}%
\toprule%
\multirow{2}{*}{TEST BATCHES}&\multicolumn{2}{c}{ vanilla}&\multicolumn{2}{c}{ selu}&\multicolumn{2}{c}{ selu sh}&\multicolumn{2}{c}{ lrelu}&\multicolumn{2}{c}{ lrelu sh}&\multicolumn{2}{c}{ vanilla }&\multicolumn{2}{c}{ selu }&\multicolumn{2}{c}{ selu sh }&\multicolumn{2}{c}{ lrelu }&\multicolumn{2}{c}{ lrelu sh }&\multicolumn{2}{c}{Benchmark}\\%
&STD&RES&STD&RES&STD&RES&STD&RES&STD&RES&NOPP&RES&NOPP&RES&NOPP&RES&NOPP&RES&NOPP&RES&STD&RES\\%
\midrule%
\EndTableHeader%
97&1.279&0.762&0.464&0.351&0.934&0.428&0.580&0.411&0.266&0.401&1.173&1.020&1.439&0.876&0.828&0.836&1.361&0.853&1.308&0.723&0.327&0.352\\%
98&0.714&0.416&0.269&0.500&0.487&0.343&0.546&0.382&0.658&0.392&1.468&0.956&1.899&2.126&1.331&0.947&0.935&1.649&0.984&1.611&0.710&0.609\\%
99&1.614&0.870&0.946&0.884&0.951&0.596&0.844&0.730&0.771&0.910&1.757&2.262&1.952&1.188&2.946&0.926&1.377&0.725&1.881&0.881&0.340&0.366\\%
101&1.419&0.506&0.710&0.801&0.658&0.643&0.640&0.658&0.686&0.734&1.062&1.178&2.043&3.177&1.741&3.189&1.225&5.208&0.928&5.891&0.466&0.432\\%
152&0.805&0.430&0.887&0.758&0.711&0.632&0.733&0.386&0.804&0.453&0.933&1.201&1.116&1.097&1.144&1.185&1.043&1.066&0.862&1.182&0.231&0.208\\%
154&0.873&0.403&0.494&0.243&0.609&0.374&0.608&0.297&0.498&0.368&0.954&0.701&1.044&0.581&1.094&0.643&1.215&1.041&0.770&0.780&0.231&0.132\\%
AVG BATCHES&1.117&0.565&0.628&0.589&0.725&0.503&0.659&0.477&0.614&0.543&1.225&1.220&1.582&1.508&1.514&1.288&1.193&1.757&1.122&1.845&0.384&0.350\\%
AVG OVERALL&0.988&0.514&0.592&0.549&0.666&0.463&0.592&0.417&0.573&0.472&1.091&1.083&1.410&1.491&1.314&1.295&1.071&1.919&1.003&2.098&0.378&0.342\\\bottomrule%
\end{tabular}%
\end{adjustbox}%
\label{tab:TOCI_CNN_Lactate}
\end{table}

\clearpage
\subsection{DL Models - VAE}

VAE models performed exceptionally well on the preprocessed spectrum for Lactate on DS1 (Tab. \ref{tab:DS13_VAE_Lac}). On the raw spectrum, the error was lower compared to Titer, but below average compared to the benchmark. On DS2 the VAE models did not beat the benchmark (Tab. \ref{tab:TOCI_VAE_Lac}).
\begin{table}[h]%
\caption{VAE models N RMSE comparison for (STD, STD+RES) on DS1 Lactate}%
\renewcommand*{\MinNumber}{0.16399999999999998}%
\renewcommand*{\MidNumber}{0.17800000000000002}%
\renewcommand*{\MaxNumber}{0.205}%
\centering%
\begin{adjustbox}{width=0.98\textwidth}%
\begin{tabular}{@{}lR@{\hskip-1.0\tabcolsep}RR@{\hskip-1.0\tabcolsep}RR@{\hskip-1.0\tabcolsep}RR@{\hskip-1.0\tabcolsep}RR@{\hskip-1.0\tabcolsep}RR@{\hskip-1.0\tabcolsep}RR@{\hskip-1.0\tabcolsep}RR@{\hskip-1.0\tabcolsep}RR@{\hskip-1.0\tabcolsep}R@{}}%
\toprule%
\multirow{2}{*}{TEST BATCHES}&\multicolumn{2}{c}{ vanilla}&\multicolumn{2}{c}{ convT}&\multicolumn{2}{c}{ fc dec}&\multicolumn{2}{c}{ fc}&\multicolumn{2}{c}{ vanilla }&\multicolumn{2}{c}{ convT }&\multicolumn{2}{c}{ fc dec }&\multicolumn{2}{c}{ fc }&\multicolumn{2}{c}{Benchmark}\\%
&STD&RES&STD&RES&STD&RES&STD&RES&NOPP&RES&NOPP&RES&NOPP&RES&NOPP&RES&STD&RES\\%
\midrule%
\EndTableHeader%
test split 1&0.148&0.149&0.145&0.155&0.147&0.157&0.144&0.154&0.156&0.227&0.178&0.176&0.178&0.187&0.808&0.769&0.164&0.164\\%
test split 2&0.183&0.166&0.168&0.173&0.167&0.169&0.186&0.180&0.238&0.194&0.195&0.187&0.194&0.177&0.733&0.674&0.170&0.173\\%
test split 3&0.195&0.202&0.186&0.186&0.200&0.199&0.218&0.198&0.237&0.271&0.243&0.277&0.226&0.261&0.783&0.716&0.177&0.187\\%
test split 4&0.151&0.155&0.148&0.173&0.139&0.161&0.147&0.139&0.155&0.193&0.150&0.183&0.157&0.177&0.857&0.395&0.169&0.170\\%
test split 5&0.164&0.151&0.181&0.168&0.166&0.165&0.180&0.150&0.195&0.194&0.203&0.193&0.190&0.194&0.616&1.459&0.205&0.210\\%
AVG BATCHES&0.168&0.165&0.166&0.171&0.164&0.170&0.175&0.164&0.196&0.216&0.194&0.203&0.189&0.199&0.759&0.803&0.177&0.181\\%
AVG OVERALL&0.170&0.167&0.167&0.171&0.167&0.172&0.179&0.167&0.201&0.220&0.199&0.210&0.192&0.204&0.761&0.883&0.178&0.182\\\bottomrule%
\end{tabular}%
\end{adjustbox}%
\vspace{-5mm}
\label{tab:DS13_VAE_Lac}
\end{table}
\begin{table}[h]%
\caption{VAE models N RMSE comparison for (STD, STD+RES) on DS2 Lactate}%
\renewcommand*{\MinNumber}{0.231}%
\renewcommand*{\MidNumber}{0.37799999999999995}%
\renewcommand*{\MaxNumber}{0.71}%
\centering%
\begin{adjustbox}{width=0.98\textwidth}%
\begin{tabular}{@{}lR@{\hskip-1.0\tabcolsep}RR@{\hskip-1.0\tabcolsep}RR@{\hskip-1.0\tabcolsep}RR@{\hskip-1.0\tabcolsep}RR@{\hskip-1.0\tabcolsep}RR@{\hskip-1.0\tabcolsep}RR@{\hskip-1.0\tabcolsep}RR@{\hskip-1.0\tabcolsep}RR@{\hskip-1.0\tabcolsep}R@{}}%
\toprule%
\multirow{2}{*}{TEST BATCHES}&\multicolumn{2}{c}{ vanilla}&\multicolumn{2}{c}{ convT}&\multicolumn{2}{c}{ fc dec}&\multicolumn{2}{c}{ fc}&\multicolumn{2}{c}{ vanilla }&\multicolumn{2}{c}{ convT }&\multicolumn{2}{c}{ fc dec }&\multicolumn{2}{c}{ fc }&\multicolumn{2}{c}{Benchmark}\\%
&STD&RES&STD&RES&STD&RES&STD&RES&NOPP&RES&NOPP&RES&NOPP&RES&NOPP&RES&STD&RES\\%
\midrule%
\EndTableHeader%
97&0.765&0.912&0.801&0.772&0.753&0.772&0.339&0.294&0.990&1.418&0.729&0.840&0.587&0.825&0.917&0.943&0.327&0.352\\%
98&0.233&0.236&0.236&0.190&0.211&0.271&0.347&0.353&1.151&1.849&0.871&1.000&0.889&0.684&1.040&1.480&0.710&0.609\\%
99&0.577&0.532&0.543&0.493&0.536&0.450&1.298&0.756&1.199&1.541&0.896&0.674&1.004&0.777&1.141&1.813&0.340&0.366\\%
101&0.430&0.461&0.397&0.468&0.393&0.476&0.613&0.641&1.313&0.991&1.798&2.497&2.112&2.488&1.292&1.795&0.466&0.432\\%
152&0.334&0.268&0.321&0.299&0.353&0.312&0.365&0.295&1.169&1.089&1.158&1.149&1.177&1.120&1.143&1.181&0.231&0.208\\%
154&0.360&0.336&0.243&0.279&0.235&0.293&0.396&0.257&1.016&0.813&0.668&0.625&0.598&0.790&0.960&0.953&0.231&0.132\\%
AVG BATCHES&0.450&0.458&0.423&0.417&0.413&0.429&0.560&0.433&1.140&1.283&1.020&1.131&1.061&1.114&1.082&1.361&0.384&0.350\\%
AVG OVERALL&0.436&0.472&0.429&0.419&0.415&0.426&0.493&0.377&1.026&1.207&0.957&1.118&1.012&1.090&0.975&1.201&0.378&0.342\\\bottomrule%
\end{tabular}%
\end{adjustbox}%
\vspace{-5mm}
\label{tab:TOCI_VAE_Lac}
\end{table}

\subsection{Fused models}

\begin{figure}[h]
\centering
\begin{minipage}{.98\textwidth}
    \centering
    \includegraphics[width=.98\textwidth]{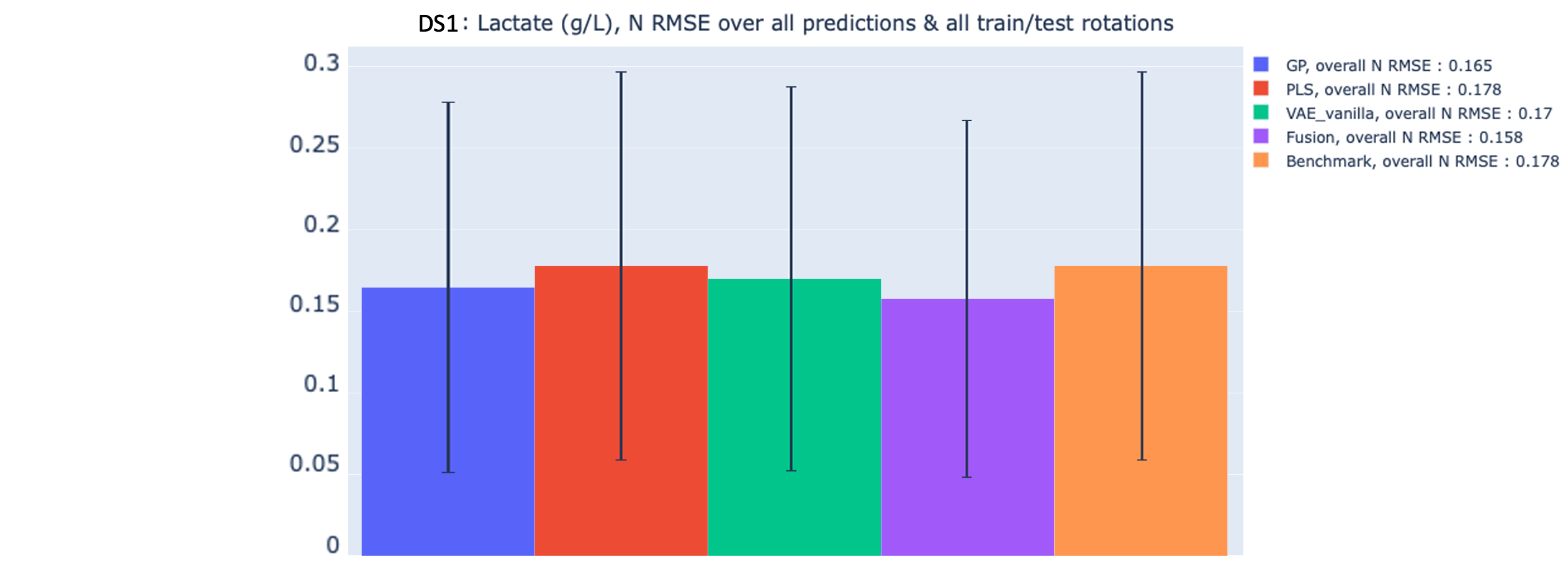}
    \captionof{figure}{Average N RMSE of GP, PLS, VAE vanilla and FWLS for DS1 Titer. Error bars correspond to 1 standard deviation of the errors (normalized), from left to right: 0.1135, 0.1190, 0.1177, 0.1094, 0.1190. The FWLS model showed the lowest N RMSE and standard deviation. XGB showed similar, but slightly higher values (not shown here).}
    \label{fig:fusionLactate_rmse}
\end{minipage}%
\end{figure}

\begin{figure}[h]
\centering
\begin{minipage}{.98\textwidth}
    \centering
    \includegraphics[width=.98\textwidth]{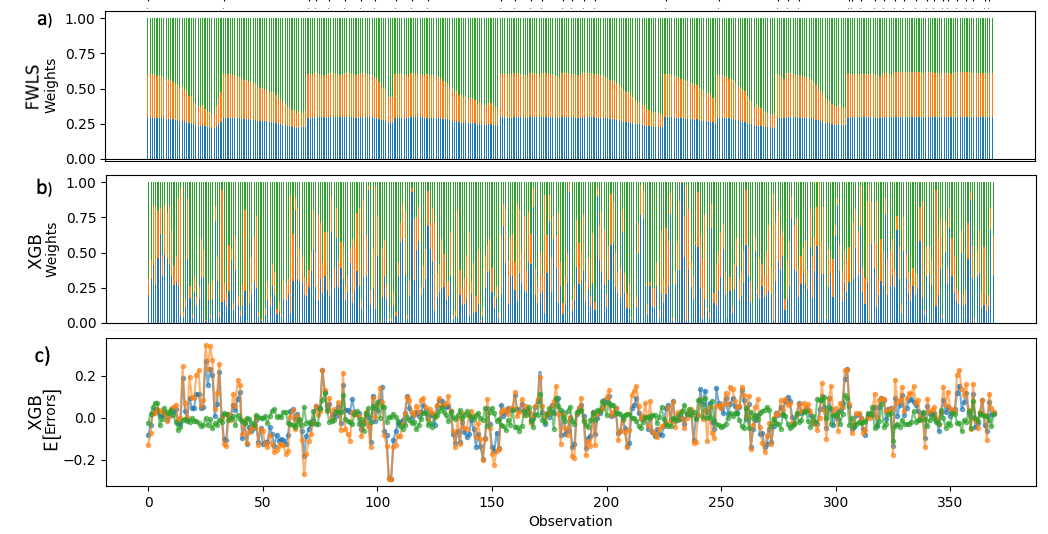}
    \captionof{figure}{Fused model performance on DS 1 test split 4 for Titer: Fusion weights obtained from a) FWLS and b) XGB. v) Estimated prediction uncertainty for the base models by XGB.}
    \label{fig:fusionLactate}
\end{minipage}%
\end{figure}

For Lactate, it was decided to test the fusion of GP, PLS and VAE vanilla models (Tab. \ref{tab:fusionDS13Lactate}). For both fusion models the resulting N RMSE was l0\% lower compared to the benchmark model. Furthermore, the per-batch error of the fused models was lower than the per-batch error averaged over the base models. Furthermore, the standard deviation of the errors was reduced by up to 10\% compared to the worst performing base model (PLS), as shown in Tab. \ref{fig:fusionLactate_rmse}. Both models produced the same average N RMSE on test split 4, therefore it seemed appropriate to compare the resulting fusion weights produced by each model. Fig. \ref{fig:fusionLactate} Fig. \ref{fig:fusionLactate} a) and b) show the weights for FWLS and XGB describing the contribution of each base model to the fused prediction. It can be clearly noticed that FWLS learned to reduce the influence of PLS for low concentration values, whereas for higher and medium concentrations, the models are almosz equally weighted. The relationship learned by XGB seems to have a more complex pattern. The trajectories of the base models' uncertainties learned by XGB, shown in Fig. \ref{fig:fusionLactate} c) suggest that the meta-estimator seems to favor predictions of the VAE and GP models, which indeed perform better than PLS in this case.\\

Overall, the fused models were the best performing models on DS1 Lactate for the STD setting, followed closely by GP and VAE models.

\begin{table}[h!]%
\caption{N RMSE comparison for FWLS and BLUE fusion of GP, PLS and VAE vanilla on DS1 Lactate}%
\renewcommand*{\MinNumber}{0.164}%
\renewcommand*{\MidNumber}{0.178}%
\renewcommand*{\MaxNumber}{0.205}%
\centering%
\begin{adjustbox}{width=0.68\textwidth}%
\begin{tabular}{lRRRRRR}
\toprule
TEST BATCHES &   GP &   PLS &  VAE vanilla & FWLS & BLUE & Benchmark\\

\midrule
\EndTableHeader
test split 1     & 0.145  & 0.164 &  0.148 &    0.140 &  0.139 &  0.164 \\
test split 2     & 0.161  & 0.170 &  0.183 &    0.162 &  0.167 &  0.170 \\
test split 3     & 0.175  & 0.177 &  0.195 &    0.179 &  0.180 &  0.177 \\
test split 4     & 0.164  & 0.169 &  0.151 &    0.144 &  0.144 &  0.169 \\
test split 5     & 0.175  & 0.205 &  0.164 &    0.157 &  0.156 &  0.205 \\
 AVG BATCHES     & 0.164  & 0.177 &  0.168 &    0.156 &  0.157 &  0.177 \\
 AVG OVERALL     & 0.165  & 0.178 &  0.170 &    0.158 &  0.159 &  0.178 \\
\bottomrule
\end{tabular}
\end{adjustbox}%
\label{tab:fusionDS13Lactate}
\end{table}

\clearpage
\section{Glucose}

\subsection{ML models}

Augmentation and resampling of DS1 Glucose yielded mixed results for the ML models (Tab. \ref{tab:DS13_ML_Glucose_AUG}). XGB,GP and SVR benefited most from resampling, while PLS performed better for AUG. Overall, GP showed the best results for RES.\\
Unlike for previous target variables, VIP gave better results for XGB and GP with VIP 256 and VIP 128 for SVR (Tab. \ref{tab:DS13_ML_Glucose_VIP}). PLS did not improve after dimensionality reduction. The ranking of the ML models was the same on DS2 (Tab. \ref{tab:TOCI_ML_Glucose}), but overall performance was lower.


\begin{table}[h]%
\caption{ML models N RMSE comparison for (STD, STD+AUG, STD+RES) on DS1 Glucose}%
\renewcommand*{\MinNumber}{0.225}%
\renewcommand*{\MidNumber}{0.263}%
\renewcommand*{\MaxNumber}{0.29100000000000004}%
\centering%
\begin{adjustbox}{width=0.98\textwidth}%
\begin{tabular}{@{}lR@{\hskip-1.5\tabcolsep}R@{\hskip-1.5\tabcolsep}RR@{\hskip-1.5\tabcolsep}R@{\hskip-1.5\tabcolsep}RR@{\hskip-1.5\tabcolsep}R@{\hskip-1.5\tabcolsep}RR@{\hskip-1.5\tabcolsep}R@{\hskip-1.5\tabcolsep}RR@{\hskip-1.5\tabcolsep}R@{\hskip-1.5\tabcolsep}R@{}}%
\toprule%
\multirow{2}{*}{TEST BATCHES}&\multicolumn{3}{c}{XGB}&\multicolumn{3}{c}{GP}&\multicolumn{3}{c}{SVR}&\multicolumn{3}{c}{PLS}&\multicolumn{3}{c}{Benchmark}\\%
&STD&AUG&RES&STD&AUG&RES&STD&AUG&RES&STD&AUG&RES&STD&AUG&RES\\%
\midrule%
\EndTableHeader%
test split 1&0.254&0.266&0.246&0.235&0.228&0.235&0.241&0.220&0.218&0.271&0.244&0.272&0.271&0.244&0.272\\%
test split 2&0.219&0.261&0.232&0.222&0.207&0.202&0.252&0.257&0.241&0.265&0.232&0.266&0.265&0.232&0.266\\%
test split 3&0.222&0.239&0.210&0.222&0.294&0.218&0.249&0.237&0.234&0.225&0.214&0.225&0.225&0.214&0.225\\%
test split 4&0.322&0.327&0.320&0.263&0.253&0.250&0.268&0.253&0.258&0.291&0.264&0.302&0.291&0.264&0.302\\%
test split 5&0.273&0.295&0.273&0.229&0.225&0.237&0.265&0.264&0.240&0.265&0.254&0.266&0.265&0.254&0.266\\%
AVG BATCHES&0.258&0.277&0.256&0.234&0.241&0.228&0.255&0.246&0.238&0.263&0.241&0.266&0.263&0.241&0.266\\%
AVG OVERALL&0.256&0.276&0.255&0.233&0.241&0.226&0.255&0.248&0.238&0.263&0.240&0.266&0.263&0.240&0.266\\\bottomrule%
\end{tabular}%
\end{adjustbox}%
\label{tab:DS13_ML_Glucose_AUG}
\end{table}

\begin{table}[h]%
\caption{ML models N RMSE comparison for (STD, STD+512, STD+256, STD+128) on DS1 Glucose}%
\renewcommand*{\MinNumber}{0.225}%
\renewcommand*{\MidNumber}{0.263}%
\renewcommand*{\MaxNumber}{0.29100000000000004}%
\centering%
\begin{adjustbox}{width=0.98\textwidth}%
\begin{tabular}{@{}lR@{\hskip-2.0\tabcolsep}R@{\hskip-2.0\tabcolsep}R@{\hskip-2.0\tabcolsep}RR@{\hskip-2.0\tabcolsep}R@{\hskip-2.0\tabcolsep}R@{\hskip-2.0\tabcolsep}RR@{\hskip-2.0\tabcolsep}R@{\hskip-2.0\tabcolsep}R@{\hskip-2.0\tabcolsep}RR@{\hskip-2.0\tabcolsep}R@{\hskip-2.0\tabcolsep}R@{\hskip-2.0\tabcolsep}RR@{\hskip-2.0\tabcolsep}R@{\hskip-2.0\tabcolsep}R@{\hskip-2.0\tabcolsep}R@{}}%
\toprule%
\multirow{2}{*}{TEST BATCHES}&\multicolumn{4}{c}{XGB}&\multicolumn{4}{c}{GP}&\multicolumn{4}{c}{SVR}&\multicolumn{4}{c}{PLS}&\multicolumn{4}{c}{Benchmark}\\%
&STD&512&256&128&STD&512&256&128&STD&512&256&128&STD&512&256&128&STD&512&256&128\\%
\midrule%
\EndTableHeader%
test split 1&0.254&0.235&0.223&0.248&0.235&0.235&0.236&0.245&0.241&0.232&0.224&0.213&0.271&0.265&0.258&0.286&0.271&0.265&0.258&0.286\\%
test split 2&0.219&0.204&0.202&0.218&0.222&0.216&0.197&0.193&0.252&0.245&0.237&0.201&0.265&0.248&0.241&0.253&0.265&0.248&0.241&0.253\\%
test split 3&0.222&0.212&0.206&0.215&0.222&0.226&0.198&0.199&0.249&0.215&0.218&0.219&0.225&0.246&0.262&0.270&0.225&0.246&0.262&0.270\\%
test split 4&0.322&0.311&0.300&0.277&0.263&0.264&0.232&0.235&0.268&0.292&0.264&0.258&0.291&0.322&0.318&0.281&0.291&0.322&0.318&0.281\\%
test split 5&0.273&0.266&0.262&0.251&0.229&0.235&0.221&0.240&0.265&0.284&0.283&0.251&0.265&0.291&0.270&0.304&0.265&0.291&0.270&0.304\\%
AVG BATCHES&0.258&0.246&0.238&0.242&0.234&0.235&0.217&0.223&0.255&0.254&0.245&0.228&0.263&0.274&0.270&0.279&0.263&0.274&0.270&0.279\\%
AVG OVERALL&0.256&0.244&0.237&0.240&0.233&0.233&0.215&0.221&0.255&0.254&0.246&0.227&0.263&0.273&0.268&0.277&0.263&0.273&0.268&0.277\\\bottomrule%
\end{tabular}%
\end{adjustbox}%
\label{tab:DS13_ML_Glucose_VIP}
\end{table}

\begin{table}[h]%
\caption{ML models N RMSE comparison for (STD) on DS2 Glucose}%
\renewcommand*{\MinNumber}{0.28600000000000003}%
\renewcommand*{\MidNumber}{0.415}%
\renewcommand*{\MaxNumber}{0.488}%
\centering%
\begin{adjustbox}{width=0.48\textwidth}%
\begin{tabular}{@{}lRRRRR@{}}%
\toprule%
\multirow{2}{*}{TEST BATCHES}&\multicolumn{1}{c}{XGB}&\multicolumn{1}{c}{GP}&\multicolumn{1}{c}{SVR}&\multicolumn{1}{c}{PLS}&\multicolumn{1}{c}{Benchmark}\\%
&STD&STD&STD&STD&STD\\%
\midrule%
\EndTableHeader%
97&0.479&0.434&0.636&0.488&0.488\\%
98&0.375&0.493&0.658&0.425&0.425\\%
99&0.353&0.287&0.454&0.286&0.286\\%
101&0.623&0.386&0.603&0.484&0.484\\%
152&0.424&0.382&0.473&0.391&0.391\\%
154&0.500&0.389&0.430&0.406&0.406\\%
AVG BATCHES&0.459&0.395&0.542&0.413&0.413\\%
AVG OVERALL&0.456&0.397&0.554&0.415&0.415\\\bottomrule%
\end{tabular}%
\end{adjustbox}%
\label{tab:TOCI_ML_Glucose}
\end{table}

\clearpage
\subsection{DL Models - KT}

KT VIP 512 models showed above average results on DS1 Glucose only after training on the RES set, as shown in Tab. \ref{tab:DS13_KT_Glucose} (except for KT VIP 512 Top 1, exhibiting same poor performance as shown in earlier sections)
Performance on DS2 was mixed, but overall much better compared to the benchmark on the standard setting (Tab. \ref{tab:TOCI_KT_Glucose}).
\begin{table}[h]%
\caption{KT models N RMSE comparison for (512, 512+RES) on DS1 Glucose}%
\renewcommand*{\MinNumber}{0.254}%
\renewcommand*{\MidNumber}{0.281}%
\renewcommand*{\MaxNumber}{0.322}%
\centering%
\begin{adjustbox}{width=0.98\textwidth}%
\begin{tabular}{@{}lR@{\hskip-1.0\tabcolsep}RR@{\hskip-1.0\tabcolsep}RR@{\hskip-1.0\tabcolsep}RR@{\hskip-1.0\tabcolsep}RR@{\hskip-1.0\tabcolsep}RR@{\hskip-1.0\tabcolsep}R@{}}%
\toprule%
\multirow{2}{*}{TEST BATCHES}&\multicolumn{2}{c}{top 1}&\multicolumn{2}{c}{top 2}&\multicolumn{2}{c}{top 3}&\multicolumn{2}{c}{top 4}&\multicolumn{2}{c}{top 5}&\multicolumn{2}{c}{Benchmark}\\%
&512&RES&512&RES&512&RES&512&RES&512&RES&512&RES\\%
\midrule%
\EndTableHeader%
test split 1&0.911&0.806&0.305&0.277&0.338&0.235&0.372&0.234&0.284&0.242&0.268&0.276\\%
test split 2&1.038&0.731&0.314&0.241&0.330&0.216&0.445&0.225&0.406&0.219&0.257&0.273\\%
test split 3&0.959&0.806&0.283&0.229&0.283&0.216&0.319&0.211&0.335&0.239&0.254&0.232\\%
test split 4&0.973&0.898&0.308&0.247&0.358&0.265&0.334&0.273&0.276&0.274&0.322&0.302\\%
test split 5&1.074&0.752&0.345&0.231&0.341&0.242&0.377&0.243&0.334&0.253&0.310&0.283\\%
AVG BATCHES&0.991&0.799&0.311&0.245&0.330&0.235&0.369&0.237&0.327&0.245&0.282&0.273\\%
AVG OVERALL&0.996&0.794&0.312&0.245&0.330&0.234&0.377&0.236&0.337&0.244&0.281&0.273\\\bottomrule%
\end{tabular}%
\end{adjustbox}%
\label{tab:DS13_KT_Glucose}
\vspace{-5mm}
\end{table}
\begin{table}[h]%
\caption{KT models N RMSE comparison for (512, 512+RES) on DS2 Glucose}%
\renewcommand*{\MinNumber}{0.28600000000000003}%
\renewcommand*{\MidNumber}{0.415}%
\renewcommand*{\MaxNumber}{0.488}%
\centering%
\begin{adjustbox}{width=0.98\textwidth}%
\begin{tabular}{@{}lR@{\hskip-1.0\tabcolsep}RR@{\hskip-1.0\tabcolsep}RR@{\hskip-1.0\tabcolsep}RR@{\hskip-1.0\tabcolsep}RR@{\hskip-1.0\tabcolsep}RR@{\hskip-1.0\tabcolsep}R@{}}%
\toprule%
\multirow{2}{*}{TEST BATCHES}&\multicolumn{2}{c}{top 1}&\multicolumn{2}{c}{top 2}&\multicolumn{2}{c}{top 3}&\multicolumn{2}{c}{top 4}&\multicolumn{2}{c}{top 5}&\multicolumn{2}{c}{Benchmark}\\%
&512&RES&512&RES&512&RES&512&RES&512&RES&STD&RES\\%
\midrule%
\EndTableHeader%
97&0.510&0.434&0.607&0.390&0.436&0.374&0.416&0.476&0.387&0.576&0.488&0.486\\%
98&0.688&0.850&0.215&0.426&0.116&0.372&0.148&0.346&0.102&0.485&0.425&0.585\\%
99&0.613&0.631&0.126&0.406&0.054&0.321&0.120&0.348&0.144&0.486&0.286&0.399\\%
101&0.713&0.579&0.266&0.635&0.174&0.383&0.104&0.446&0.103&0.630&0.484&0.434\\%
152&0.751&0.400&0.386&0.281&0.371&0.223&0.375&0.256&0.728&0.623&0.391&0.815\\%
154&0.824&0.630&0.392&0.308&0.375&0.374&0.293&0.378&0.625&0.419&0.406&0.396\\%
AVG BATCHES&0.683&0.587&0.332&0.408&0.254&0.341&0.243&0.375&0.348&0.537&0.413&0.519\\%
AVG OVERALL&0.658&0.604&0.373&0.421&0.286&0.348&0.269&0.387&0.377&0.530&0.415&0.506\\\bottomrule%
\end{tabular}%
\end{adjustbox}%
\label{tab:TOCI_KT_Glucose}
\vspace{-5mm}
\end{table}

\subsection{DL Models - CNN}

Tab. \ref{tab:DS13_CNN_Glucose} shows how CNN models performed better than the benchmark on DS1, especially for the RES setting. Tab. \ref{tab:TOCI_CNN_Glucose} shows the contrary for DS2.

\begin{table}[h!]%
\caption{CNN models N RMSE comparison for (STD, STD+RES) on DS1 Glucose}%
\renewcommand*{\MinNumber}{0.23199999999999998}%
\renewcommand*{\MidNumber}{0.271}%
\renewcommand*{\MaxNumber}{0.29100000000000004}%
\centering%
\begin{adjustbox}{width=0.98\textwidth}%
\begin{tabular}{@{}lR@{\hskip-1.0\tabcolsep}RR@{\hskip-1.0\tabcolsep}RR@{\hskip-1.0\tabcolsep}RR@{\hskip-1.0\tabcolsep}RR@{\hskip-1.0\tabcolsep}RR@{\hskip-1.0\tabcolsep}RR@{\hskip-1.0\tabcolsep}RR@{\hskip-1.0\tabcolsep}RR@{\hskip-1.0\tabcolsep}RR@{\hskip-1.0\tabcolsep}RR@{\hskip-1.0\tabcolsep}R@{}}%
\toprule%
\multirow{2}{*}{TEST BATCHES}&\multicolumn{2}{c}{ vanilla}&\multicolumn{2}{c}{ selu}&\multicolumn{2}{c}{ selu sh}&\multicolumn{2}{c}{ lrelu}&\multicolumn{2}{c}{ lrelu sh}&\multicolumn{2}{c}{ vanilla }&\multicolumn{2}{c}{ selu }&\multicolumn{2}{c}{ selu sh }&\multicolumn{2}{c}{ lrelu }&\multicolumn{2}{c}{ lrelu sh }&\multicolumn{2}{c}{Benchmark}\\%
&STD&RES&STD&RES&STD&RES&STD&RES&STD&RES&STD&RES&NOPP&RES&NOPP&RES&NOPP&RES&NOPP&RES&STD&RES\\%
\midrule%
\EndTableHeader%
test split 1&0.269&0.254&0.351&0.248&0.384&0.226&0.399&0.282&0.332&0.251&0.608&0.497&0.357&0.261&0.357&0.524&0.854&0.309&0.711&0.253&0.275&0.276\\%
test split 2&0.225&0.239&0.570&0.238&0.452&0.211&0.289&0.262&0.303&0.228&0.717&0.271&0.590&0.271&0.664&0.287&0.940&0.294&1.007&0.267&0.273&0.273\\%
test split 3&0.250&0.203&0.232&0.240&0.291&0.231&0.307&0.238&0.261&0.240&0.370&0.365&0.331&0.287&0.249&0.286&0.944&0.273&0.954&0.325&0.232&0.232\\%
test split 4&0.292&0.281&0.329&0.310&0.352&0.291&0.266&0.244&0.292&0.289&0.342&1.608&0.316&0.354&0.344&0.303&0.357&0.339&0.896&0.339&0.291&0.302\\%
test split 5&0.287&0.261&0.484&0.311&0.533&0.298&0.344&0.266&0.294&0.306&0.765&0.339&0.542&0.725&0.609&0.298&0.864&0.436&0.623&0.410&0.282&0.283\\%
AVG BATCHES&0.265&0.248&0.393&0.269&0.402&0.252&0.321&0.258&0.296&0.263&0.560&0.616&0.427&0.380&0.444&0.340&0.792&0.330&0.838&0.319&0.271&0.273\\%
AVG OVERALL&0.263&0.247&0.424&0.269&0.414&0.251&0.322&0.259&0.297&0.262&0.598&0.749&0.456&0.413&0.490&0.345&0.836&0.333&0.863&0.321&0.271&0.273\\\bottomrule%
\end{tabular}%
\end{adjustbox}%
\label{tab:DS13_CNN_Glucose}
\vspace{-5mm}
\end{table}
\begin{table}[h!]%
\caption{CNN models N RMSE comparison for (STD, STD+RES) on DS2 Glucose}%
\renewcommand*{\MinNumber}{0.28600000000000003}%
\renewcommand*{\MidNumber}{0.415}%
\renewcommand*{\MaxNumber}{0.488}%
\centering%
\begin{adjustbox}{width=0.98\textwidth}%
\begin{tabular}{@{}lR@{\hskip-1.0\tabcolsep}RR@{\hskip-1.0\tabcolsep}RR@{\hskip-1.0\tabcolsep}RR@{\hskip-1.0\tabcolsep}RR@{\hskip-1.0\tabcolsep}RR@{\hskip-1.0\tabcolsep}RR@{\hskip-1.0\tabcolsep}RR@{\hskip-1.0\tabcolsep}RR@{\hskip-1.0\tabcolsep}RR@{\hskip-1.0\tabcolsep}RR@{\hskip-1.0\tabcolsep}R@{}}%
\toprule%
\multirow{2}{*}{TEST BATCHES}&\multicolumn{2}{c}{ vanilla}&\multicolumn{2}{c}{ selu}&\multicolumn{2}{c}{ selu sh}&\multicolumn{2}{c}{ lrelu}&\multicolumn{2}{c}{ lrelu sh}&\multicolumn{2}{c}{ vanilla }&\multicolumn{2}{c}{ selu }&\multicolumn{2}{c}{ selu sh }&\multicolumn{2}{c}{ lrelu }&\multicolumn{2}{c}{ lrelu sh }&\multicolumn{2}{c}{Benchmark}\\%
&STD&RES&STD&RES&STD&RES&STD&RES&STD&RES&STD&RES&STD&RES&STD&RES&STD&RES&STD&RES&STD&RES\\%
\midrule%
\EndTableHeader%
97&0.536&0.556&0.762&0.601&0.850&0.495&0.983&0.735&0.702&0.689&0.886&1.031&0.821&0.869&0.843&0.965&0.885&0.902&0.862&0.917&0.488&0.486\\%
98&1.097&0.849&0.737&0.470&0.432&0.556&0.311&0.325&0.362&0.419&0.869&0.864&1.062&1.132&0.874&1.213&1.724&1.789&1.345&1.969&0.425&0.585\\%
99&0.389&0.325&0.560&0.432&0.510&0.461&0.443&0.379&0.379&0.340&0.849&0.811&0.805&0.951&0.824&0.988&0.695&1.003&0.821&0.956&0.286&0.399\\%
101&0.489&0.446&0.572&0.535&0.622&0.612&0.502&0.517&0.478&0.623&1.004&1.960&1.889&1.922&1.456&1.262&2.140&1.426&1.131&1.318&0.484&0.434\\%
152&0.468&0.634&0.705&0.672&0.575&0.542&0.451&0.356&0.574&0.359&1.105&0.989&1.052&1.130&1.223&1.162&0.970&1.057&1.009&1.006&0.391&0.815\\%
154&0.493&0.385&0.485&0.624&0.474&0.395&0.520&0.465&0.525&0.453&0.901&1.117&0.868&1.212&0.958&1.064&0.945&1.392&0.972&0.912&0.406&0.396\\%
AVG BATCHES&0.579&0.533&0.637&0.556&0.577&0.510&0.535&0.463&0.503&0.480&0.936&1.129&1.083&1.203&1.030&1.109&1.226&1.262&1.023&1.180&0.413&0.519\\%
AVG OVERALL&0.630&0.553&0.644&0.541&0.604&0.505&0.605&0.499&0.513&0.507&0.906&1.147&1.096&1.193&0.999&1.082&1.293&1.267&1.014&1.232&0.415&0.506\\\bottomrule%
\end{tabular}%
\end{adjustbox}%
\label{tab:TOCI_CNN_Glucose}
\vspace{-5mm}
\end{table}

\clearpage
\subsection{DL Models - VAE}

Tab. \ref{tab:DS13_VAE_Glucose} shows how VAE models performed significantly better than the benchmark on DS1, especially for the RES setting, but also in the NOPP and NOPP+RES settings, which is the first example of VAE Models outperforming the benchmark on the raw spectrum.
Tab. \ref{tab:TOCI_VAE_Glucose} shows the contrary for DS2. This poor performance on DS2 Glucose for all DL models suggests that these models require more training data for the prediction of this target variable (which was not the case for Titer and Lactate).

\begin{table}[h]%
\caption{VAE models N RMSE comparison for (STD, STD+RES) on DS1 Glucose}%
\renewcommand*{\MinNumber}{0.23199999999999998}%
\renewcommand*{\MidNumber}{0.271}%
\renewcommand*{\MaxNumber}{0.29100000000000004}%
\centering%
\begin{adjustbox}{width=0.98\textwidth}%
\begin{tabular}{@{}lR@{\hskip-1.0\tabcolsep}RR@{\hskip-1.0\tabcolsep}RR@{\hskip-1.0\tabcolsep}RR@{\hskip-1.0\tabcolsep}RR@{\hskip-1.0\tabcolsep}RR@{\hskip-1.0\tabcolsep}RR@{\hskip-1.0\tabcolsep}RR@{\hskip-1.0\tabcolsep}RR@{\hskip-1.0\tabcolsep}R@{}}%
\toprule%
\multirow{2}{*}{TEST BATCHES}&\multicolumn{2}{c}{ vanilla}&\multicolumn{2}{c}{ convT}&\multicolumn{2}{c}{ fc dec}&\multicolumn{2}{c}{ fc}&\multicolumn{2}{c}{ vanilla }&\multicolumn{2}{c}{ convT }&\multicolumn{2}{c}{ fc dec }&\multicolumn{2}{c}{ fc }&\multicolumn{2}{c}{Benchmark}\\%
&STD&RES&STD&RES&STD&RES&STD&RES&NOPP&RES&NOPP&RES&NOPP&RES&NOPP&RES&STD&RES\\%
\midrule%
\EndTableHeader%
test split 1&0.231&0.227&0.227&0.234&0.233&0.235&0.228&0.261&0.258&0.248&0.245&0.282&0.265&0.253&0.547&0.658&0.275&0.276\\%
test split 2&0.206&0.215&0.215&0.195&0.206&0.199&0.204&0.214&0.249&0.229&0.236&0.260&0.234&0.264&0.525&0.690&0.273&0.273\\%
test split 3&0.222&0.208&0.202&0.213&0.209&0.198&0.199&0.200&0.236&0.230&0.218&0.241&0.231&0.237&0.584&0.672&0.232&0.232\\%
test split 4&0.261&0.272&0.252&0.252&0.265&0.261&0.255&0.257&0.252&0.277&0.264&0.240&0.270&0.251&0.507&0.529&0.291&0.302\\%
test split 5&0.249&0.245&0.244&0.235&0.249&0.244&0.242&0.253&0.274&0.269&0.255&0.293&0.290&0.289&0.536&0.524&0.282&0.283\\%
AVG BATCHES&0.234&0.233&0.228&0.226&0.232&0.228&0.226&0.237&0.254&0.251&0.244&0.263&0.258&0.259&0.540&0.614&0.271&0.273\\%
AVG OVERALL&0.232&0.232&0.227&0.224&0.231&0.226&0.224&0.236&0.254&0.249&0.243&0.264&0.257&0.260&0.540&0.624&0.271&0.273\\\bottomrule%
\end{tabular}%
\end{adjustbox}%
\label{tab:DS13_VAE_Glucose}
\vspace{-5mm}
\end{table}
\begin{table}[h]%
\caption{VAE models N RMSE comparison for (STD, STD+RES) on DS2 Glucose}%
\renewcommand*{\MinNumber}{0.28600000000000003}%
\renewcommand*{\MidNumber}{0.415}%
\renewcommand*{\MaxNumber}{0.488}%
\centering%
\begin{adjustbox}{width=0.98\textwidth}%
\begin{tabular}{@{}lR@{\hskip-1.0\tabcolsep}RR@{\hskip-1.0\tabcolsep}RR@{\hskip-1.0\tabcolsep}RR@{\hskip-1.0\tabcolsep}RR@{\hskip-1.0\tabcolsep}RR@{\hskip-1.0\tabcolsep}RR@{\hskip-1.0\tabcolsep}RR@{\hskip-1.0\tabcolsep}RR@{\hskip-1.0\tabcolsep}R@{}}%
\toprule%
\multirow{2}{*}{TEST BATCHES}&\multicolumn{2}{c}{ vanilla}&\multicolumn{2}{c}{ convT}&\multicolumn{2}{c}{ fc dec}&\multicolumn{2}{c}{ fc}&\multicolumn{2}{c}{ vanilla }&\multicolumn{2}{c}{ convT }&\multicolumn{2}{c}{ fc dec }&\multicolumn{2}{c}{ fc }&\multicolumn{2}{c}{Benchmark}\\%
&STD&RES&STD&RES&STD&RES&STD&RES&STD&RES&STD&RES&STD&RES&STD&RES&STD&RES\\%
\midrule%
\EndTableHeader%
97&0.574&0.600&0.483&0.611&0.591&0.652&0.348&0.538&0.838&0.897&1.007&0.889&0.904&0.781&1.270&0.926&0.488&0.486\\%
98&0.551&0.574&0.570&0.477&0.590&0.483&0.508&0.392&0.824&0.826&0.817&0.897&1.030&0.957&0.920&0.972&0.425&0.585\\%
99&0.406&0.332&0.394&0.348&0.394&0.366&0.425&0.450&0.967&0.916&0.845&0.684&0.855&0.493&0.963&0.820&0.286&0.399\\%
101&0.523&0.512&0.461&0.469&0.529&0.512&0.553&0.592&1.121&1.569&1.005&2.391&1.352&1.481&1.021&1.478&0.484&0.434\\%
152&0.347&0.888&0.319&0.493&0.346&0.718&0.340&0.727&1.166&1.249&1.032&1.025&1.072&1.054&1.980&1.099&0.391&0.815\\%
154&0.561&0.453&0.455&0.413&0.454&0.400&0.413&0.422&1.009&0.963&0.890&0.868&0.911&0.758&0.918&1.020&0.406&0.396\\%
AVG BATCHES&0.494&0.560&0.447&0.469&0.484&0.522&0.431&0.520&0.987&1.070&0.933&1.126&1.020&0.921&1.179&1.052&0.413&0.519\\%
AVG OVERALL&0.504&0.549&0.457&0.476&0.501&0.521&0.432&0.506&0.952&1.045&0.916&1.200&0.999&0.920&1.152&1.030&0.415&0.506\\\bottomrule%
\end{tabular}%
\end{adjustbox}%
\label{tab:TOCI_VAE_Glucose}
\vspace{-5mm}
\end{table}

\subsection{Fused Models}

For Glucose, FWLS fusion yielded overalll a lower N RMSE than the individual models and performed better than BLUE fusion.

\begin{table}[h!]%
\caption{N RMSE comparison for FWLS and BLUE fusion of XGB, GP and VAE vanilla on DS1 Glucose}%
\renewcommand*{\MinNumber}{0.225}%
\renewcommand*{\MidNumber}{0.263}%
\renewcommand*{\MaxNumber}{0.291}%
\centering%
\begin{adjustbox}{width=0.68\textwidth}%
\begin{tabular}{lRRRRRR}
\toprule
TEST BATCHES &   XGB &   GP &  VAE vanilla & FWLS & BLUE & Benchmark\\
\midrule
\EndTableHeader
test split 1      & 0.254 & 0.235 &  0.229 &    0.232 & 0.253 & 0.271 \\
test split 2      & 0.219 & 0.222 &  0.209 &    0.197 & 0.217 & 0.265 \\
test split 3      & 0.222 & 0.222 &  0.218 &    0.208 & 0.219 & 0.225 \\
test split 4      & 0.322 & 0.263 &  0.261 &    0.287 & 0.320 & 0.291 \\
test split 5      & 0.273 & 0.229 &  0.257 &    0.244 & 0.269 & 0.265 \\
 AVG BATCHES      & 0.258 & 0.234 &  0.235 &    0.233 & 0.256 & 0.263 \\
 AVG OVERALL      & 0.256 & 0.233 &  0.233 &    0.231 & 0.254 & 0.263 \\
\bottomrule
\end{tabular}
\end{adjustbox}%
\label{tab:fusionDS13Glucose}
\end{table}

\bigbreak

Overall, the FWLS fused model was once again amongst the best performing models on DS1 in the STD setting, this time for Glucose, together with GP and VAE models, achieving an improvement of 15\% vs the benchmark.

\clearpage
\section{VCD}

\subsection{ML models}

VCD was a particularly difficult prediction target, especially for ML models. In this group, XGB and GP showed best results for DS1. Tab. \ref{tab:DS13_ML_VCD_AUG} shows that augmentation did not influence the results for this taget variable.
Dimensional reduction was only beneficial for SVR, allowing it to beat the benchmark, but not the other models in the STD setting (Tab. \ref{tab:DS13_ML_VCD_VIP}).\\
On DS2, GP showed best results, followed by SVR.


\begin{table}[h!]%
\caption{ML models N RMSE comparison for (STD, STD+AUG, STD+RES) on DS1 VCD}%
\renewcommand*{\MinNumber}{0.287}%
\renewcommand*{\MidNumber}{0.354}%
\renewcommand*{\MaxNumber}{0.429}%
\centering%
\begin{adjustbox}{width=0.98\textwidth}%
\begin{tabular}{@{}lR@{\hskip-1.5\tabcolsep}R@{\hskip-1.5\tabcolsep}RR@{\hskip-1.5\tabcolsep}R@{\hskip-1.5\tabcolsep}RR@{\hskip-1.5\tabcolsep}R@{\hskip-1.5\tabcolsep}RR@{\hskip-1.5\tabcolsep}R@{\hskip-1.5\tabcolsep}RR@{\hskip-1.5\tabcolsep}R@{\hskip-1.5\tabcolsep}R@{}}%
\toprule%
\multirow{2}{*}{TEST BATCHES}&\multicolumn{3}{c}{XGB}&\multicolumn{3}{c}{GP}&\multicolumn{3}{c}{SVR}&\multicolumn{3}{c}{PLS}&\multicolumn{3}{c}{Benchmark}\\%
&STD&AUG&RES&STD&AUG&RES&STD&AUG&RES&STD&AUG&RES&STD&AUG&RES\\%
\midrule%
\EndTableHeader%
test split 1&0.317&0.313&0.307&0.257&0.255&0.251&0.392&0.374&0.373&0.350&0.351&0.350&0.350&0.351&0.350\\%
test split 2&0.396&0.413&0.392&0.342&0.345&0.336&0.427&0.420&0.435&0.429&0.428&0.436&0.429&0.428&0.436\\%
test split 3&0.199&0.205&0.211&0.270&0.268&0.277&0.257&0.254&0.309&0.346&0.329&0.348&0.346&0.329&0.348\\%
test split 4&0.246&0.258&0.265&0.305&0.309&0.297&0.383&0.381&0.375&0.287&0.302&0.279&0.287&0.302&0.279\\%
test split 5&0.257&0.264&0.268&0.366&0.369&0.361&0.291&0.292&0.303&0.316&0.308&0.316&0.316&0.308&0.316\\%
AVG BATCHES&0.283&0.291&0.289&0.308&0.309&0.304&0.350&0.344&0.359&0.346&0.343&0.346&0.346&0.343&0.346\\%
AVG OVERALL&0.302&0.310&0.304&0.311&0.313&0.307&0.363&0.356&0.368&0.354&0.352&0.355&0.354&0.352&0.355\\\bottomrule%
\end{tabular}%
\end{adjustbox}%
\label{tab:DS13_ML_VCD_AUG}
\vspace{-5mm}
\end{table}
\begin{table}[h!]%
\caption{ML models N RMSE comparison for (STD, STD+512, STD+256, STD+128) on DS1 VCD}%
\renewcommand*{\MinNumber}{0.287}%
\renewcommand*{\MidNumber}{0.354}%
\renewcommand*{\MaxNumber}{0.429}%
\centering%
\begin{adjustbox}{width=0.98\textwidth}%
\begin{tabular}{@{}lR@{\hskip-2.0\tabcolsep}R@{\hskip-2.0\tabcolsep}R@{\hskip-2.0\tabcolsep}RR@{\hskip-2.0\tabcolsep}R@{\hskip-2.0\tabcolsep}R@{\hskip-2.0\tabcolsep}RR@{\hskip-2.0\tabcolsep}R@{\hskip-2.0\tabcolsep}R@{\hskip-2.0\tabcolsep}RR@{\hskip-2.0\tabcolsep}R@{\hskip-2.0\tabcolsep}R@{\hskip-2.0\tabcolsep}RR@{\hskip-2.0\tabcolsep}R@{\hskip-2.0\tabcolsep}R@{\hskip-2.0\tabcolsep}R@{}}%
\toprule%
\multirow{2}{*}{TEST BATCHES}&\multicolumn{4}{c}{XGB}&\multicolumn{4}{c}{GP}&\multicolumn{4}{c}{SVR}&\multicolumn{4}{c}{PLS}&\multicolumn{4}{c}{Benchmark}\\%
&STD&512&256&128&STD&512&256&128&STD&512&256&128&STD&512&256&128&STD&512&256&128\\%
\midrule%
\EndTableHeader%
test split 1&0.317&0.300&0.355&0.357&0.257&0.257&0.305&1.533&0.392&0.368&0.343&0.341&0.350&0.376&0.377&0.394&0.350&0.376&0.377&0.394\\%
test split 2&0.396&0.375&0.415&0.405&0.342&0.368&0.449&1.607&0.427&0.398&0.400&0.392&0.429&0.415&0.429&0.430&0.429&0.415&0.429&0.430\\%
test split 3&0.199&0.180&0.244&0.272&0.270&0.287&0.296&1.627&0.257&0.244&0.227&0.228&0.346&0.425&0.329&0.362&0.346&0.425&0.329&0.362\\%
test split 4&0.246&0.247&0.313&0.322&0.305&0.308&0.295&1.593&0.383&0.376&0.320&0.331&0.287&0.279&0.384&0.409&0.287&0.279&0.384&0.409\\%
test split 5&0.257&0.270&0.256&0.283&0.366&0.421&0.484&0.508&0.291&0.294&0.329&0.293&0.316&0.384&0.341&0.371&0.316&0.384&0.341&0.371\\%
AVG BATCHES&0.283&0.274&0.317&0.328&0.308&0.328&0.366&1.374&0.350&0.336&0.324&0.317&0.346&0.376&0.372&0.393&0.346&0.376&0.372&0.393\\%
AVG OVERALL&0.302&0.291&0.332&0.338&0.311&0.334&0.380&1.440&0.363&0.347&0.336&0.329&0.354&0.380&0.377&0.396&0.354&0.380&0.377&0.396\\\bottomrule%
\end{tabular}%
\end{adjustbox}%
\vspace{-5mm}
\label{tab:DS13_ML_VCD_VIP}
\end{table}
\begin{table}[h!]%
\caption{ML models N RMSE comparison for (STD) on DS2 VCD}%
\renewcommand*{\MinNumber}{0.262}%
\renewcommand*{\MidNumber}{0.505}%
\renewcommand*{\MaxNumber}{0.75}%
\centering%
\begin{adjustbox}{width=0.48\textwidth}%
\begin{tabular}{@{}lRRRRR@{}}%
\toprule%
\multirow{2}{*}{TEST BATCHES}&\multicolumn{1}{c}{XGB}&\multicolumn{1}{c}{GP}&\multicolumn{1}{c}{SVR}&\multicolumn{1}{c}{PLS}&\multicolumn{1}{c}{Benchmark}\\%
&STD&STD&STD&STD&STD\\%
\midrule%
\EndTableHeader%
97&0.356&0.222&0.266&0.453&0.453\\%
98&0.240&0.331&0.420&0.609&0.609\\%
99&0.212&0.176&0.345&0.262&0.262\\%
101&1.177&0.771&0.693&0.750&0.750\\%
152&0.340&0.463&0.718&0.344&0.344\\%
154&0.360&0.449&0.560&0.563&0.563\\%
AVG BATCHES&0.447&0.402&0.500&0.497&0.497\\%
AVG OVERALL&0.514&0.412&0.489&0.505&0.505\\\bottomrule%
\end{tabular}%
\end{adjustbox}%
\vspace{-5mm}
\label{tab:TOCI_ML_VCD}
\end{table}

\clearpage
\vspace{-5mm}
\subsection{DL Models - KT}

Tab. \ref{tab:DS13_KT_VCD}: KT VIP 512 Top 1 once again failed to converge, while the other top models showed significantly better performance than the benchmark (20\% lower error for STD and 25\% lower error for RES). Tab \ref{tab:TOCI_KT_VCD}, shows similar behaviour for DS2 (up to 65\% lower error for STD and 50\% lower error for RES)

\begin{table}[h]%
\caption{KT models N RMSE comparison for (512, 512+RES) on DS1 VCD}%
\renewcommand*{\MinNumber}{0.27899999999999997}%
\renewcommand*{\MidNumber}{0.38}%
\renewcommand*{\MaxNumber}{0.425}%
\centering%
\begin{adjustbox}{width=0.98\textwidth}%
\begin{tabular}{@{}lR@{\hskip-1.0\tabcolsep}RR@{\hskip-1.0\tabcolsep}RR@{\hskip-1.0\tabcolsep}RR@{\hskip-1.0\tabcolsep}RR@{\hskip-1.0\tabcolsep}RR@{\hskip-1.0\tabcolsep}R@{}}%
\toprule%
\multirow{2}{*}{TEST BATCHES}&\multicolumn{2}{c}{top 1}&\multicolumn{2}{c}{top 2}&\multicolumn{2}{c}{top 3}&\multicolumn{2}{c}{top 4}&\multicolumn{2}{c}{top 5}&\multicolumn{2}{c}{Benchmark}\\%
&512&RES&512&RES&512&RES&512&RES&512&RES&512&RES\\%
\midrule%
\EndTableHeader%
test split 1&0.660&0.509&0.228&0.268&0.245&0.227&0.258&0.268&0.215&0.196&0.372&0.340\\%
test split 2&0.684&0.567&0.344&0.362&0.391&0.337&0.360&0.348&0.412&0.337&0.415&0.436\\%
test split 3&0.560&0.413&0.230&0.250&0.244&0.231&0.264&0.256&0.291&0.261&0.425&0.348\\%
test split 4&0.546&0.534&0.278&0.234&0.216&0.234&0.230&0.244&0.254&0.260&0.279&0.279\\%
test split 5&0.570&0.501&0.300&0.286&0.355&0.260&0.326&0.239&0.420&0.403&0.384&0.316\\%
AVG BATCHES&0.604&0.505&0.276&0.280&0.290&0.258&0.288&0.271&0.318&0.291&0.375&0.344\\%
AVG OVERALL&0.614&0.512&0.284&0.289&0.306&0.266&0.297&0.279&0.334&0.302&0.380&0.353\\\bottomrule%
\end{tabular}%
\end{adjustbox}%
\label{tab:DS13_KT_VCD}
\vspace{-5mm}
\end{table}

\begin{table}[h!]%
\caption{KT models N RMSE comparison for (512, 512+RES) on DS2 VCD}%
\renewcommand*{\MinNumber}{0.262}%
\renewcommand*{\MidNumber}{0.505}%
\renewcommand*{\MaxNumber}{0.75}%
\centering%
\begin{adjustbox}{width=0.98\textwidth}%
\begin{tabular}{@{}lR@{\hskip-1.0\tabcolsep}RR@{\hskip-1.0\tabcolsep}RR@{\hskip-1.0\tabcolsep}RR@{\hskip-1.0\tabcolsep}RR@{\hskip-1.0\tabcolsep}RR@{\hskip-1.0\tabcolsep}R@{}}%
\toprule%
\multirow{2}{*}{TEST BATCHES}&\multicolumn{2}{c}{top 1}&\multicolumn{2}{c}{top 2}&\multicolumn{2}{c}{top 3}&\multicolumn{2}{c}{top 4}&\multicolumn{2}{c}{top 5}&\multicolumn{2}{c}{Benchmark}\\%
&512&RES&512&RES&512&RES&512&RES&512&RES&STD&RES\\%
\midrule%
\EndTableHeader%
97&0.927&1.057&0.201&0.193&0.259&0.209&0.313&0.195&0.346&0.259&0.453&0.408\\%
98&0.894&0.826&0.066&0.317&0.069&0.319&0.109&0.350&0.085&0.285&0.609&0.371\\%
99&0.867&0.851&0.079&0.126&0.095&0.129&0.124&0.173&0.009&0.128&0.262&0.255\\%
101&0.663&0.690&0.086&0.599&0.116&0.579&0.078&0.576&0.113&0.422&0.750&0.579\\%
152&0.941&1.126&0.199&0.224&0.162&0.172&0.169&0.269&1.127&0.350&0.344&0.430\\%
154&1.147&1.000&0.368&0.082&0.470&0.045&0.477&0.074&1.099&0.008&0.563&0.585\\%
AVG BATCHES&0.906&0.925&0.166&0.257&0.195&0.242&0.212&0.273&0.463&0.242&0.497&0.438\\%
AVG OVERALL&0.899&0.915&0.185&0.291&0.226&0.283&0.244&0.301&0.584&0.264&0.505&0.430\\\bottomrule%
\end{tabular}%
\end{adjustbox}%
\label{tab:TOCI_KT_VCD}
\vspace{-5mm}
\end{table}

\subsection{DL Models - CNN}

CNN models have shown better performance than the KT VIP 512 models for STD+RES, and slightly better results for the NOPP+RES setting than the benchmark in the STD setting (Tab. \ref{tab:DS13_CNN_VCD}).\\
On DS2, surprisingly, the CNN performed better on the raw spectrum (Tab. \ref{tab:TOCI_CNN_VCD}, outperforming the benchmark for the NOPP and NOPP+RES settings, but not the KT VIP 512 models.

\begin{table}[h]%
\caption{CNN models N RMSE comparison for (STD, STD+RES) on DS1 VCD}%
\renewcommand*{\MinNumber}{0.287}%
\renewcommand*{\MidNumber}{0.35200000000000004}%
\renewcommand*{\MaxNumber}{0.429}%
\centering%
\begin{adjustbox}{width=0.98\textwidth}%
\begin{tabular}{@{}lR@{\hskip-1.0\tabcolsep}RR@{\hskip-1.0\tabcolsep}RR@{\hskip-1.0\tabcolsep}RR@{\hskip-1.0\tabcolsep}RR@{\hskip-1.0\tabcolsep}RR@{\hskip-1.0\tabcolsep}RR@{\hskip-1.0\tabcolsep}RR@{\hskip-1.0\tabcolsep}RR@{\hskip-1.0\tabcolsep}RR@{\hskip-1.0\tabcolsep}RR@{\hskip-1.0\tabcolsep}R@{}}%
\toprule%
\multirow{2}{*}{TEST BATCHES}&\multicolumn{2}{c}{ vanilla}&\multicolumn{2}{c}{ selu}&\multicolumn{2}{c}{ selu sh}&\multicolumn{2}{c}{ lrelu}&\multicolumn{2}{c}{ lrelu sh}&\multicolumn{2}{c}{ vanilla }&\multicolumn{2}{c}{ selu }&\multicolumn{2}{c}{ selu sh }&\multicolumn{2}{c}{ lrelu }&\multicolumn{2}{c}{ lrelu sh }&\multicolumn{2}{c}{Benchmark}\\%
&STD&RES&STD&RES&STD&RES&STD&RES&STD&RES&NOPP&RES&NOPP&RES&NOPP&RES&NOPP&RES&NOPP&RES&STD&RES\\%
\midrule%
\EndTableHeader%
test split 1&0.302&0.203&0.258&0.240&0.269&0.221&0.349&0.248&0.272&0.226&0.442&0.332&0.425&0.332&0.337&0.325&0.456&0.421&0.476&0.432&0.340&0.340\\%
test split 2&0.449&0.380&0.411&0.362&0.383&0.364&0.374&0.343&0.459&0.362&0.563&0.385&0.481&0.457&0.543&0.489&0.547&0.567&0.640&0.595&0.429&0.436\\%
test split 3&0.233&0.224&0.299&0.240&0.257&0.252&0.248&0.238&0.267&0.266&0.421&0.395&0.299&0.250&0.432&0.293&0.469&0.438&0.478&0.424&0.346&0.348\\%
test split 4&0.268&0.245&0.267&0.255&0.242&0.290&0.258&0.241&0.284&0.265&0.409&0.326&0.363&0.293&0.311&0.317&0.432&0.424&0.446&0.434&0.287&0.279\\%
test split 5&0.365&0.241&0.248&0.198&0.263&0.205&0.266&0.235&0.281&0.213&0.481&0.351&0.514&0.311&0.312&0.254&0.456&0.363&0.551&0.331&0.316&0.316\\%
AVG BATCHES&0.323&0.259&0.297&0.259&0.283&0.266&0.299&0.261&0.313&0.266&0.463&0.358&0.416&0.329&0.387&0.335&0.472&0.442&0.518&0.443&0.344&0.344\\%
AVG OVERALL&0.342&0.273&0.309&0.271&0.294&0.277&0.311&0.270&0.331&0.276&0.473&0.358&0.431&0.347&0.405&0.356&0.478&0.454&0.531&0.460&0.352&0.353\\\bottomrule%
\end{tabular}%
\end{adjustbox}%
\label{tab:DS13_CNN_VCD}
\vspace{-5mm}
\end{table}

\begin{table}[h!]%
\caption{CNN models N RMSE comparison for (STD, STD+RES) on DS2 VCD}%
\renewcommand*{\MinNumber}{0.262}%
\renewcommand*{\MidNumber}{0.505}%
\renewcommand*{\MaxNumber}{0.75}%
\centering%
\begin{adjustbox}{width=0.98\textwidth}%
\begin{tabular}{@{}lR@{\hskip-1.0\tabcolsep}RR@{\hskip-1.0\tabcolsep}RR@{\hskip-1.0\tabcolsep}RR@{\hskip-1.0\tabcolsep}RR@{\hskip-1.0\tabcolsep}RR@{\hskip-1.0\tabcolsep}RR@{\hskip-1.0\tabcolsep}RR@{\hskip-1.0\tabcolsep}RR@{\hskip-1.0\tabcolsep}RR@{\hskip-1.0\tabcolsep}RR@{\hskip-1.0\tabcolsep}R@{}}%
\toprule%
\multirow{2}{*}{TEST BATCHES}&\multicolumn{2}{c}{ vanilla}&\multicolumn{2}{c}{ selu}&\multicolumn{2}{c}{ selu sh}&\multicolumn{2}{c}{ lrelu}&\multicolumn{2}{c}{ lrelu sh}&\multicolumn{2}{c}{ vanilla }&\multicolumn{2}{c}{ selu }&\multicolumn{2}{c}{ selu sh }&\multicolumn{2}{c}{ lrelu }&\multicolumn{2}{c}{ lrelu sh }&\multicolumn{2}{c}{Benchmark}\\%
&STD&RES&STD&RES&STD&RES&STD&RES&STD&RES&NOPP&RES&NOPP&RES&NOPP&RES&NOPP&RES&NOPP&RES&STD&RES\\%
\midrule%
\EndTableHeader%
97&0.200&0.242&0.245&0.180&0.246&0.169&0.243&0.255&0.166&0.175&0.288&0.311&0.290&0.189&0.344&0.157&0.245&0.227&0.271&0.205&0.453&0.408\\%
98&0.328&1.633&0.161&0.166&0.147&0.171&0.509&0.383&0.515&0.290&0.702&0.459&0.237&0.360&0.272&0.313&0.745&0.538&0.484&0.765&0.609&0.371\\%
99&0.386&0.200&0.219&0.154&0.213&0.134&0.162&0.205&0.210&0.213&0.290&0.361&0.176&0.151&0.198&0.173&0.163&0.184&0.142&0.171&0.262&0.255\\%
101&0.588&0.509&0.719&0.677&0.671&0.633&0.702&0.814&0.732&0.816&0.776&0.690&0.735&0.707&0.735&0.782&1.076&1.141&1.105&1.045&0.750&0.579\\%
152&1.764&0.165&0.574&0.295&0.695&0.326&0.611&0.537&0.604&0.685&0.552&0.568&0.265&0.238&0.246&0.289&0.319&0.324&0.276&0.259&0.344&0.430\\%
154&0.720&0.639&0.545&0.435&0.588&0.384&0.559&0.294&0.982&0.341&0.626&0.488&0.409&0.279&0.378&0.229&0.485&0.227&0.459&0.162&0.563&0.585\\%
AVG BATCHES&0.664&0.565&0.410&0.318&0.427&0.303&0.465&0.415&0.535&0.420&0.539&0.480&0.352&0.321&0.362&0.324&0.506&0.440&0.456&0.434&0.497&0.438\\%
AVG OVERALL&0.734&0.775&0.420&0.339&0.436&0.320&0.467&0.429&0.559&0.439&0.543&0.469&0.371&0.348&0.381&0.360&0.567&0.517&0.516&0.531&0.505&0.430\\\bottomrule%
\end{tabular}%
\end{adjustbox}%
\label{tab:TOCI_CNN_VCD}
\vspace{-5mm}
\end{table}

\clearpage
\subsection{DL Models - VAE}

For STD and RES settings, VAEs outperformed all previous models on both DS1 (Tab. \ref{tab:DS13_VAE_VCD}) and DS2 (Tab. \ref{tab:TOCI_VAE_VCD}). On DS1 resampling did not seem to influence the N RMSE as much as on DS2. Furthermore, the VAE models (except fc) performed well on the raw spectrum data, beating the CNN models and the benchmark for the non-resampled training set.

\begin{table}[h]%
\caption{VAE models N RMSE comparison for (STD, STD+RES) on DS1 VCD}%
\renewcommand*{\MinNumber}{0.287}%
\renewcommand*{\MidNumber}{0.35200000000000004}%
\renewcommand*{\MaxNumber}{0.429}%
\centering%
\begin{adjustbox}{width=0.98\textwidth}%
\begin{tabular}{@{}lR@{\hskip-1.0\tabcolsep}RR@{\hskip-1.0\tabcolsep}RR@{\hskip-1.0\tabcolsep}RR@{\hskip-1.0\tabcolsep}RR@{\hskip-1.0\tabcolsep}RR@{\hskip-1.0\tabcolsep}RR@{\hskip-1.0\tabcolsep}RR@{\hskip-1.0\tabcolsep}RR@{\hskip-1.0\tabcolsep}R@{}}%
\toprule%
\multirow{2}{*}{TEST BATCHES}&\multicolumn{2}{c}{ vanilla}&\multicolumn{2}{c}{ convT}&\multicolumn{2}{c}{ fc dec}&\multicolumn{2}{c}{ fc}&\multicolumn{2}{c}{ vanilla }&\multicolumn{2}{c}{ convT }&\multicolumn{2}{c}{ fc dec }&\multicolumn{2}{c}{ fc }&\multicolumn{2}{c}{Benchmark}\\%
&STD&RES&STD&RES&STD&RES&STD&RES&NOPP&RES&NOPP&RES&NOPP&RES&NOPP&RES&STD&RES\\%
\midrule%
\EndTableHeader%
test split 1&0.246&0.246&0.219&0.226&0.230&0.227&0.252&0.224&0.257&0.293&0.294&0.301&0.272&0.362&0.467&0.447&0.340&0.340\\%
test split 2&0.343&0.359&0.359&0.354&0.368&0.354&0.354&0.364&0.451&0.412&0.431&0.438&0.471&0.486&0.958&0.730&0.429&0.436\\%
test split 3&0.256&0.249&0.250&0.241&0.239&0.239&0.252&0.231&0.229&0.249&0.233&0.275&0.228&0.274&0.858&0.536&0.346&0.348\\%
test split 4&0.226&0.236&0.226&0.224&0.234&0.215&0.200&0.217&0.274&0.308&0.312&0.278&0.291&0.288&1.046&0.447&0.287&0.279\\%
test split 5&0.208&0.230&0.200&0.209&0.206&0.227&0.237&0.251&0.274&0.259&0.244&0.254&0.254&0.278&0.435&0.970&0.316&0.316\\%
AVG BATCHES&0.256&0.264&0.251&0.251&0.255&0.252&0.259&0.258&0.297&0.304&0.303&0.309&0.303&0.338&0.753&0.626&0.344&0.344\\%
AVG OVERALL&0.265&0.274&0.263&0.263&0.269&0.264&0.271&0.270&0.318&0.317&0.320&0.325&0.327&0.359&0.787&0.664&0.352&0.353\\\bottomrule%
\end{tabular}%
\end{adjustbox}%
\label{tab:DS13_VAE_VCD}
\vspace{-5mm}
\end{table}

\begin{table}[h]%
\caption{VAE models N RMSE comparison for (STD, STD+RES) on DS2 VCD}%
\renewcommand*{\MinNumber}{0.262}%
\renewcommand*{\MidNumber}{0.505}%
\renewcommand*{\MaxNumber}{0.75}%
\centering%
\begin{adjustbox}{width=0.98\textwidth}%
\begin{tabular}{@{}lR@{\hskip-1.0\tabcolsep}RR@{\hskip-1.0\tabcolsep}RR@{\hskip-1.0\tabcolsep}RR@{\hskip-1.0\tabcolsep}RR@{\hskip-1.0\tabcolsep}RR@{\hskip-1.0\tabcolsep}RR@{\hskip-1.0\tabcolsep}RR@{\hskip-1.0\tabcolsep}RR@{\hskip-1.0\tabcolsep}R@{}}%
\toprule%
\multirow{2}{*}{TEST BATCHES}&\multicolumn{2}{c}{ vanilla}&\multicolumn{2}{c}{ convT}&\multicolumn{2}{c}{ fc dec}&\multicolumn{2}{c}{ fc}&\multicolumn{2}{c}{ vanilla }&\multicolumn{2}{c}{ convT }&\multicolumn{2}{c}{ fc dec }&\multicolumn{2}{c}{ fc }&\multicolumn{2}{c}{Benchmark}\\%
&STD&RES&STD&RES&STD&RES&STD&RES&NOPP&RES&NOPP&RES&NOPP&RES&NOPP&RES&STD&RES\\%
\midrule%
\EndTableHeader%
97&0.163&0.295&0.235&0.216&0.166&0.191&0.147&0.187&0.188&0.185&0.145&0.148&0.161&0.133&1.349&1.407&0.453&0.408\\%
98&0.133&0.308&0.121&0.310&0.132&0.233&0.471&0.412&0.440&0.439&0.410&0.392&0.405&0.393&0.523&0.870&0.609&0.371\\%
99&0.171&0.189&0.155&0.147&0.162&0.174&0.212&0.167&0.178&0.130&0.192&0.138&0.183&0.136&1.010&1.669&0.262&0.255\\%
101&0.592&0.561&0.577&0.623&0.590&0.595&1.250&0.698&0.954&0.595&0.713&0.728&0.717&0.698&0.738&1.416&0.750&0.579\\%
152&0.183&0.352&0.198&0.249&0.197&0.212&0.274&1.010&0.194&0.225&0.197&0.225&0.190&0.175&0.350&1.938&0.344&0.430\\%
154&0.234&0.239&0.221&0.240&0.210&0.225&0.420&0.466&0.387&0.314&0.330&0.289&0.367&0.296&2.000&1.706&0.563&0.585\\%
AVG BATCHES&0.246&0.324&0.251&0.297&0.243&0.272&0.462&0.490&0.390&0.315&0.331&0.320&0.337&0.305&0.995&1.501&0.497&0.438\\%
AVG OVERALL&0.272&0.327&0.274&0.315&0.268&0.289&0.548&0.509&0.445&0.336&0.363&0.355&0.367&0.343&1.116&1.483&0.505&0.430\\\bottomrule%
\end{tabular}%
\end{adjustbox}%
\label{tab:TOCI_VAE_VCD}
\vspace{-5mm}
\end{table}

\subsection{Fusion}

For VCD, model fusion was again performed with XGB, GP adn VAE vanilla models in the STD setting for DS1, but it did not produce better results than the best base model (VAE vanilla). The results are shown in Tab. \ref{tab:fusionDS13VCD}. Nevertheless, it can be noticed that FWLS fusion produced better results than BLUE, but did not manage to beat the best base model. It is possible that a different choice of base models would yield better results. 

\begin{table}[h!]%
\caption{N RMSE comparison for FWLS and BLUE fusion of XGB, GP and VAE vanilla on DS1 Glucose}%
\renewcommand*{\MinNumber}{0.287}%
\renewcommand*{\MidNumber}{0.344}%
\renewcommand*{\MaxNumber}{0.429}%
\centering%
\begin{adjustbox}{width=0.68\textwidth}%
\begin{tabular}{lRRRRRR}
\toprule
TEST BATCHES &   XGB &   GP &  VAE vanilla & FWLS & BLUE & Benchmark\\
\midrule
\EndTableHeader
test split 1      & 0.298 & 0.250 &         0.246 &    0.251    & 0.253     & 0.340  \\
test split 2      & 0.396 & 0.342 &         0.343 &    0.340    & 0.338     & 0.429  \\
test split 3      & 0.199 & 0.270 &         0.256 &    0.219    & 0.205     & 0.346  \\
test split 4      & 0.246 & 0.305 &         0.226 &    0.232    & 0.241     & 0.287  \\
test split 5      & 0.257 & 0.366 &         0.208 &    0.272    & 0.342     & 0.316  \\
 AVG BATCHES      & 0.279 & 0.307 &         0.256 &    0.263    & 0.276     & 0.344  \\
 AVG OVERALL      & 0.297 & 0.311 &         0.265 &    0.272    & 0.287     & 0.352  \\
\bottomrule
\end{tabular}
\end{adjustbox}%
\label{tab:fusionDS13VCD}
\end{table}

Overall, VAEs showed once again the best results on DS1 with up to 25\% lower N RMSE compared to the benchmark in the standard setting, closely followed by CNNs and KT models, which were best on DS2 (65\% lower N RMSE compared to benchmark and 55\% lower compared to the best ML model).

\cleardoublepage
\chapter{Improvements}\label{sec:improvements}

A number of improvement suggestions arose after carefully evaluating each models performance on the different datasets and target variables.
Firstly, the quality of certain batches of DS1 needs to be reassessed. The outlier removal proposed in this work had a positive effect on overall model performance, but it is suggested to enforce a stricter selection for the training, validation and test data. Due to the particular distribution profiles of each target variable and limited amount of training data, random splits for training, validation and tests are not admissible, especially during the hyper parameter search. This randomness may favor models that happen to randomly, i.e. by chance, fit well even during cross-validation but still surpass well generalizing models in the overall ranking, merely because of the high number of hyper parameter trials. This is why the model selection process during hyper parameter tuning should be reevaluated.\\
After that, a new hyper parameter search should be performed for CNN and VAE models. While the hyperparameter space for CNNs is large, but well defined (there exists an excessive amount of guidelines for CNN architectures), VAEs require much more research in this area. An educated guess would be the simplification of the current model, which should reduce the degrees of freedom for the regressor and make the whole regression problem more dependent on the latent part. At first it was assumed that non-linearities in the regressor and latent part of the network would be beneficial for the regression because of the flexibility of non-linear models. However, it was realized that this may quickly lead back to the curse of dimensionality. In fact, the previously shown representations of the test data in the latent space clearly show that the concentrations are linearly separable, therefore the simplest linear regression model should yield decent performance on the latent space.\\
Finally, an additional exploration path may be offered by the application of Generative Adversarial Networks (GANs), which may be implemented using the fundamentals of the developed VAEs.\cite{NIPS2014_5423}

\cleardoublepage
\chapter{Conclusion and Outlook}

The modeling capabilities of selected neural network architectures have been tested on two distinct datasets of Raman spectra for four different target variables, essential to the monitoring of bioprocesses. The majority of the developed models proved to outperform the PLS benchmark as well as other Machine Learning models for selected settings. Tab. 7.1 shows that every proposed model type excelled at least once for a given target variable and setting.\\
Due to the limited amount of training data, which is unusual for classic deep learning applications, different data augmentation and resampling methods were tested.\\
Furthermore, an input feature importance analysis was performed for each target variable which allowed to significantly improve the performance of DL models, and in some cases also of the ML models.\\
Most promising results were achieved for Feed Forward Neural Nets and Variational Auto Encoders. The Feed Forward Networks architecture was optimized using distributed hyperparameter tuning, which allowed to test a total of over 50.000 models in a limited time frame, which allowed to outperform manually designed networks by far.\\
The regression problem was facilitated through VAEs by regressing on a low dimensional latent space, which was regularized by a variational constraint in form of a target-dependent prior. This allowed to obtain a latent space where points with different concentrations were well linearly separated, which suggests a disentanglement of the concentration from other latent variables.\\
Finally, the robustness of the predictions could be further improved with fused models. A Neural Network was used to implement the FWLS method by weighting the contribution of each base model depending on the spectral input.\\
In conclusion, this work demonstrated how Deep Learning methods can be efficiently used for modeling the concentration of different target molecules from Raman spectra and outperform state of the art methods.

\newcommand{\ApplyGradientInv}[1]{%
  \iftoggle{inTableHeader}{#1}{
    \ifdim #1 pt > \MidNumber pt
        \pgfmathsetmacro{\PercentColor}{max(min(100.0*(#1 - \MidNumber)/(\MaxNumber-\MidNumber),100.0),0.00)} %
        \pgfsetfillopacity{0.3}\colorbox{green!\PercentColor!yellow}{\pgfsetfillopacity{1}#1}
    \else
        \pgfmathsetmacro{\PercentColor}{max(min(100.0*(\MidNumber - #1)/(\MidNumber-\MinNumber),100.0),0.00)} %
        \pgfsetfillopacity{0.3}\colorbox{red!\PercentColor!yellow}{\pgfsetfillopacity{1}#1}
    \fi
  }}

\newcolumntype{L}{>{\collectcell\ApplyGradientInv}c<{\endcollectcell}}

\begin{table}[h!]%
\caption{Overall best Model's improvements in \% relative to PLS Benchmark on DS1 and DS2}%
\renewcommand*{\MinNumber}{-100}%
\renewcommand*{\MidNumber}{0}%
\renewcommand*{\MaxNumber}{25}%
\centering%
\begin{adjustbox}{width=0.48\textwidth}%
\begin{tabular}{@{}lLLLLLLLLLLL@{}}%
\toprule%
\multirow{2}{*}{Dataset}&
\multicolumn{2}{c}{VAE}&
\multicolumn{2}{c}{CNN}&
\multicolumn{2}{c}{NN}&
\multicolumn{2}{c}{GP}&
\multicolumn{2}{c}{XGB}&
\multicolumn{1}{c}{Fusion}\\%
&DS1&DS2&DS1&DS2&DS1&DS2&DS1&DS2&DS1&DS2&DS1\\
\midrule%
\EndTableHeader
Titer   & 3&   28&  10&    40&      18&     56&   0&  -4&   -18&  25&   20\\
Lactate & 6&   0&  -2&   -12&        0&     3&    7&   7&   -14&  -50&  11\\
Glucose & 17&  14&   9&   -22&      15&     36&  15&  15&     3&  -9&   12\\
VCD     & 25&  47&  23&    36&      24&     66&  12&  12&    14&  -2&   22\\
\bottomrule
\end{tabular}%
\end{adjustbox}%
\vspace{-5mm}
\label{tab:TOCI_ML_VCD}
\end{table}

\cleardoublepage

\appendix

\chapter{Figures}

\begin{figure}[h!]
    \centering
    \includegraphics[width =0.6\textwidth]{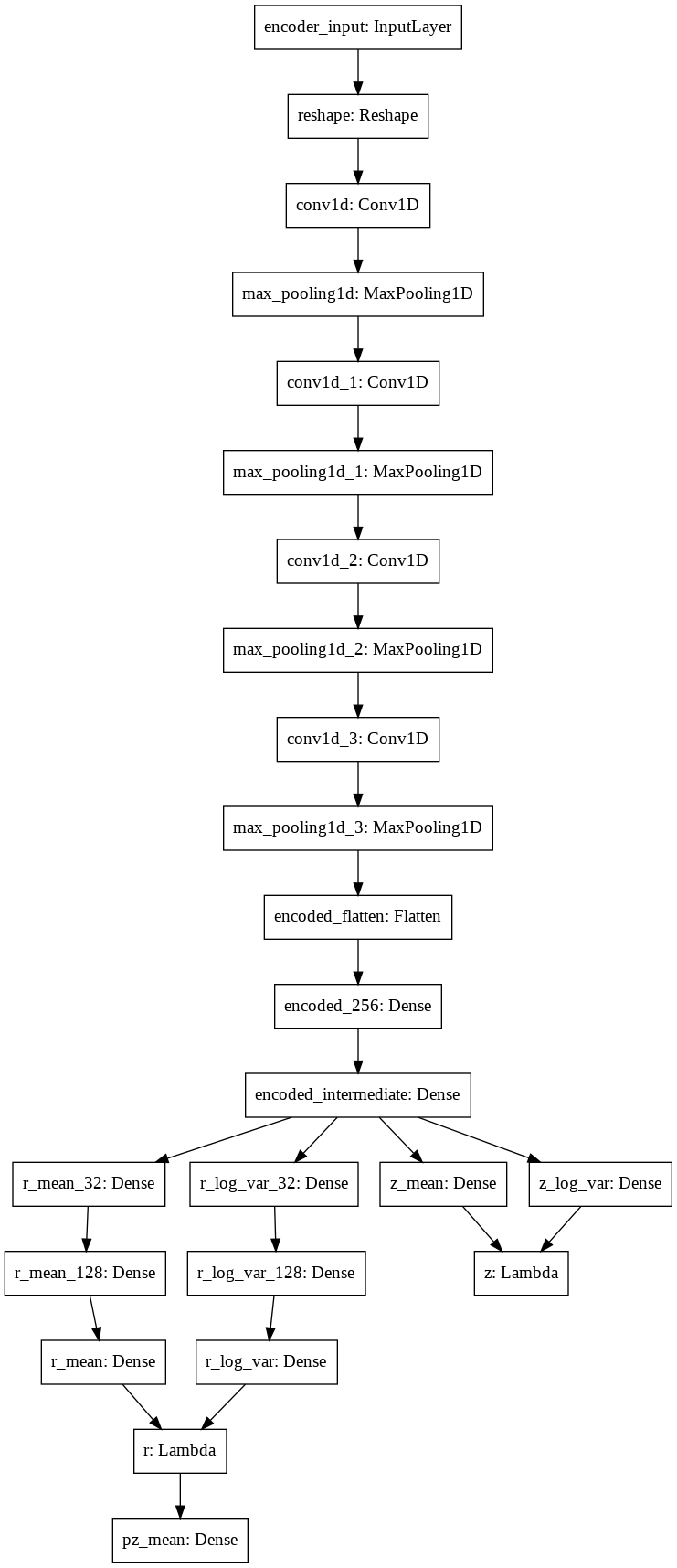}
 \label{img:computation_encoder}
 \caption{The computation graph of the encoder subnetwork of the VAE Regressor model}
\end{figure}

\begin{figure}[h!]
    \centering
    \includegraphics[width =0.3\textwidth]{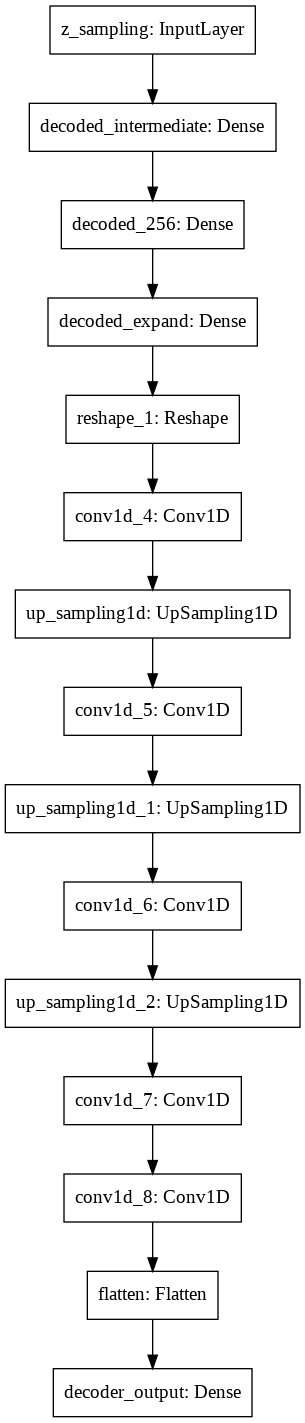}
 \label{img:computation_decoder}
 \caption{The computation graph of the decoder subnetwork of the VAE Regressor model}
\end{figure}

\begin{figure}[h!]
    \includegraphics[width =0.98\textwidth]{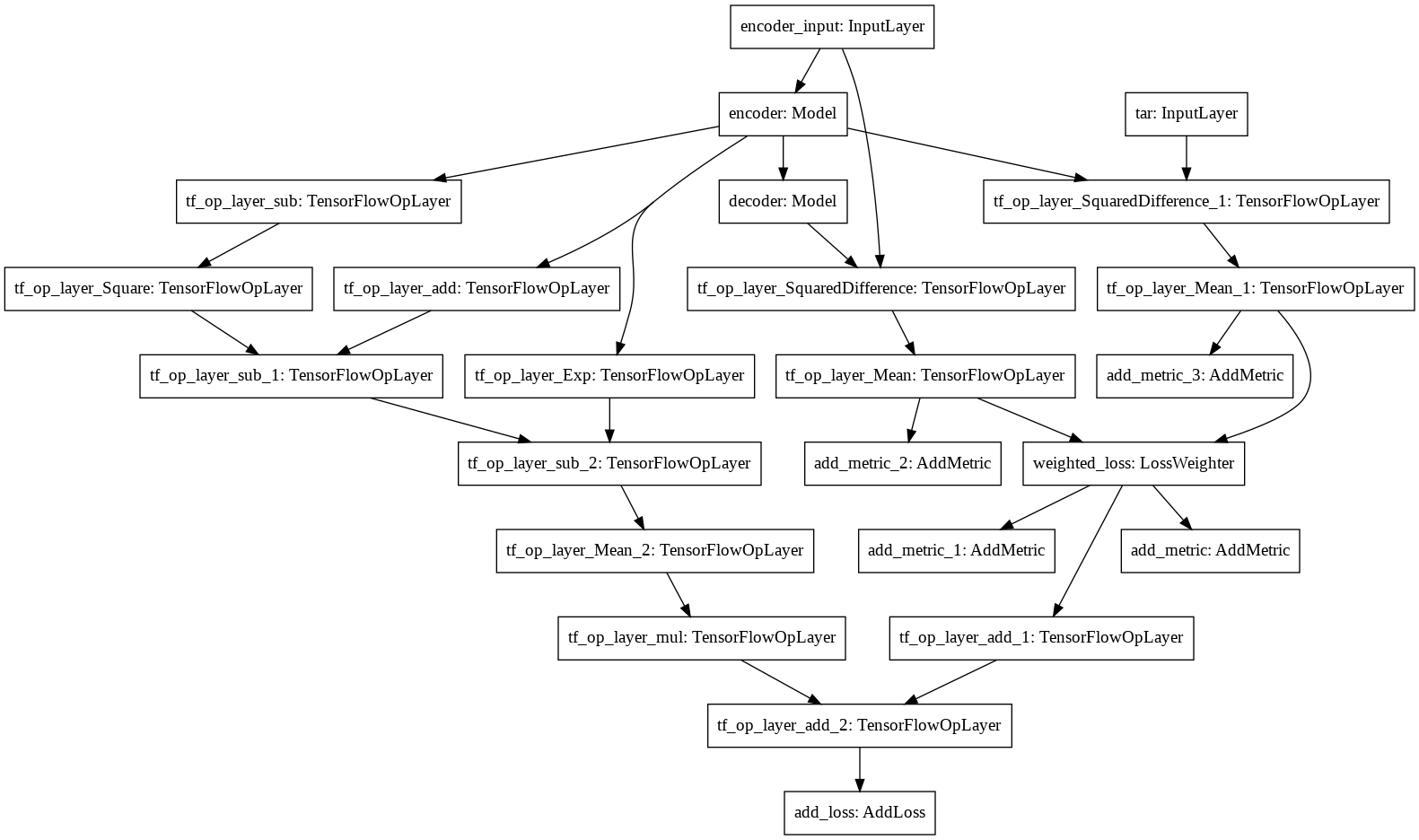}
 \label{img:computation_all}
 \caption{The computation graph of the full VAE Regressor model network including losses and metrics}
\end{figure}


\bibliographystyle{unsrt}
\bibliography{bibliography}

\clearpage
\pagestyle{plain}

\section*{Acknowledgements}
\addcontentsline{toc}{section}{Acknowledgements}

I would like to thank Prof. Gonzalo Guillen Gosalbez and the Institute for Chemical and Bioengineering for hosting me during this project.
Furthermore, I would like to express my gratitude to the whole DataHow team, for allowing me to work on this challenging and multidisciplinary project and sharing there know-how with me.\\
Especially, I would like to thank my supervisors Dr. Michael Sokolov and Dr. Fabian Feidl for their steady support throughout the project.

Additionally, I would like to thank the ID SIS HPC group of ETH for providing the computational power for this work, it would not have been possible to achieve these results without the Euler HPC cluster. At this point, it is also necessary to mention the creators of Cyberduck for their beautiful piece of software, making server storage browsing so much easier. Also, I would like to thank the keras-team for their swift support with keras-tuner.

I express my gratitude to Dr. Jason Brownlee who's guideline articles helped me a lot with design of the deep models. Thanks to him, I learned more about applied deep learning than from 3 years of graduate courses at a top 10 university.

Finally, I would like to thank my family members, for their support and love during this time.

\cleardoublepage



\begin{figure}[p]
    \vspace*{-2cm}
    \makebox[\linewidth]{
        \includegraphics[width=1.3\linewidth]{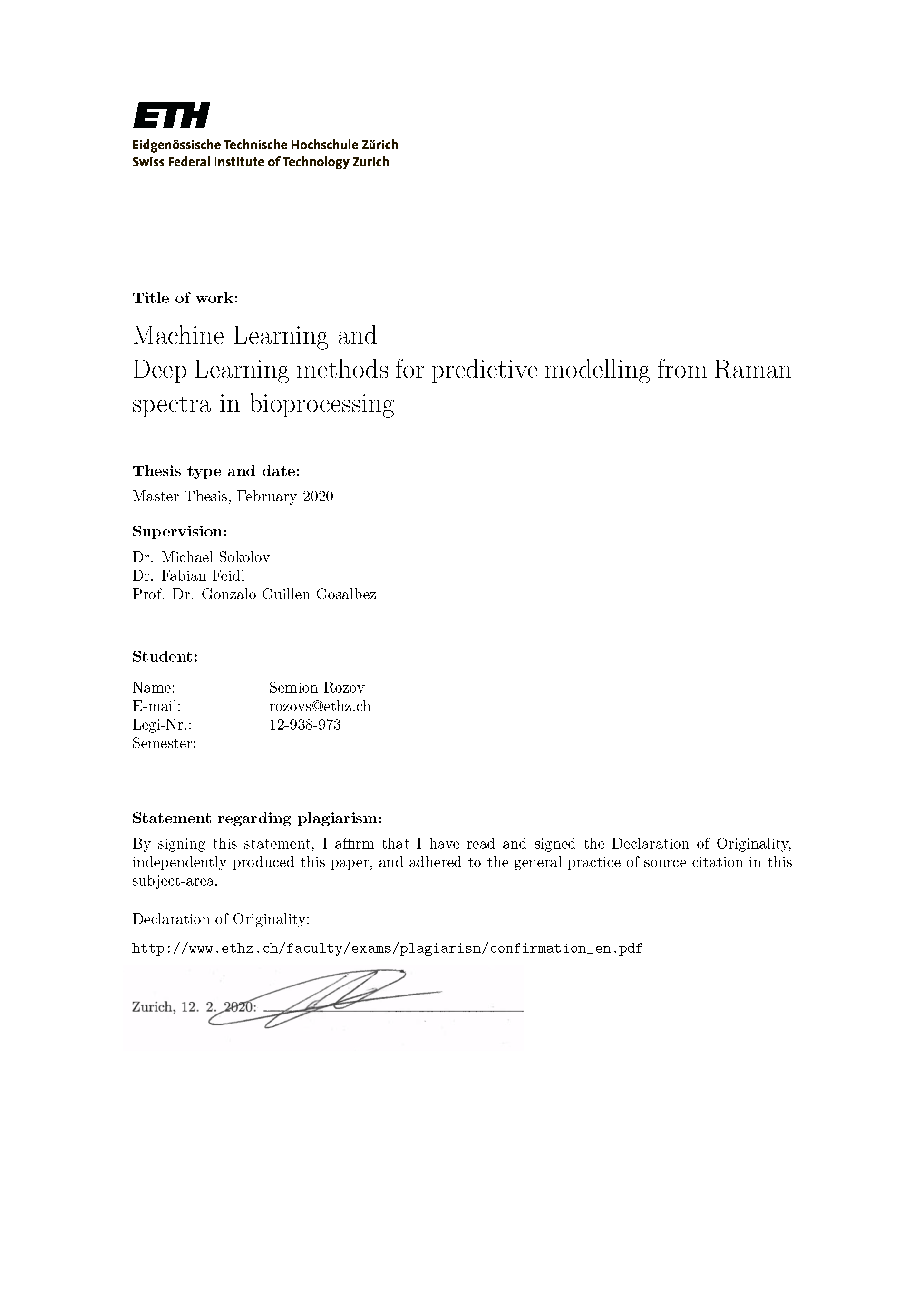}
    }
\end{figure}

\end{document}